%% file: art_fission.tex
\documentclass[dvips]{elsart}
\usepackage{epsfig}
\usepackage{longtable}
\usepackage{graphics}
\usepackage{latexsym}
\usepackage{amssymb}
\begin{document}
\textheight=23 cm
\topmargin=-1cm
\textwidth=16.5 cm
\oddsidemargin=0.5 cm
\setlongtables
\evensidemargin=0.5 cm
\journal{Nucl. Phys. A}
\begin{frontmatter}
\title{ FISSION-RESIDUES PRODUCED IN THE SPALLATION  
REACTION  $^{238}$U~+~p at 1~A~GeV.}
\author[ipn]{M.~Bernas},
\author[gsi]{P.~Armbruster},
\author[gsi,usc]{J.~Benlliure},
\author[sac]{A.~Boudard},
\author[usc]{E.~Casarejos},
\author[bor]{ S. Czajkowski},
\author[gsi]{T.~Enqvist}\footnote{Present address:  University of
Jyv\"askyl\"a, 40351 Jyv\"askyl\"a,  Finland},
\author[sac]{R.~Legrain}$^\dagger$,
\author[sac]{S.~Leray},
\author[ipn]{B.~Mustapha}\footnote{Present address: Argonne National Laboratory, Argonne, IL 60439, USA}
\author[gsi,ipn]{P.~Napolitani},
\author[usc]{J.~Pereira},
\author[gsi,ipn]{F.~Rejmund},
\author[gsi]{M.-V.~Ricciardi},
\author[gsi]{K.-H.~Schmidt},
\author[ipn]{C.~St\'ephan},
\author[gsi,ipn]{J.~Taieb}\footnote{Present address: CEA/Saclay
DM2S/SERMA/LENR, 91191 Gif/Yvette CEDEX, France},
\author[ipn]{L.~Tassan-Got},
\author[sac]{C.~Volant}.

\address[ipn]{Institut de Physique Nucl\'eaire, 91406 Orsay Cedex, France}
\address[gsi]{Gesellschaft f\"ur Schwerionenforschung, Planckstr.~1,
64291~Darmstadt, Germany}
\address[usc]{Universidad de Santiago de Compostela, 15706 Santiago de
Compostela, Spain}
\address[sac]{DAPNIA/SPhN, CEA/Saclay, 91191 Gif sur Yvette Cedex, France}
\address[bor]{CEN Bordeaux-Gradignan, Le Haut-Vigneau, 33175 Gradignan
Cedex, France}

\date{}

\maketitle


\begin{abstract}
Fission fragments from 1 A GeV $^{238}$U projectiles irradiating a hydrogen target 
were investigated by using
the fragment separator FRS for magnetic selection of reaction products
including ray-tracing and $\Delta$E-ToF techniques. The momentum spectra of
733 identified fragments were analysed to provide  isotopic production cross
sections, fission-fragment velocities and recoil
momenta of the fissioning parent nuclei.
Besides their general relevance, these quantities are also demanded for 
applications. Calculations  and simulations with 
codes commonly used and recently developed or improved  are compared to
the data.
\end{abstract}
\end{frontmatter}

\section{Introduction}
Fission of heavy nuclei excited at high collision energies has been investigated for
the last decades. Produced by fragmentation or proton collision, the excited 
nuclei release their energy by emitting neutrons and charged particles, and 
eventually by fission. At low excitation energy, close to the fission barrier
the probability of fission for nuclei around U stays constant or
decreases at increasing excitation energy (first chance fission).
Increasing further the excitation
energy, two opposite tendancies are observed. High excitation energy and
emission of 
charged particles reduce the probability of fission. On the
other hand, neutron emission produces more and more fissile nuclei, enhancing
fission.
For the highest excitation energies, lighter elements which are less
fissile, are created and finally fission vanishes.
The competition between the various 
decay channels provides keys to understand the structure of hot nuclei,
level densities, dissipation and disappearance of shell effects with excitation
energy. 

One of the first experimental approaches to the question of competition of
the different decay-channels was opened with radiochemical and off-line
mass separator 
techniques. G. Friedlander et al. \cite{Frie} studied isotopic production of 
the alkaline
elements rubidium and cesium in p + U and p + W reactions at various energies.
This pioneering work already   characterises the 
evolution of the symmetric breaking of the excited fragmentation residues 
 as a
function of the incident energy. A decade later, similar measurements on
alkalines were
undertaken using on-line mass separators \cite{Klap,Beli}
which we will  use as  a 
basis for comparison in the following.
Meanwhile many results  evaluated in  reference \cite{Prok} 
were obtained to complete these first studies.

The energy dependence of the reactions already studied in 
reference \cite{Frie} must be known and controlled for all
applications. Until now this aspect has been explored by radiochemistry
and $\gamma$-spectroscopy. The systems p + Pb \cite{Glor} and 
p + Bi \cite{Tita} were 
extensively
investigated by using off-line $\gamma$-spectrometry in a wide
range of energies. 

Why do we start new studies on this question ? The new experimental
equipment operating at GSI offers an efficient way to
separate isotopes in inverse kinematics at relativistic energies.
A 1 A GeV U-beam is
delivered at the heavy-ion synchrotron (SIS), and in-flight
separation techniques become applicable to all the energetic U residues.
The inverse kinematics at relativistic energies provides fully ionised 
forward-focussed  fragments which are separated by using the
FRagment Separator (FRS) and identified by energy loss ($\Delta$E) and
time-of-flight (ToF) measurements. Each
 of the isotopic cross-sections can be determined, and the kinematics of every
residue  can also be reconstructed. Based only on physical properties of the
ions, the method does not depend on chemistry. As beta-decay
half-lives are much longer than the 0.3 $\mu$s separation time, primary
residues are observed.

One of the important findings of the combination of the 
relativistic
U beam at SIS with the FRS separation techniques was the identification and
yield measurement of 117 new neutron-rich nuclei down to the sub-nanobarn level,
among which 
the doubly-magic nucleus $^{78}$Ni \cite{Bern,Enge}. 
The efficiency, selectivity and high sensitivity of the method was clearly 
demonstrated.  It
triggered new efforts to produce beams of exotic nuclei by
in-flight techniques. 

Apart from the scientific interest of proton induced spallation reactions, 
applications were  soon looked upon. 
For the purpose of Accelerator Driven Systems (ADS), for energy
production and transmutation of nuclear waste   \cite{Rubb}, a great number 
of data sets were missing.
Spallation targets for  neutron-sources 
require extensive data in the same domain.  Patents were approved in 
Germany and France  \cite{pate}. Thus a campaign of systematic 
measurements was
undertaken to obtain isotopic production cross-sections for a number of systems 
with an accuracy  of about 10\% and
an improved understanding of reaction mechanisms in order to interpolate 
to other  systems and energies below 1 A GeV.

For this purpose,  at 800 A MeV Au + p \cite{Must,Rejm,Benl1},  Pb + p  at 1
A GeV \cite{Wlaz,Enqv1} and at 0.5 A GeV,  Pb + d  at 1 A GeV \cite{Enqv2} 
and spallation from 1 A GeV U + p \cite{Taie}
and 1 A GeV U + d \cite{Ruiz} were
already studied and published. The missing part is still under analysis. 
 Spallation of Fe
on p and d  was measured between 0.3 and 1.5 A GeV.
Very recently, the systems Xe + p and Xe + d  were
measured between 0.2 and 1 A GeV to conclude this experimental program.

In this article we report on the (U + p)-fission products measured at 1 A GeV.
The system plays the role of a prototype for high-energy nucleon-induced
fission of an actinide.
 At high energies, data on fission of $^{238}$U induced by  protons can
 be used as a benchmark for other actinide nuclei present in
 incineration situations.

The results on fission-residues in spallation  of 1 A GeV $^{338}$U on
protons together with our results on evaporation-residues \cite{Taie}
 bring a comprehensive overview on this system.
Earlier  results on U-projectile fission were obtained on  Pb
\cite{Armb,Donz,Schw,Enqv3} and on Be \cite{Enge2} targets.
Altogether they should lead to
a coherent reconstruction of the intermediate systems after the primary cascade
and nucleon evaporation preceeding fission.
 The reproduction of the data with 
standard codes, the
intra-nuclear cascade-evaporation model LAHET
\cite{LAHE} is far from being satisfactory. Improvements of codes have been 
published \cite{Gaim,Jung,Benl,Boud}, and are applied here. 

Finally, the new data on isotopic yield measurements for $^{238}$U and
$^{208}$Pb together with present and coming data on their energy 
dependence should provide a reference in order to control
nuclide production in the forthcoming applications of  proton-induced 
spallation.

\section{Experiment}

\subsection{Experimental set-up}

Uranium-ions delivered by the Penning source are accelerated in the UNILAC and
post-accelerated in the SIS of GSI-Darmstadt. The beam energy of the
$^{238}$U$^{76+}$ ions  on target is 1 A GeV and their velocity $\beta_0$ = 0.876. 
The average 
time structure is a pulse of 7 s every 13 s. The beam intensity delivered is
 10$^6$ to 10$^8$ ions/pulse. It is  controlled and
monitored prior to the target by means of a secondary electron chamber, 
called SEETRAM \cite{Jura}.

A plastic target could not be chosen, since the production yields from C 
prevail upon the H yields in the mass region
 intermediate between fragmentation and fission. The uncertainties in 
 the sharing between C and H in a plastic-target would not be consistent with 
 our accuracy  requirements.

The liquid hydrogen target used in the present experiment consist of a 
(1.25 $\pm$ 0.03) cm  thick circular
cell closed by two thin parallel titanium windows. The cell is located at the
bottom of an helium cryostat. Surrounding the cell, thin aluminised
mylar reflectors and external titanium windows ensure thermal and vacuum
insulation \cite{Ches}.
The hydrogen thickness of (87.2 $\pm$2.2) mg/cm$^2$ \cite{Must} is chosen to
obtain a 10\% reaction probability. 
The Ti target walls 
have a total thickness of (36.3) mg/cm$^2$. Their contribution to the
counting rates remains below 10\%. It has been 
measured systematically using a ``dummy'' target of the same composition as
the empty target cell.

\subsection{Separation and identification in A and Z}

At energies close to 1 A GeV, fission fragments are observed up to Z = 72 and
they are 
nearly totally
stripped. In the worst case at Z = 72, less than 15\% of the ions still carry electrons
\cite{Sche}.
They are momentum analysed with the fragment
separator FRS, a double-stage achromatic spectrometer with an
intermediate dispersive image plane S2 and a final image plane S4 \cite{Geis}.
A schematic view of the FRS and detectors is shown in Fig. 1.\\

\begin{figure*}[ht]
\begin{center}
\epsfig{file=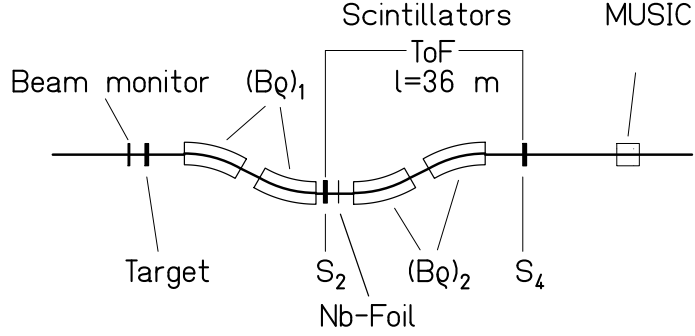, angle=-90, scale= 0.5}  
\caption{ Schematic view of the fragment separator with the detector 
equipment.}
\end{center}
\end{figure*}

In S4 the separated fragments are identified in Z by their energy losses
measured in a four-stage ionization chamber, MUSIC \cite{Pfut}. The mass 
number A of the analysed
fragments is determined from a measurement of velocity and magnetic
rigidity between
S2 and S4. The measurement is performed by two position sensitive plastic
scintillators  \cite{Voss}. The times and positions in S2 and S4 have 
to be known accurately
in order to correlate the time of flight with the precise length of the 
flight-path.
Fission fragments fill the full phase space defined by the acceptance of 
the FRS. The
presence of the 5 mm thick scintillator in S2 slightly alters the
achromaticity of the FRS. Nevertheless 36 elements from Z = 28 to 64 are
simultaneously transmitted to S4. Signals from the S4 scintillator
trigger the data-acquisition system.
Mass and charge numbers are calibrated by comparison to the parameters of
the primary uranium beam 
and by using the known structure of fission yields in the asymmetric fission
domain. 

\subsection{Constitution of the momentum spectra}

The momentum acceptance of the FRS is 3 \%. Applying the basic relation 

\begin{eqnarray}
B\rho  =  \frac {\beta\gamma\cdot A} {Z} \times \frac{c\cdot m_0} {e} 
\end{eqnarray}

with
 Z and A numbers
determined,  a  3 \% fraction
of the $\beta\gamma$ spectrum is obtained for each measurement.
The total spectrum of transmitted fission
fragments covers a range of 7 \% to 20\% of the reduced momentum
 of the
projectile, $\beta_0\gamma_0$.
The measurements consist of a 2 \% step-by-step scanning of the FRS
magnetic fields tuned on cadmium, (Z = 48).
After normalisation to the incoming dose given by the SEETRAM,
and dead-time corrections, complete $\beta\gamma$ distributions are
reconstructed
by assembling about 10 files of overlapping B$\rho$ settings to produce a 
$\beta\gamma$ spectrum for each of the isotopes;

\begin{eqnarray}
 \frac {dn (\beta\gamma)}{ d(\beta\gamma)}
 = \frac {n_i (\beta\gamma)} {N_i (1-T_i)}                                
\end{eqnarray}

 where n$_i$ ($\beta\gamma$) is the i$^{th}$ bin of the $\beta\gamma$
 spectrum, T$_i$ is the dead time fraction
 and N$_i$ the number of incident ions  for
 this measurement.
All in all 4$\cdot$10$^6$ events were
accumulated.\\
Fragments are spread over 20 cm in the image plane at S2 covered by the
first
position-sensitive scintillator. For an identified isotope the
distribution in S2
depends only upon $\beta\gamma$, and the accuracy on S2 position governs
the accuracy of the magnetic rigidity.
In the present case, a precision in  position of 2.5 mm (FWHM) leads to a
resolving power of
$\beta\gamma$ / $\Delta\beta\gamma$ = 2600, given the dispersion
 in momentum  of the first stage of the FRS by percent,
 D = 6.5 cm/$\%$.
In this experiment the  response  of the scintillator was slightly
non-linear,  and a routine was systematically used in the analysis 
software to
correct for this, \cite{Must}.\\
The FRS magnetic fields tuned on Z = 48 guide this element to the 
center of S4 
and display on both sides the neighbouring elements, gradually apart 
from the center of
S4. The 36 elements are transmitted together to the final focus S4, and 
momentum distributions of the produced isotopes are 
simultaneously measured in one series of B$\rho$ settings of the FRS.
The $\beta\gamma$ spectra can be converted to the projectile frame and 
are presented in cm/ns as velocity spectra in the system of the 
projectile.
This coordinate system is appropriate for comparison with previous results 
concerning fission from target nuclei. 
Moreover, at high excitation energy, fission fragments are emitted
isotropically in the fissioning system  \cite{Armb,Boch}. This allows for 
analytical formulations
to be used in the analysis of the $\beta\gamma$ distributions.

\subsection{Description of the velocity spectra} 

\begin{figure*}[ht]
\begin{center}
\epsfig{file=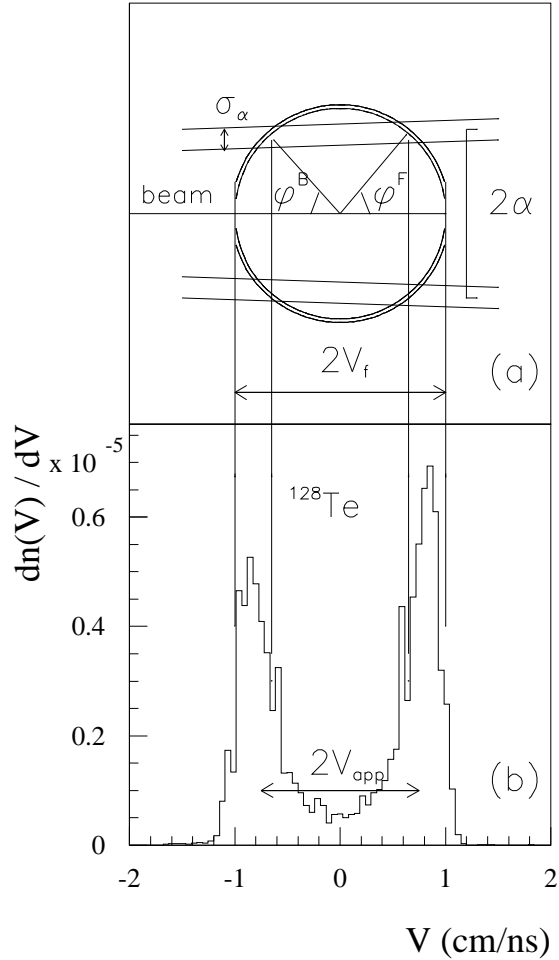,scale=0.8}  
\caption{ a) Schematic view of the experimental parameters shaping the
measured
velocity spectrum in the frame of the fissioning system.  V$_f$ is
the fission-fragment velocity, $\alpha$ is the
angular acceptance of the FRS, and $\sigma_{\alpha}$ its variance. 
$\varphi^{B,F}$ are the corresponding
emission angles of fission fragments  in the fissioning system
in forward and backward directions,
respectively.
b) Velocity spectrum of $^{128}$Te in the frame of the fissioning system.
The velocity V = 0 refers to the projectile frame. V$_{app}$ is the apparent
fission velocity defined in the text.}
\end{center}
\end{figure*}

The reconstructed velocity spectra are the corner stones of our 
analysis. An
example of the velocity spectrum is shown in Fig. 2b. Two narrow 
asymmetric
peaks are observed roughly equidistant from velocity zero, i. e. the
projectile velocity.  Velocities are defined in reference to the beam 
direction i. e. forwardly
 and backwardly emitted fragments have positive and negative velocities,
 respectively.
The velocities of the
fission fragments arise from their Coulomb repulsion at scission.
Velocity vectors of a specific isotope populate a thin spherical
shell in the fissioning system, which may be slightly shifted compared
to
the projectile system by the primary reaction recoil.
The sphere is pictured in Fig. 2a by a circle,
 the cut of the sphere by a plane which contains the beam.
 Only forward and backward cups of the sphere, defined by the angular
 acceptance of the FRS $\alpha$, are transmitted, and
the  longitudinal projections of their velocity
 distributions are shaping the two peaks (Fig. 2b).
 
 For selected elements the velocity spectra of all isotopes are 
 presented on scatter-plots
in Fig. 3. Rates are given in terms of 
velocities and of neutron number N.
They all show strongly populated areas along two parallel lines 
associated
to backward and forward emitted fission fragments. The lines, almost 
symmetric  in respect to 
the axis (V = 0, i.e. projectile frame), show a fission fragment velocity 
decreasing with 
increasing atomic number Z
of the fragment, a trend due to momentum conservation in fission. 
The parallel lines contrast with  U + Pb scatter-plots 
\cite{Enqv3} where 1)
neutron-rich nuclei are strongly enhanced for zirconium and tellurium
due to electromagnetic fission and 2) fragmentation residues 
sit in between the forward and backward regions for 
neutron-deficient isotopes. In spite of the absence of evaporation residues,
at decreasing masses of the lighter elements,  
events with small velocities are found. Besides the small contribution
of fragmentation residues produced in target windows, 
this is a first indication for 
the contribution of secondary reactions which will be discussed in the
following section.\\

\begin{figure*}[ht]
\epsfig{file=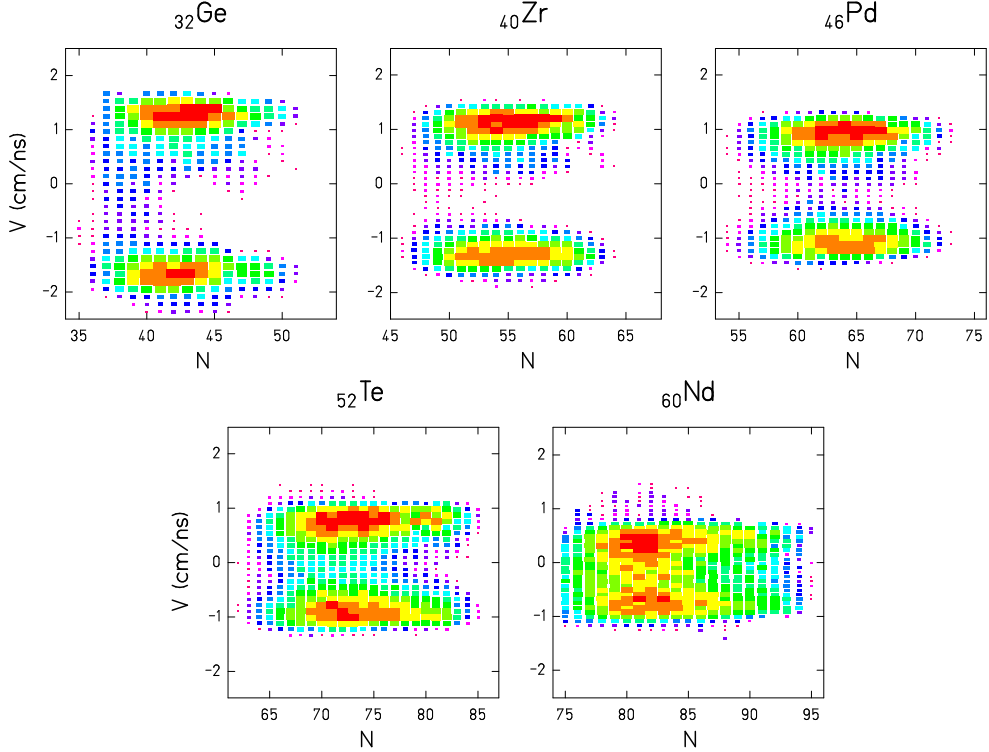,scale=0.9}  
\caption{Cluster plots of number of events measured on target 
in terms of velocity versus 
neutron number. The fission velocity coordinate is
given in the center of mass of the projectile. The intensity scale is
logarithmic with a factor of 2 between adjacent colours.}
\end{figure*}

Data taken on the ``dummy'' target show the same pattern as in Fig. 3 
with a contribution of fragmentation. Yields are defined by
the integrated velocity spectra. Isotopic distributions of yields in 
the ''dummy''
and in the target are compared in Fig. 4. As intended,  the yields in
the ``dummy'' are only
a few \% of the yields in the target, except in the neutron-deficient
region
where they reach larger values, but less than 20\% . 
Asymmetric fission, leading to neutron rich isotopes, is also clearly
enhanced for titanium relatively to hydrogen. The hydrogen  
contribution is obtained by subtracting the yields
on the ''dummy'' from the target yields.
Uncertainties are estimated
from the fluctuations of the intensity ratio of the two peaks, as will 
be discussed later. The
error bars on yields measured for the ``dummy'' are negligible when 
reported to yields on hydrogen.

\begin{figure*}[ht]
\begin{center}
\epsfig{file=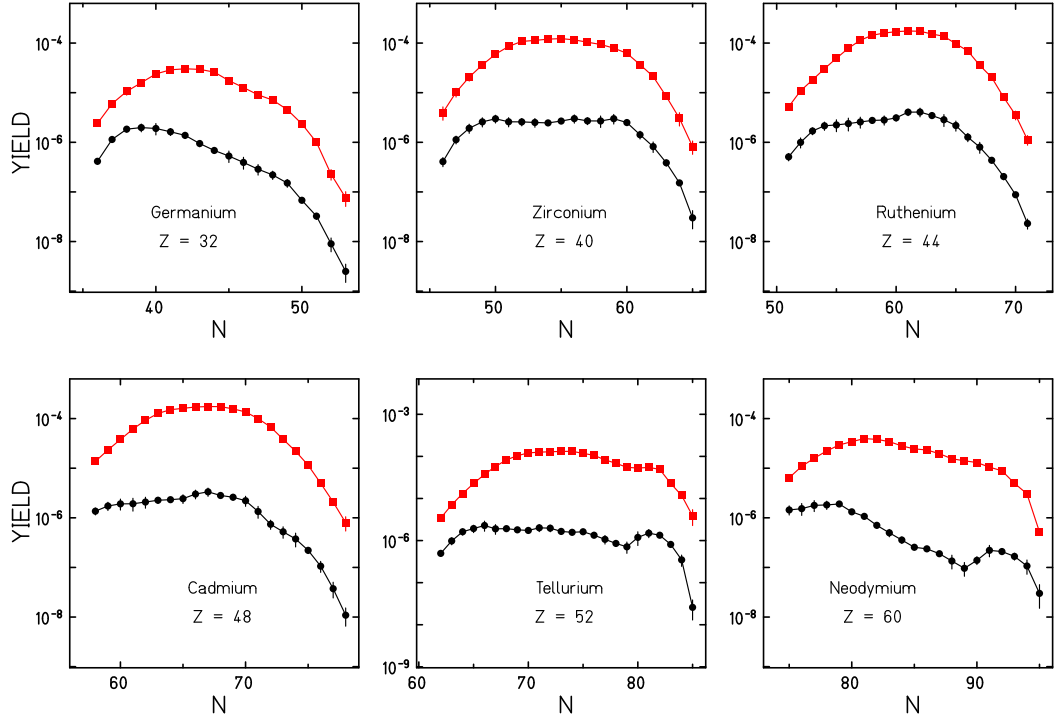,scale = 0.85}  
\caption{Comparison of transmitted yields [arb. units] of isotopes on 
hydrogen (squares)
and ``dummy'' (circles) targets as a function of their neutron number
for selected elements.}
\end{center} 
\end{figure*}
Evaluating velocities, the $\beta\gamma$ distributions on target and
``dummy'' are found to be very similar for a given isotope, and the 
yield of the ``dummy'' is on a few percent level.
Therefore, the  contribution of the titanium windows could be
ignored in the evaluation of fission-fragment velocities in U + p.   


\subsection{Fission velocity and transmission} 

 The
 geometrical aperture of the spectrometer seen from the target defines
a solid angle $\Omega$ simply related to the angular acceptance 
$\alpha$ assuming axial symmetry.
$\alpha$ is the limiting value for the emission angle $\phi$ of 
transmitted fission fragments in the laboratory
system. $\phi$ corresponds to angles $\varphi^F$ and $\varphi^B$
 in the fissioning system with F and B set for forward and backward,
 respectively, see Fig. 2a. The angular transmission T$_\Omega$ is
 defined 
 as the ratio of the yield measured for a given fission fragment to 
 the total production yield for this fragment.

 The determination of T$_\Omega$ is a key to obtain the
 cross-sections. It depends upon fission fragment velocities and it
 must be evaluated for each fragment.
 T$_\Omega$ is a function of the three variables $\alpha$, $\beta_0$
representing the beam and $\beta_f$ the fission process with
 V$_f$ =  $\beta_f$.c, the fission velocity, i.e. the radius of the 
 sphere
 shown in Fig. 2a.  As shown
in Appendix 1, for a bundle of particles with an angular 
 divergence
 $\alpha$ starting from a point-like source and entering the FRS on its
 principal trajectory, the transmission T$_\Omega$ can 
 be calculated rigourously, without any approximation,
 when setting the variances to zero.
 In the fissioning system,  T$_\Omega$ is obtained by 
integrating the
surface elements of the spherical velocity shell within the two solid
angles,
and we obtain \cite{Enqv3}:

\begin{eqnarray}
 T_\Omega = T_{ \Omega}{^B} + T_{ \Omega}{^F}  = 
 1 - \frac { \cos \varphi^F + \cos \varphi^B }{2}        
\end{eqnarray}

In Fig. 2b the velocity distribution of transmitted $^{128}$Te is 
given as an example. Three quantities are evaluated from these 
distributions:\\
 1) The spectrum dn($\beta\gamma$) / d($\beta\gamma$)
integrated over the whole range of $\beta\gamma$,  evaluated for each
of the isotopes gives the measured yields.\\
 2) A shift of the mean velocity of forward and backward emitted
 fission fragments, compared to
the projectile velocity (V = 0) gives the velocity of the fissioning
parent nucleus, the recoil velocity.\\
 3) The centroid of each of the two peaks defines a mean 
  fission-fragment velocity, the apparent
velocity  $\beta_{app}^{F,B}$ = V$_{app}$/c different for forward and 
backward emitted fragments. We define the mean value of the two 
apparent velocities, 
\begin{eqnarray}
 \beta_{app} = \frac {\mid\beta_{app}^F\mid + \mid\beta_{app}^B\mid }{2} 
\end{eqnarray}
 The two peaks are shaped by the variances of the
three variables $\alpha$, $\beta_0$ and $\beta_f$,  as will be shown
later by
a
Monte-Carlo simulation. With all variances set to
zero -Fig. 5a- the peaks become rectangles, as the phase-space 
density of a spherical homogeneously filled shell per interval 
d$\beta_\parallel$ with $\beta_\parallel$ the
longitudinal velocity, is constant.
The projection of the spherical shell
on the beam axis is a rectangle.
In this case the apparent velocity of forward and backward emitted 
fission fragments (see Fig. 2) is given by:
\begin{center}
$$ \beta^{F,B}_{app} = \beta_f - \frac {\beta_f - 
\beta_f \cos \varphi^{F,B} }{2} = 
\beta_f \frac {1 + \cos \varphi^{F,B} }{2}   $$
\end{center}
and the mean value follows:

\begin{eqnarray}
\beta_{app} = \frac { \beta_f } { 2}   (1 + \frac { cos \varphi^F + cos
\varphi^B}{2} ) 
\end{eqnarray}

With eq. (3) defining the transmission, we obtain the fission velocity
 depending on $\beta_{app}$ and T$_\Omega$:

\begin{eqnarray}
 \beta_f = \frac {2 \beta_{app} } {2 - T_\Omega} 
\end{eqnarray}

from which the limiting cases are deduced 
 $\beta_f$ = $\beta_{app}$ for T$_\Omega$ = 0, and 
 $\beta_f$ = 2$\beta_{app}$ for T$_\Omega$ = 1.

In Appendix 1 we show 
that  with
 $\varphi_\Sigma$ =
($\varphi^F$ + $\varphi^B$)/2
the fission fragment velocity $\beta_f$ and the transmission T$_\Omega$ 
are given  in very good approximation by: 
\begin{eqnarray}
 \beta_f = \beta_{app} [  1 + (\tan \varphi_\Sigma/2)^2  ] 
\end{eqnarray}
\begin{eqnarray}
 T_\Omega  = \frac {2} {1 + [ \cot  \varphi_\Sigma/2 ]^2 } 
\end{eqnarray}
with 
\begin{eqnarray}
 \tan(\varphi_\Sigma/2) = \frac {\beta_{lim}} {2  \beta_{app}} 
\end{eqnarray}
$\beta_{lim}$ is defined by $\alpha$ and the velocity of the fissioning
parent nucleus $\beta$ approximated by $\beta_0$
\begin{eqnarray}
 \beta_{lim} = 
 \alpha \sqrt { \gamma^2 - 1 } \approx \alpha\beta_0\gamma_0 
\end{eqnarray}

Eq.(7) and (8) demonstrate the central importance of the
ratio $\beta_{lim}$/2$\beta_{app}$.

Summarising the procedure:
\begin{itemize}
\item
From the $\beta\gamma$ distribution, the mean apparent velocity
$\beta_{app}$ = ($\beta^F_{app}$ + $\beta^B_{app}$)/2 is evaluated
for each isotope.

\item T$_\Omega$ and $\beta_f$ are simple functions of 
the mean opening angle $\varphi_{\Sigma}$ = 
($\varphi^F$ + $\varphi^B$)/2.

\item $\beta_{lim}$ is related to the acceptance angle $\alpha$ of the
spectrometer and to the projectile velocity $\beta_0$
\end{itemize}

\subsection{Simulation of the angular cut of the FRS }

In order to include the variances on the FRS angular acceptance and the
kinematics of
fissioning nuclei, we have simulated the physical cut in
phase space due to the angular acceptance of the FRS. The purpose is to
relate V$_f$ and the transmission to the measured V$_{app}$. \\
The Monte-Carlo calculation is based on the following assumptions:\\
1) In the system of the fissioning parent nucleus, approximated by the
system of the projectile nucleus, fission fragments are 
isotropically emitted with a given velocity V$_f$.\\
2) The spherical shell of velocity vectors is converted to the 
laboratory system moving relativistically with a velocity $\beta$.
Fragments emitted within the FRS acceptance angle  $\alpha$ are 
transmitted.
At this stage we obtain a transmission value and the momentum spectrum
in the laboratory system.\\
3) This spectrum is reconverted into the projectile system.\\
4) Variances of velocities and of the acceptance angle are
further parameters which are included.

The simulation enlightens the effects of the experimental constraints
on the measurement.
On Fig. 5a the forward and backward rectangles correspond to the
share of transmitted fission-fragments with $\sigma_{\alpha}$ = 0. 
A filled rectangle would be obtained 
for $\beta_f$ smaller than $\alpha\beta\gamma$ and T$_\Omega$ 
would then reach 100 \%. 
The second curve in the same frame results from the introduction of a 
variance of the acceptance
angle $\sigma_{\alpha}$ = 2.5 mr derived from the known geometry of
the FRS \cite{Pere,Schw3}.
It shows that the filling of the spectra at intermediate velocities 
comes mainly from the large variance of the FRS acceptance angle. \\

\begin{figure*}[ht]
\includegraphics[scale=0.32]{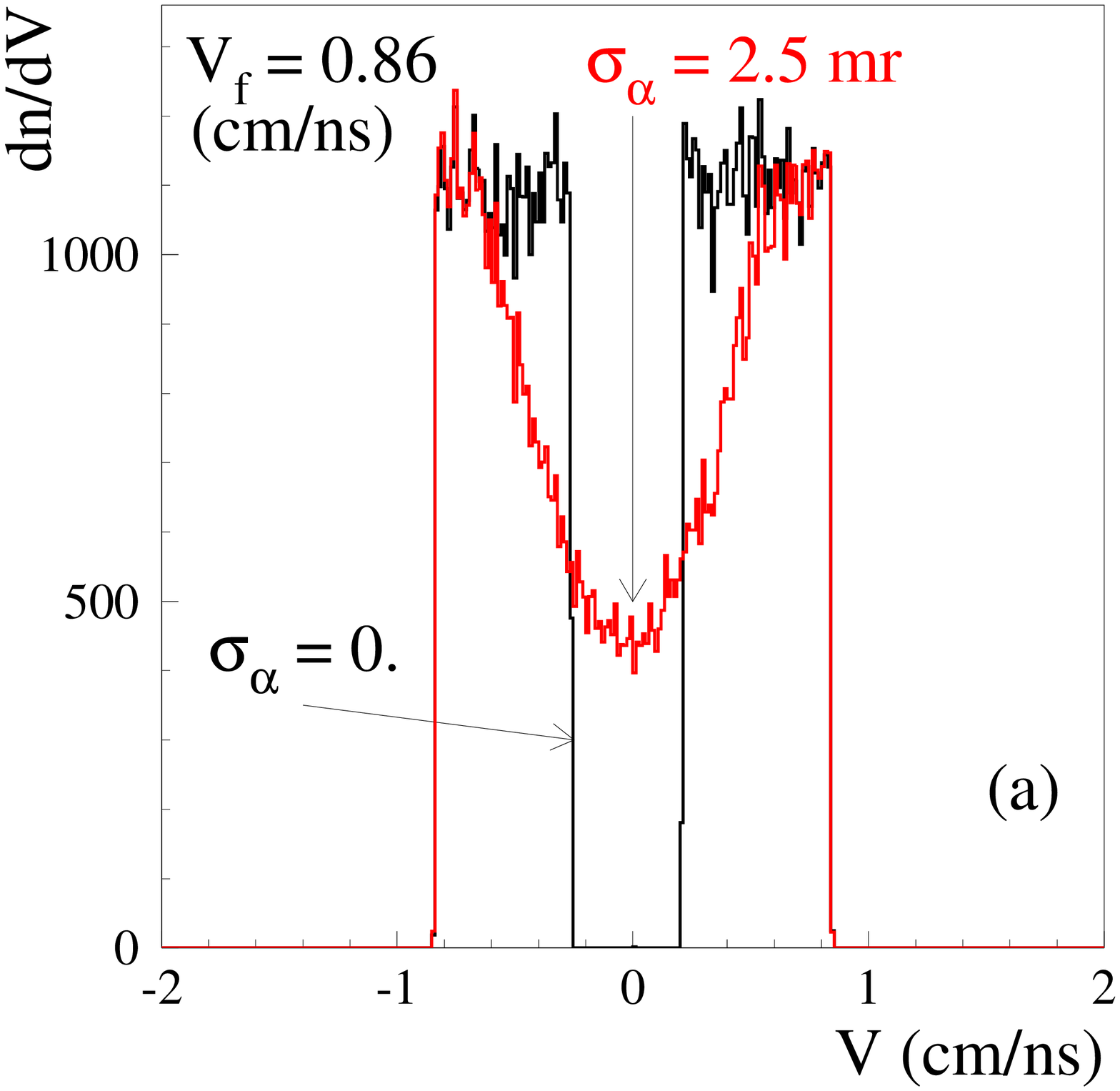} \hfill
\includegraphics[scale=0.32]{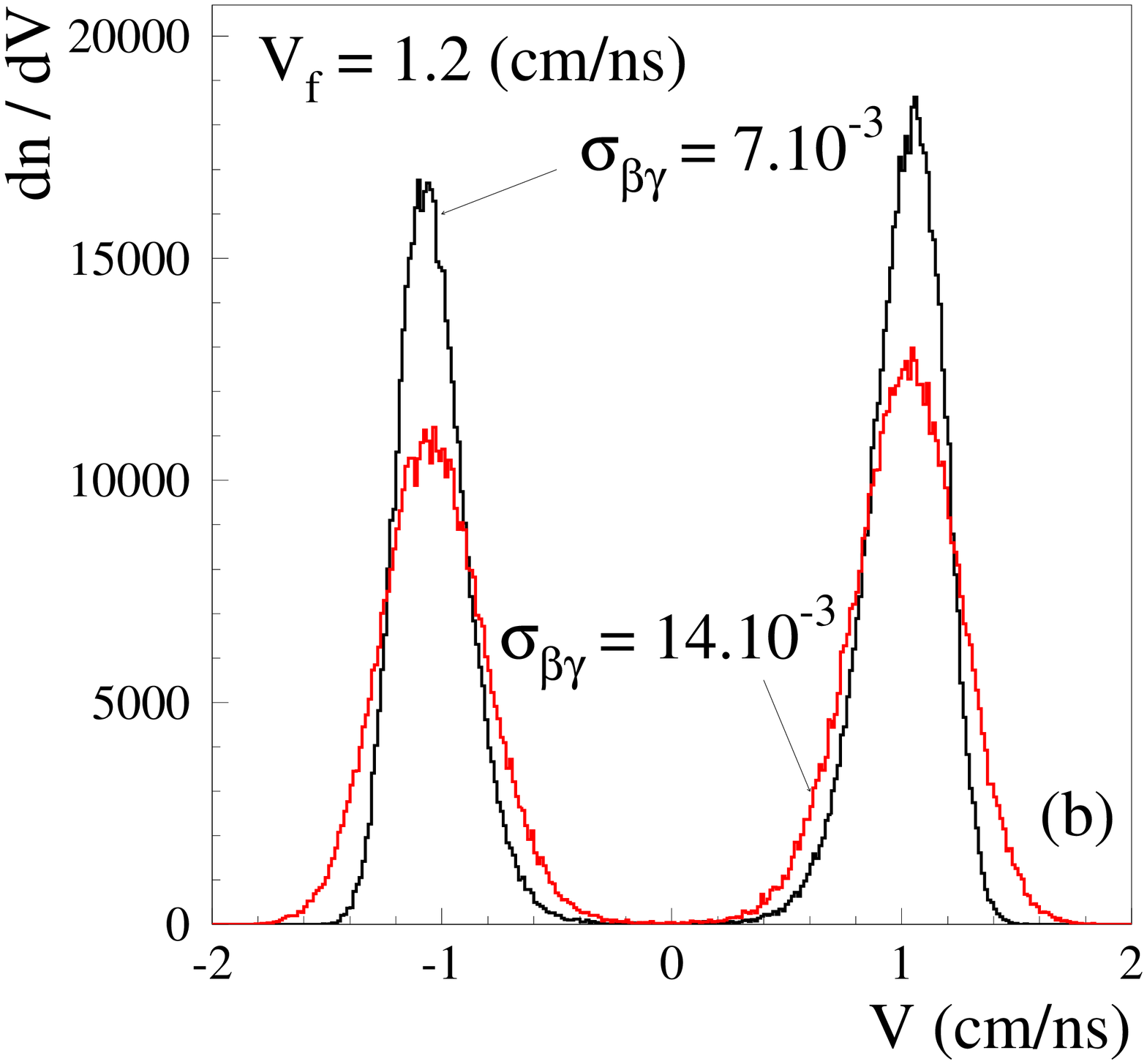} \hfill
\caption{ Impact of variances a) $\sigma_{\alpha}$ of  $\alpha$, the 
acceptance
angle of the FRS, and b) $\sigma_{\beta\gamma}$ of  $\beta\gamma$, the
momentum of the fissioning nucleus. Fission-velocities are different in
a) and b) in order
to better illustrate the contributions of both variances.}
\end{figure*}


Fig. 2b versus Fig. 2a shows that the external slopes depend only on the
fluctuations of the reduced
momenta of the fission fragments and/or of the fissioning nuclei.
Several sources contribute to these fluctuations - fission fragment
velocities, target location straggling and recoil momenta of the
fissioning
nuclei -, and lead to a broadening of the peaks as illustrated on 
Fig. 5b. The 
relative contributions of these sources are quantified in the 
discussion, see sect. 4.8.

\begin{figure}[ht]
\includegraphics[scale=0.33]{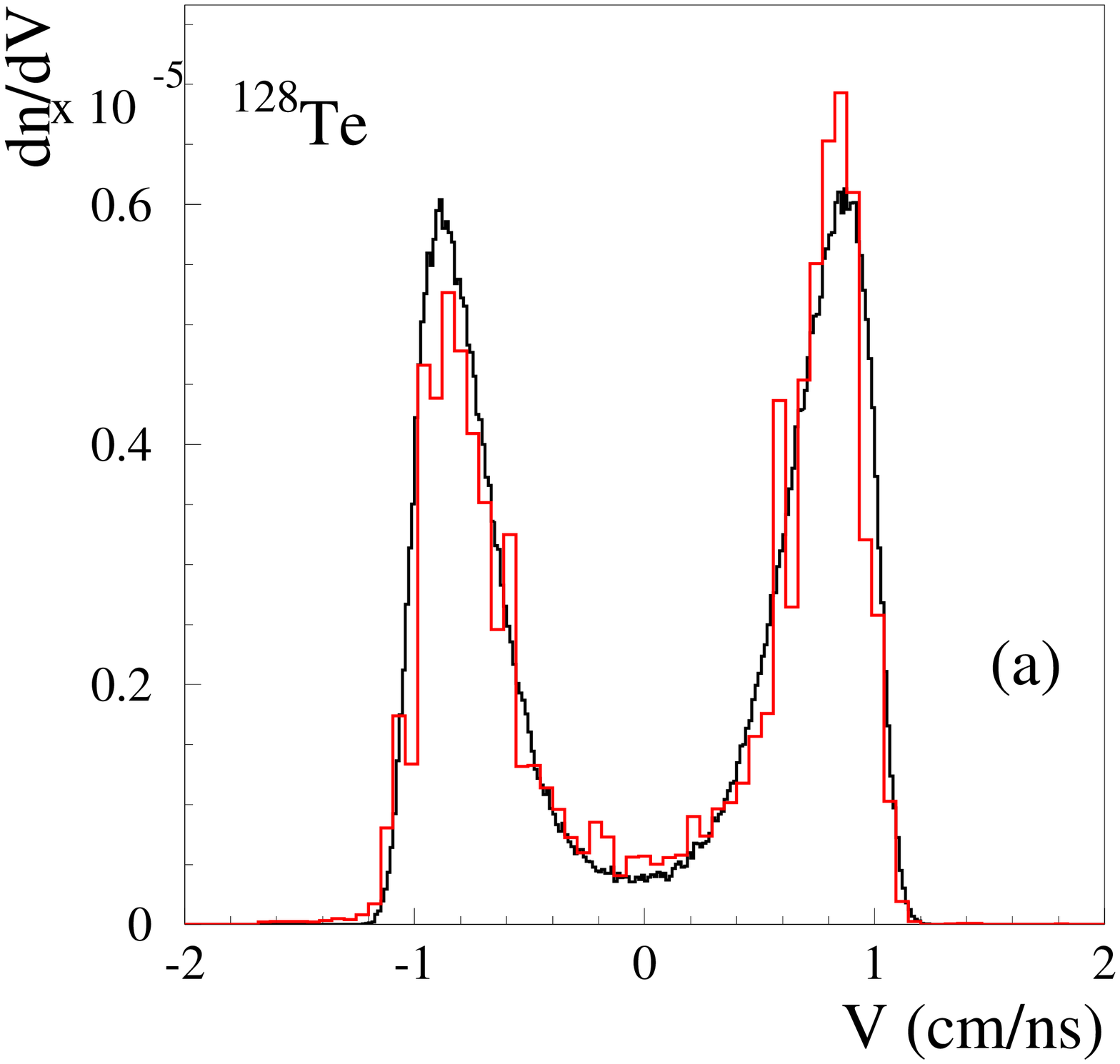} \hfill
\includegraphics[scale=0.33]{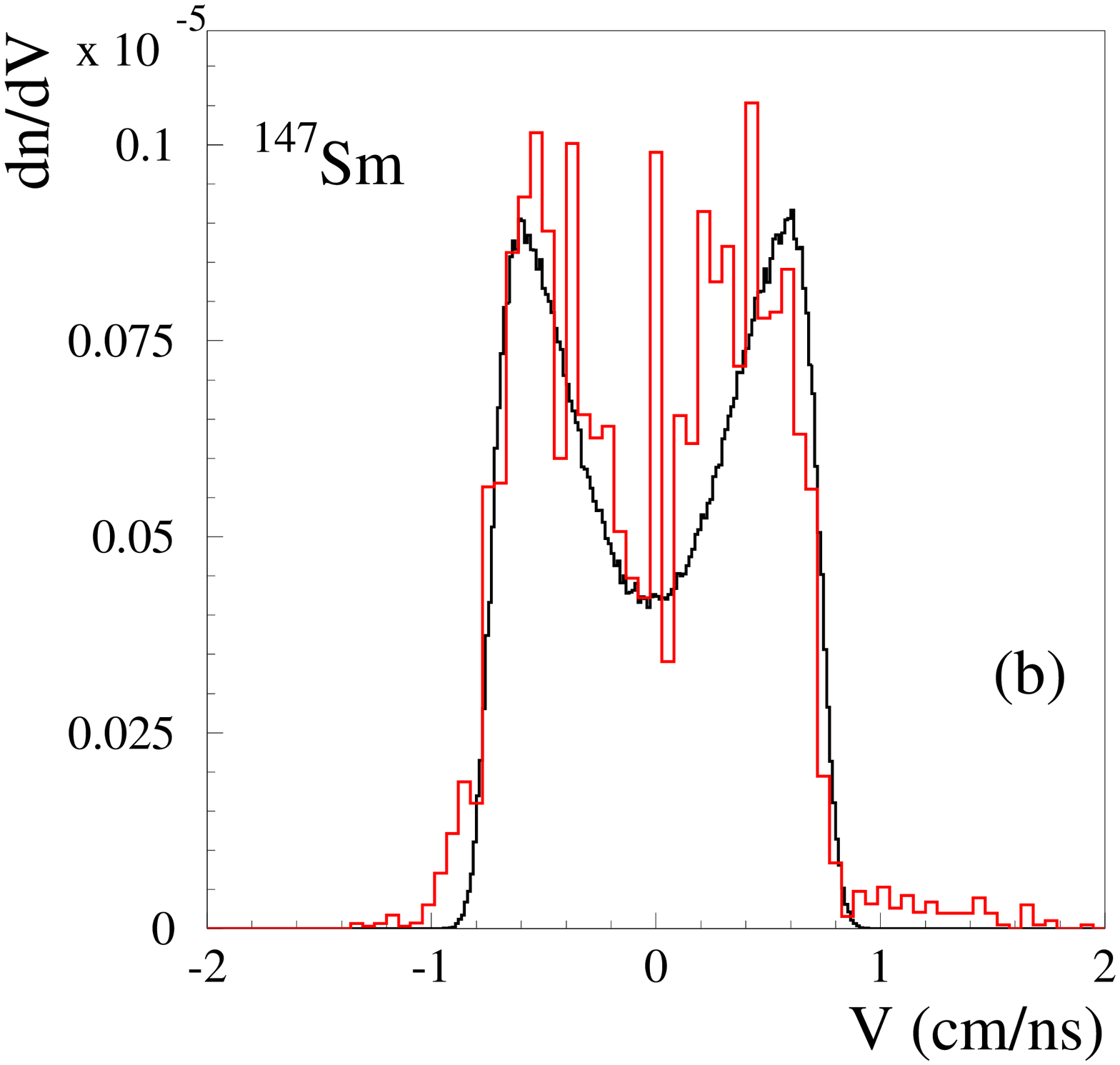} \hfill
\caption{ Examples of simulation with $\sigma_\alpha$ =2.5 mr
$\sigma_{\beta\gamma}$ = 7.10$^{-3}$ compared with the measured spectra for
$^{128}$Te and $^{147}$Sm. }
\end{figure} 

With  $\sigma_{\beta\gamma}$ = 7.10$^{-3}$ the external slopes are
almost vertical and the value of V$_f$ = 1.2 cm/ns introduced in the
simulation can be
 extracted from the distance between the external sides at half
maximum. On the contrary, if $\sigma_{\beta\gamma}$ becomes larger than
the width of the rectangles of Fig. 5a, this half-maximum rule is no
longer valid, as seen for $\sigma_{\beta\gamma}$ = 14.$10^{-3}$.
A value of $\sigma_{\beta\gamma}$ =  7.10$^{-3}$ is selected 
as an ``empirical broadening'' to optimize
the simulation. 
The result is illustrated in both examples of Fig. 6a and 6b.

\begin{figure*}[ht]
\includegraphics[scale=0.37]{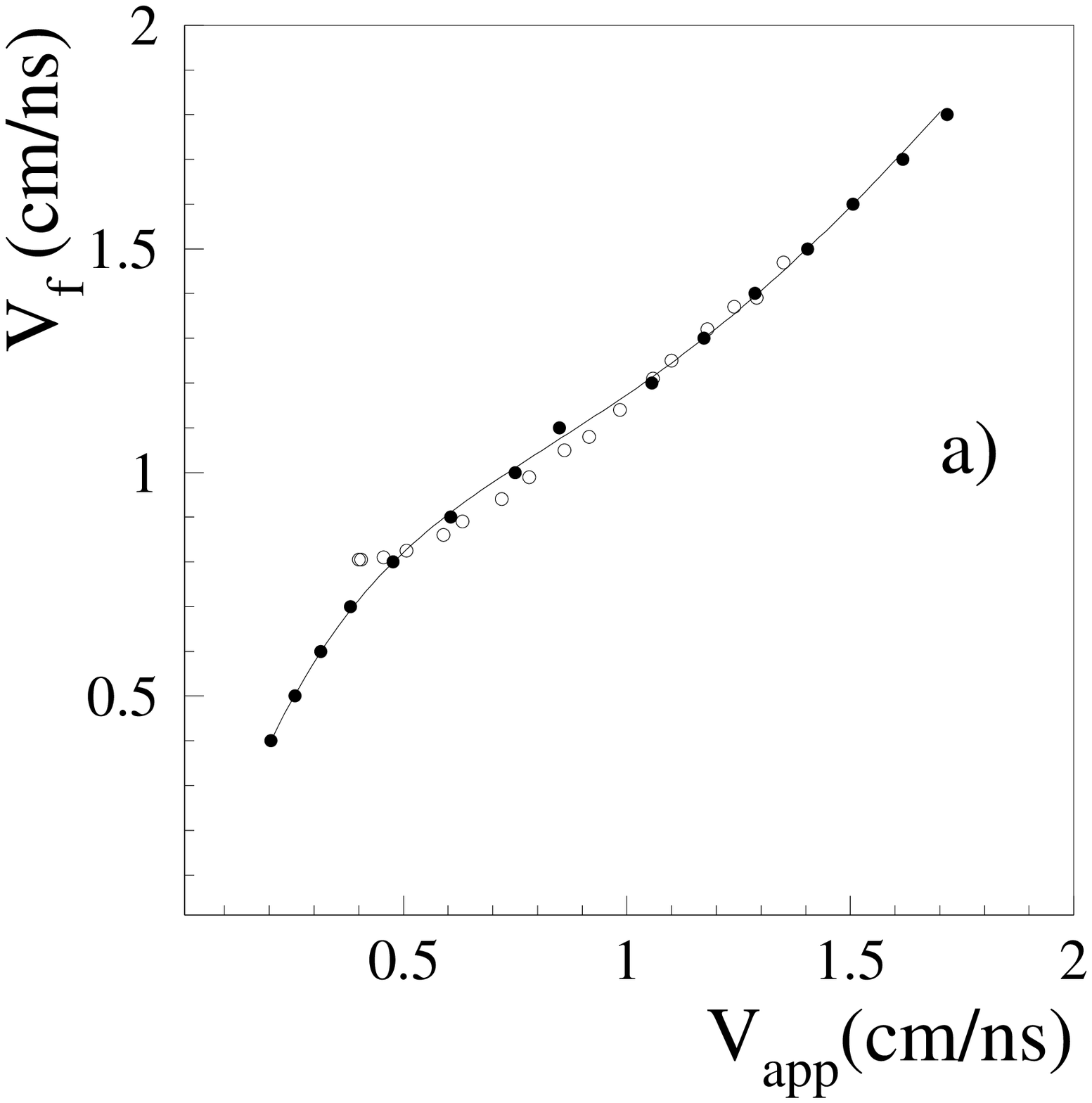} \hfill
\includegraphics[scale=0.37]{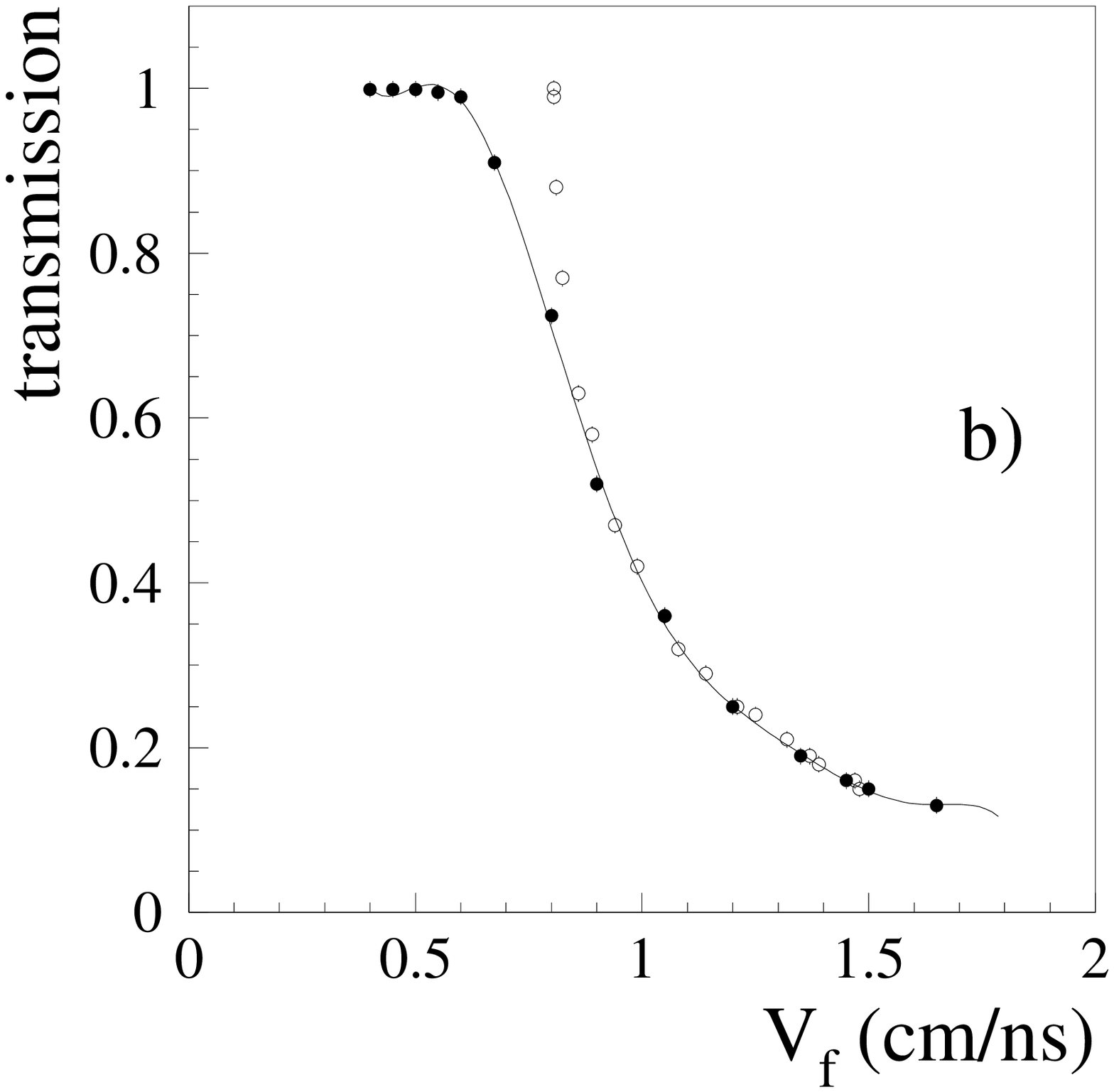} \hfill
\caption{ Fission fragment velocity as a function of apparent velocity
a) and transmission b) as a function of fission fragment velocity given
by the simulation (full points) and calculations (empty
points) using eqs.
(7) and (8). Lines are fits to the full points.}
\end{figure*} 

The relation between V$_{app}$ and V$_f$ either from
eq. 6 or from the simulation  are very close as shown in
Fig 7(a) and the trends at T$_{\Omega}$ = 0 or 1
are correct in both cases. Simulated and calculated values of T$_\Omega$
are the same for fission velocities larger than 0.8 cm/ns ( Fig. 7b).
For smaller
velocities the calculations, which do not include the large variance of
the acceptance angle, cannot describe the transmission. Approaching
T$_{\Omega}$ = 1 the simulation accounting for the variances provides better 
results. It is seen on the velocity spectra of isotopes of neodymium to
gadolinium where a few isotopes (the heaviest isotopes showing
the largest fission
velocities) still show a residual valley between forward and  backward 
peaks. In the following transmissions and fission fragment velocities 
are obtained 
using  the values of V$_{app}$  determined for each 
isotope and 
the fitted curves of Fig. 7.

\section{Experimental results}

\subsection{Fission fragment velocities}

\begin{figure}[!h]
\includegraphics[scale=0.8]{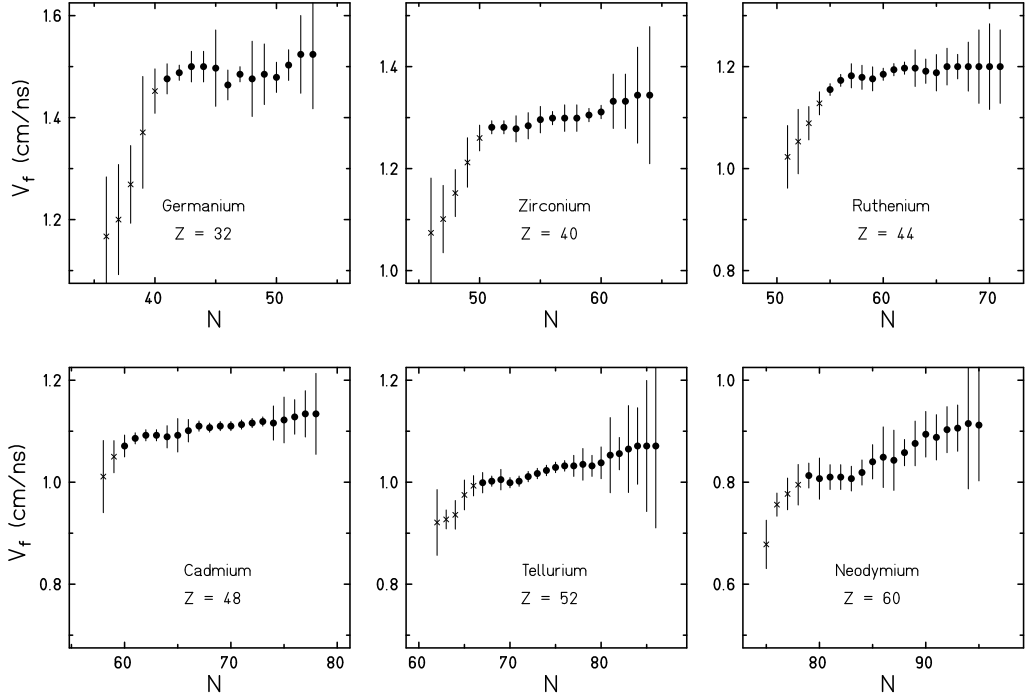} \hfill
\caption{ Fission velocities for selected elements.
 Cross symbols correspond to regions where secondary-reaction
 products are mixed with direct fission fragments.}
\end{figure} 

The velocity spectrum of each isotope has been analysed to extract a
value of the apparent velocity. The apparent velocities have been
converted into fission fragment velocities.
For a given element, the fission fragment velocity  decreases 
smoothly from a maximum value for the heaviest isotopes down to values 
smaller by 10\%  for the lighter isotopes, Fig. 8.
Further down to the most neutron-deficient isotopes, the velocities fall 
abruptly.
 In case of U+p,
this fall cannot be related to the occurence of evaporation residues 
since this process does not populate the neutron-deficient region of 
fission  fragments.
The small contribution from fragmentation yields in the Ti-windows 
(Fig. 4) cannot explain the observed fall in velocity of 20\%.
This fall is due to secondary reactions in which heavier primary
fission fragments loose
nucleons in a secondary fragmentation reaction in the target.
Atomic numbers and masses are reduced, whereas the velocity spectra
are only blurred.
 These contributions - due to larger Z with smaller fission-fragment
 velocities - are superimposed with
the spectrum of the primary isotope. This contamination reduces
the apparent velocity of the isotope. It 
becomes significant in the neutron-deficient region where productions
by direct fission is low and secondary-reaction products are abundant.
However, the small velocities of the lightest isotopes of neodymium 
cannot be explained by this argument, as secondary reactions of fission
fragments contribute little.
 Here secondary reactions of spallation products could pollute the 
 yields of neutron-deficient fission fragments.


\begin{figure*}[ht]
\begin{center}
\includegraphics[scale=0.6]{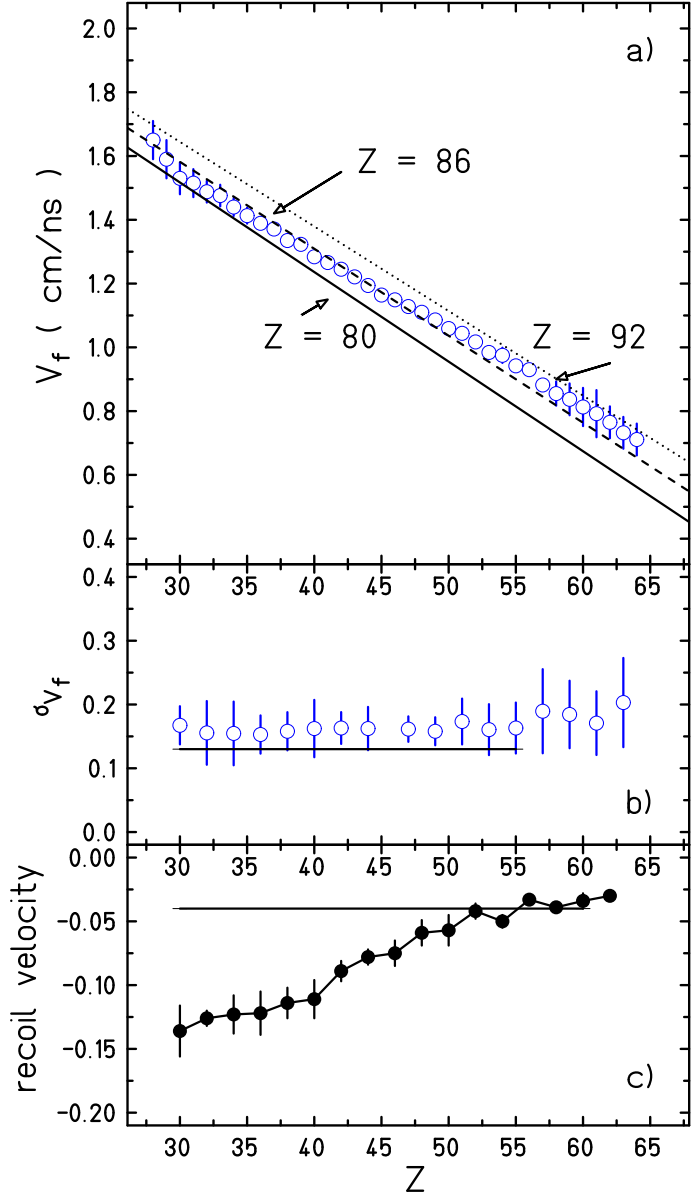} \hfill
\caption{ Fission fragment velocities as a function of Z. a) The lines
are obtained using the calculation of velocity for fissioning parent
nuclei  of Z$_0$ = 80 (full line), Z$_0$ = 86 (dashed line) and Z$_0$
= 92 (dotted line) with the parametrisation  of \cite{Wilk} \cite{Bock}
and a deformation of $\beta$~=~0.65 for both fragments. 
b) Measured variances of the fission velocities in cm/ns. 
 c) Measured displacement of the mean velocity 
of forward and backward emitted fragments.
The lines indicate the
effective broadening b) and the recoil velocity c) as measured 
for the mean 
parent-fissioning nucleus \cite{Taie}- see  text sect. 4.8 -.}
\end{center}
\end{figure*} 

The velocity of an element is taken as the mean value of the velocities
for the
four isotopes most abundantly populated, in order to select the dominant
symmetric fission regime.
This velocity is plotted as a
function of Z in Fig. 9a. The uncertainties are less than 4\%.
Three lines are indicated showing the velocity dependence  expected for
three values of the nuclear charge of the fissioning  parent nucleus 
Z$_0$ = 92, 86 and 80 \cite{Wilk,Bock}. 
The scission configuration of the fissioning nuclei is approximated by
the outline of two deformed nuclei ( $\beta_1$  =  $\beta_2$ = 0.65)
at a given distance. Using momentum conservation,
the fragment velocities are calculated for the total kinetic
energies released from the scission configuration.

None of the three lines coincides exactly with the data. Fissioning
nuclei with Z$_0$ between 84 and 90 can contribute to the production
of fission fragments.
The general trend towards smaller fission fragment velocities going 
from neutron-rich isotopes to neutron-deficient isotopes is observed 
for all elements (Fig. 8) corroborating that
heavy (Z = 90) and light (Z = 84) parent nuclei do contribute. 
The first ones contribute to the more neutron-rich and the last to 
less neutron-rich isotopes.

The slope of the external edges of the velocity distributions
introduced empirically in
the simulation (Fig. 5b) as a variance $\sigma_{\beta\gamma}$  are
now evaluated systematically  as a function of
Z and plotted in fig. 9b).
Unfolding these measured values and the underlying rectangle of the
transmitted momenta, fig. 5a, the effective broadening
$\sigma_{V_f}$ is obtained, and given by the line in fig. 9b.
Both variances are related by
$\sigma_{V_f}$ = $\frac{1} {\gamma}$ $\cdot$ $\sigma_{\beta\gamma}$ $\cdot$ c.
%
%
%
Within the limit of our accuracy this width does not
depend on the atomic number. Its mean value is
$\sigma_{V_f}$ = 0.13 cm/ns, which corresponds to
$\sigma_{\beta\gamma}$ $\approx$ 8.10$^{-3}$ in good agreement with the value
of 7.10$^{-3}$ empirically deduced in sect. 2.6, from the simulations
of the  measured spectra (Fig. 5b).


The recoil of the fissioning nucleus is determined from the mean value
of the forward and backward fission velocities. The values  plotted in
Fig. 9c are averaged over the four most abundant isotopes of an element.   
This recoil is small and it decreases
from -0.03 at Z = 62 to -0.13 cm/ns
for Z = 30.  We obtain a mean value of -0.08 cm/ns.

\subsection{From yields to cross sections}

Isotopic cross sections of (U + p)-fission are obtained by dividing the
production yields on hydrogen by the proper transmission, as
presented in sections (2.5) and (2.6). Our final
values are compared to those used in the analysis of the parallel
experiment U + d at 1 A GeV evaluated by J. Pereira  applying a 
different method \cite{Pere}. A good agreement is found between both
sets of transmission-values evaluated in terms of the fission-fragment 
velocities.


The correction for secondary reactions occuring in the target has to be
applied. All fission fragments, neutron-rich more than proton-rich 
undergo secondary spallation and the yield is depleted. For 
neutron-deficient isotopes, the fission yields are increased by secondary 
evaporation residues. The correction for secondary reactions enhances the 
share of neutron rich isotopes and reduces neutron-deficient ones.
 This is the reason for the ``hook'' shape of this 
correction already presented in a previous article \cite{Enqv1} and 
calculated more precisely in a forthcoming paper \cite{Paol}. 
When they exceed 50 \%,
 cross sections for corresponding isotopes are no longer considered. 
Our experimental method to determine cross sections of primary residues
reaches its limit.

 Moreover, small secondary effects need to be corrected for:    
For fragments  not centered in S4, i. e. apart from the optical axis, 
the FRS acceptance angle and consequently the transmission
are  slightly reduced. 
The FRS angular acceptance falls from 14.8 mr on the optical axis down
to 13.3 mr at 8 cm apart from the axis in S4 \cite{Pere}. A smooth 
parabolic function was used to account for this relative loss in 
transmission.
The correction given in table 1, reaches 14\% for elements like
nickel and samarium.

Fragments reacting in the scintillator SC2 at S2 escape the 
magnetic
filter of the second half of the FRS. The related loss increases 
linearly from 11 \% for nickel up to 17 \% for gadolinium. It is 
calculated using the total reaction  cross section formulation of 
Karol  \cite{Karo}. The results are given in table 1  for the range
of elements investigated.

\input{tab1.tex}

A loss by ionic charge exchange occurs, either in the target or in
the plastic scintillator S2. Eventhough the probability is small for
fission fragments at relativistic energies, this factor, negligible
up to tellurium, reaches 9 \% for gadolinium. Values, calculated using
the code GLOBAL \cite{Sche} are to be considered for  heavy elements as
shown in table 1.

\subsection{Discussion of uncertainties}

  a)  Systematic uncertainties

\begin{itemize}

\item  The uncertainty on the SEETRAM calibration is evaluated to 7 \% 
\cite{Jura}

\item The uncertainty on the transmission is evaluated to 5 \%. It is
mainly due to the  4 \% uncertainty on the fission-fragment velocity.
\item  The uncertainty on the target thickness was established in a 
dedicated experiment to 2.5 \% \cite{Must} for a beam hitting the 
target in its center.  However, an uncertainty
of 6 mm in the vertical position of the cryostat supporting the H$_2$
cell leads to an increased uncertainty of 4 \% on the target thickness
\cite{Taie}.

\item  The loss factors given in table 1 are determined with an overall
uncertainty of 3 \%. 
  
\end{itemize}

All contributions quadratically added up lead to a total systematic
uncertainty of
 10 \%. This systematic uncertainty is not included in the error bars 
 of the cross sections presented in tables of Appendix 2
 and in figures 10-12.

 b)  Statistical uncertainties

\begin{itemize}
\item  The single-event uncertainty on the angular transmission is 
large and mainly determined by the large variance of
$\alpha$. $\sigma_{\alpha}$ is evaluated to 17 \% \cite{Pere}.In 
comparison the uncertainty on $\beta_{app}$ is negligible. Taking 50
observed events corresponding to the smallest cross section (20$\mu$b)
as the worst case, the statistical uncertainty for the transmission
reaches 4 \%. Statistical uncertainties are smaller than 
the systematic uncertainty.

\item Uncertainties for the yields have been determined from the
 ratio of the rates integrated for forward and backward emitted 
 fragments.  The difference between the measured ratio
 and the kinematically expected ratio, is taken as twice the 
 relative uncertainty
 of the yields. The absolute uncertainties are given in Fig. 4, Fig. 10-12
 and in Appendix 2. 
 For the most abundant isotopes the relative uncertainty is approximately
 2 \%, that is much less than the systematic uncertainty.
For the less abundant isotopes, the statistical uncertainties are 
dominant and overcome the total systematic uncertainty.
\end{itemize}

Isotopic cross sections are given in Appendix 2 and 
their distributions for each of the elements appear in
Fig. 10-12. They show a rather regular bell shape with some enhancement
in the region of neutron-rich isotopes for elements known as asymmetric
fission products.  At the maximum of the elemental distributions, cross
sections vary from 0.5 mb to 20 mb. Values down to 20 $\mu$b
 are measured, mostly in the neutron-rich region not polluted by 
 secondary reactions. On the neutron-deficient side, high rates of 
 secondary reactions limit cross-section measurements to values larger
 than a few mb.


\begin{figure}[!h]
\begin{center}
\includegraphics[scale = 0.9]{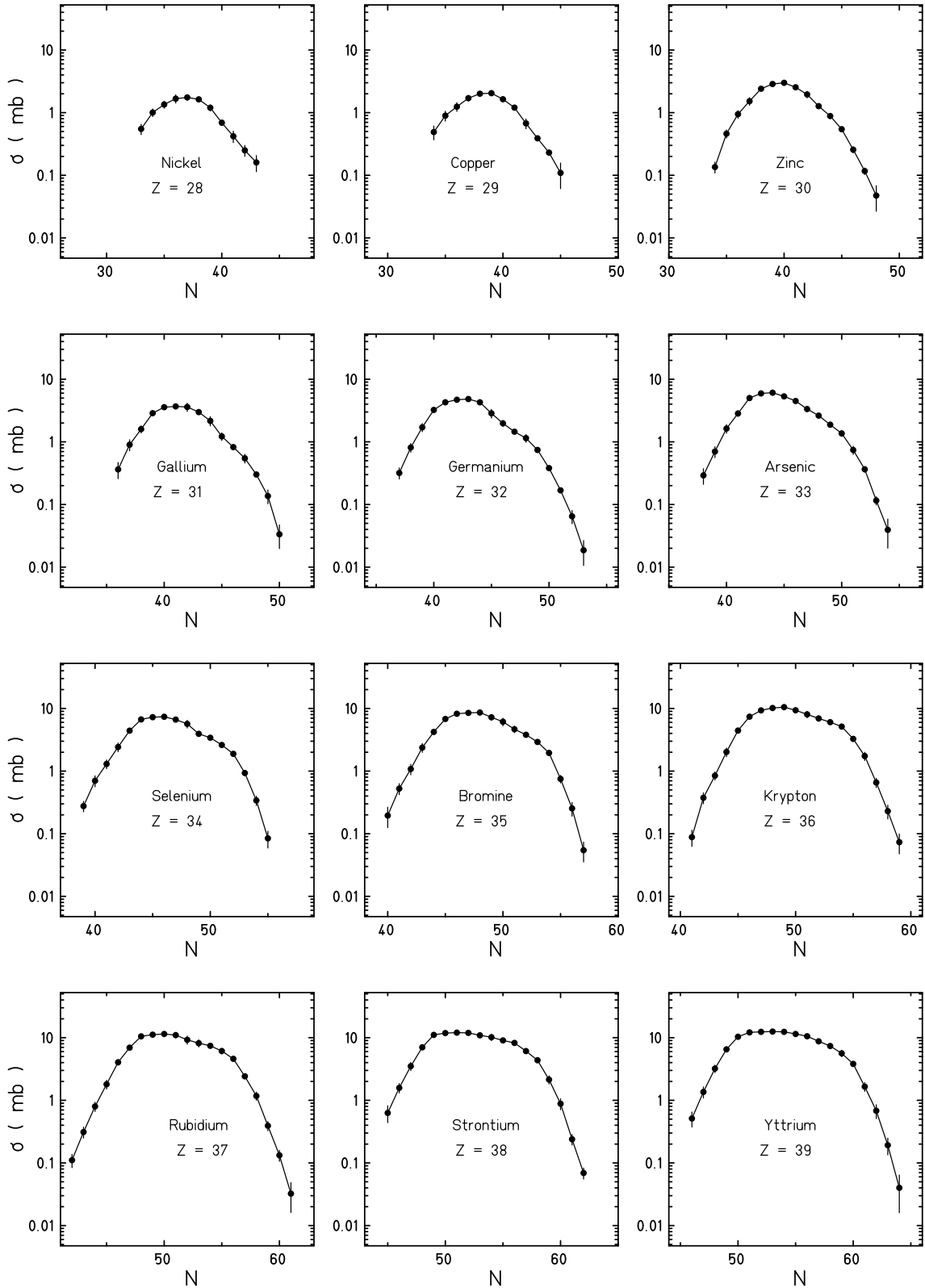} 
\caption{Isotopic cross sections for fission-fragments in  
the reaction $^{238}$U 1 A GeV + p
for elements  between $_{28}$Ni and $_{39}$Y.}
\end{center}
\end{figure}

\newpage

\begin{figure}[!h]
\begin{center}
\includegraphics[scale=0.9]{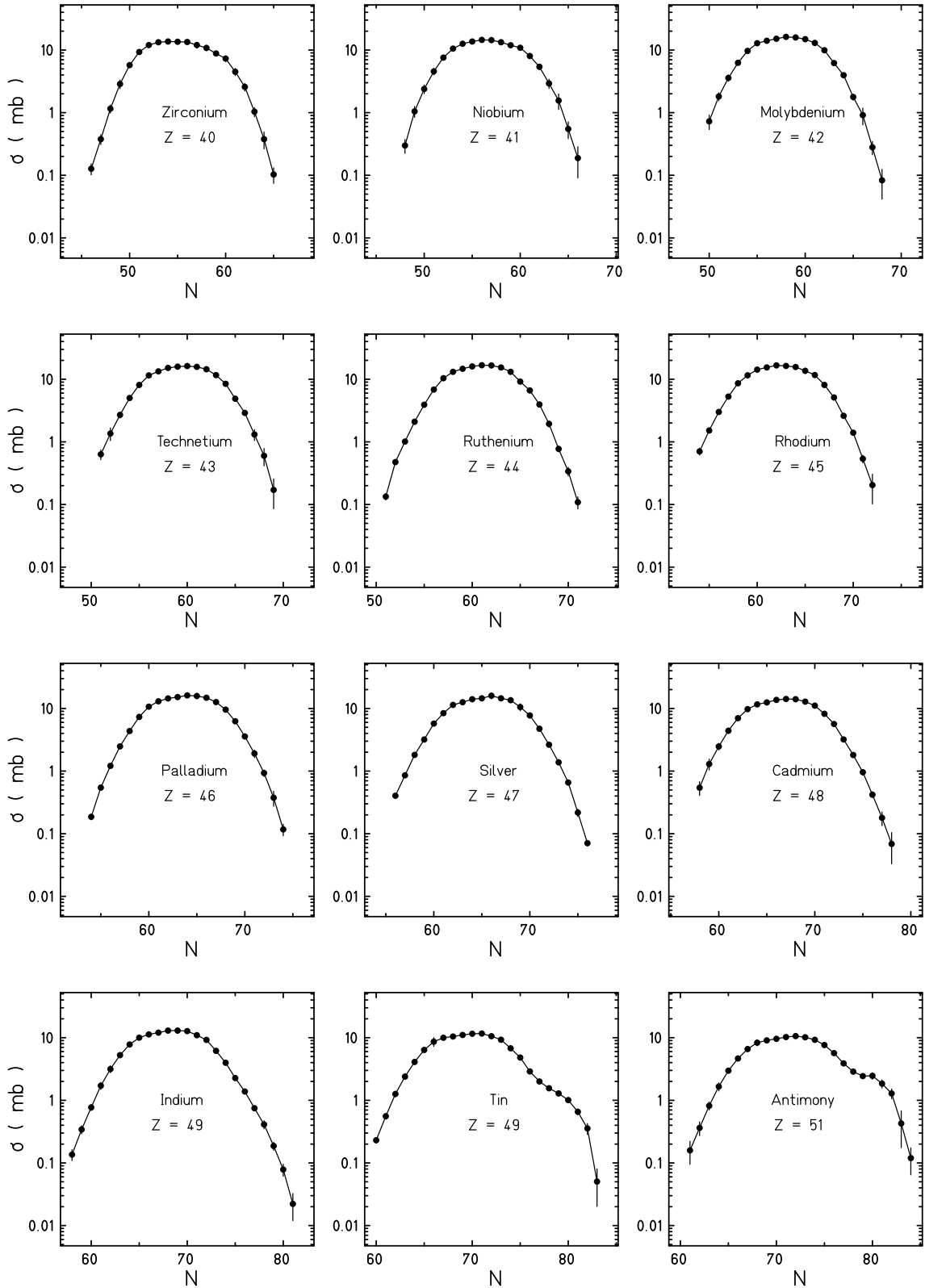} 
\caption{Isotopic cross sections for fission-fragments in  
the reaction $^{238}$U 1 A GeV + p
for elements  between $_{40}$Zr and $_{51}$Sb.}
\end{center}
\end{figure}

\newpage

\begin{figure}[!h]
\begin{center}
\includegraphics[scale=0.9]{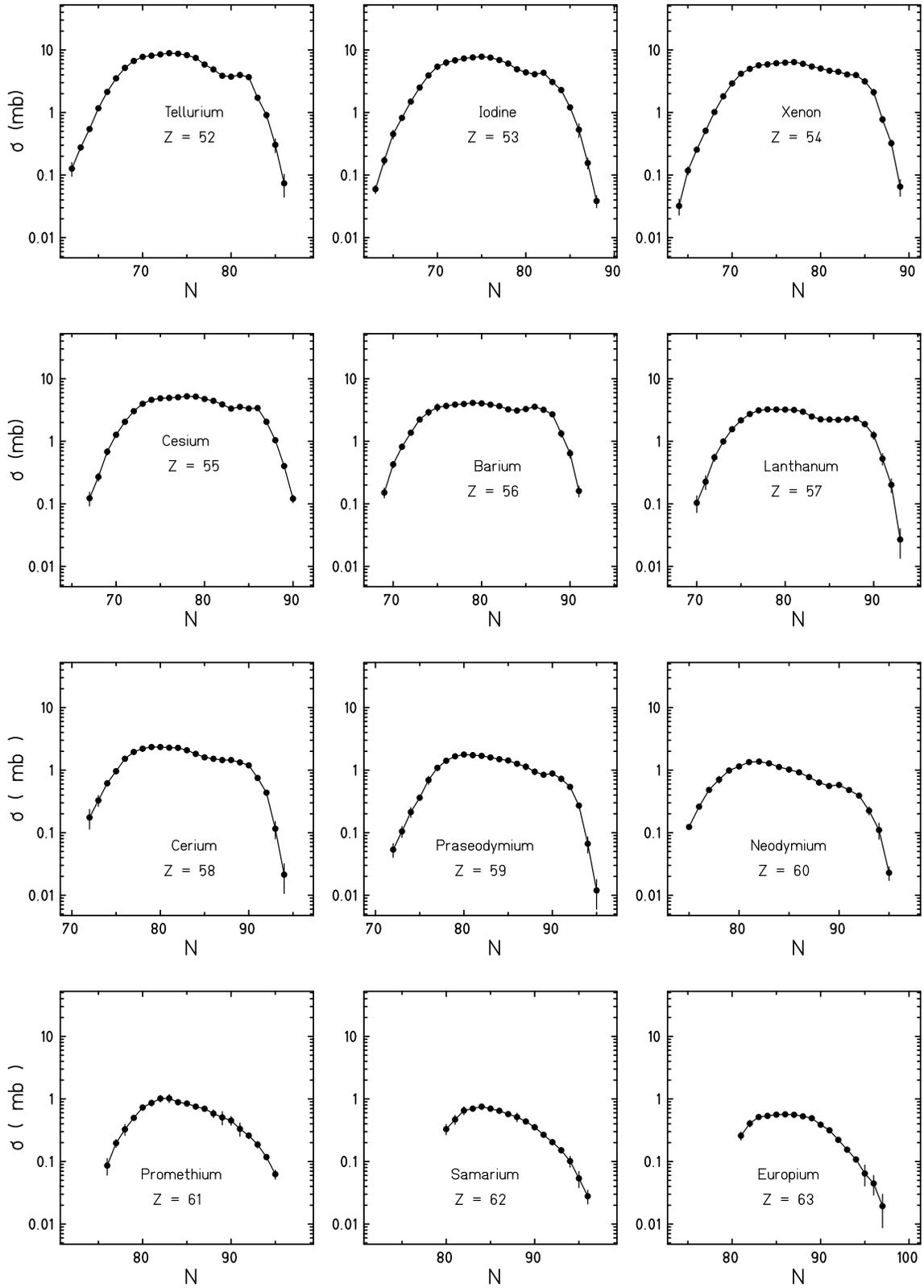}
\caption{Isotopic cross sections for fission-fragments in  
the reaction $^{238}$U 1 A GeV + p
for elements  between $_{52}$Te and $_{63}$Eu.}
\end{center}
\end{figure}

\subsection{Comparison with previous isotopic distributions} 

 Measurements of primary isotopic cross sections had been
performed for alkalines only. Using fast on-line separator techniques  
combined with the chemical selectivity of 
surface ionisation ion-sources \cite{Klap,Beli} relative cross sections
were obtained covering a 
wide range of
isotopes.
The normalisation of the yields to cross sections was based on the 
results of G. Friedlander
 \cite{Frie}. Production yields on U-fission
 were measured at a proton energy of 24 GeV for isotopes of  
 the  complementary elements rubidium and cesium  
 \cite{Chau}.
 The distributions are compared to our results in Fig. 13.
At 24 GeV, besides fission, evaporation residues populate all 
elements down to the region covered by fission.
Moreover, the excitation energy range covers a wider domain, and
evaporation of charged particles and neutrons leads to more 
neutron-deficient isotopes.
The elemental cross sections  for both rubidium and cesium 
are about 75 \% of the  values presently measured at 1 GeV, in the same
ratio as total fission cross sections \cite{Prok}.
In the region of the 8 most neutron-rich fragments, isotopic cross
sections are the same as the ones presently measured.
The probability to excite the U nucleus with a small excitation energy
to produce neutron-rich fragments is independent of the  proton energy,
as indeed mentioned already in \cite{Frie}. 

\begin{figure}[h]
\includegraphics[scale=0.8]{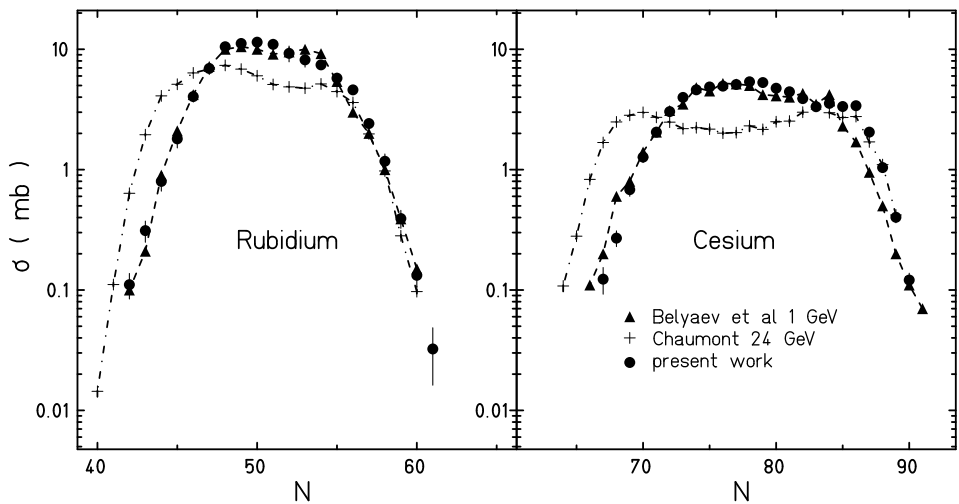} \hfill
\caption{ Comparison of  isotopic distributions  previously measured 
by J. Chaumont
 \cite{Chau} -dotted-dashed line- and by B. Belyaev \cite{Beli} -dashed
 line-with the present data -full points-.}
\end{figure} 

Later the production of the isotopes of cesium and rubidium was
investigated for the system $^{238}$U + p at  1 GeV  by
using also on-line mass separator techniques \cite{Beli}.
For rubidium our data are very consistent, but cross sections are found
larger by a factor 1.8 for the six
cesium-isotopes with N$\ge$85, as seen in
Fig. 13. In the present experiment, isotopic yields are measured
simultaneously for the 36 elements populated in fission, insuring a coherent
mass calibration for cesium as for all other elements including rubidium. 
Moreover the extraction of the asymmetric fission component of
all elements achieved in sec. 4-3 confirms the present mass calibration.
Finally
the agreement of our isotopic distributions with those of the light elements
analysed in the work of \cite{Ricc}, also validates 
our results for the cesium isotopic distribution.
In the on-line mass-separator measurement \cite{Beli} the separation
efficiency might have overestimated for these isotopes.
\section{Features of fission}

\subsection{ Elemental cross sections}

The elemental cross sections, integrated over isotopes for each element
are plotted in Fig. 14a. The distribution shows a Gaussian shape with
a shoulder around Z = 52 revealing the contribution of  asymmetric 
fission. The mean value of Z is 
$\overline{Z}$ = 45.
It suggests that $_{90}$Th is the mean element undergoing fission.
The variance of the Z-distribution is 7 charge units.

 In Fig. 14b the ratio of the mean number of neutrons 
 $\overline{N}$/Z for each element is given.  The largest mean number
 of neutrons per element and coherently the largest variances are found
 around Z = 55.  A  mean number of  neutrons $\overline{N}$ = 63 is 
 deduced.

 The local variance $\sigma^Z_N$, plotted on Fig. 14c, increases from
 2.5 to almost 5. The presence of the asymmetric 
 fission  component is revealed by the maxima at the two peaks of 
 asymmetric fission.
\subsection{Total cross sections}

The fission cross section summed over all isotopes is found to be
(1.45 $\pm$ 0.15)~b. However, a few elements are not involved in the
analysis, in the region above Z = 64 and below 28. Above Z = 64 the
cross section contributing is evaluated to 7 mb \cite{Ruiz}. Below
Z = 28, the analysis of M. V. Ricciardi \cite{Ricc} gives a
contribution of 72 mb. Thus the total fission cross section amounts to
(1.53 $\pm$ 0.15) b.
This value compares well with previous results of direct measurements
of fission cross sections (1.46 $\pm$ 0.07) b and
(1.48 $\pm$ 0.06) b \cite{Boch,Vais}, respectively. A value of 
(1.52 $\pm$ 0.12) b was obtained in a recent measurement at GSI 
\cite{Jura2}. It is worth noticing that the present technique offers a
high degree of selectivity and sensitivity to measure isotopic yields
but a lower precision for total cross section measurements than simpler
dedicated experiments.
If the cross section for evaporation residues of 0.46 b, obtained by
summing isotopic values found by J. Taieb in his analysis of the 
$^{238}$U 1 A GeV + p  reaction \cite{Taie} is added,
a reaction cross section of (1.99 $\pm$ 0.17) b is obtained. It agrees
with the value of 1.96 b obtained by using the Glauber approach as
described by P. J. Karol 
\cite{Karo} with updated nuclear-density distributions \cite{Broh},
and with the INCL-calculation 1.94 b \cite{Boud}.

\subsection{ Symmetric and asymmetric fission}

\begin{figure}[h]
\includegraphics[scale=0.8]{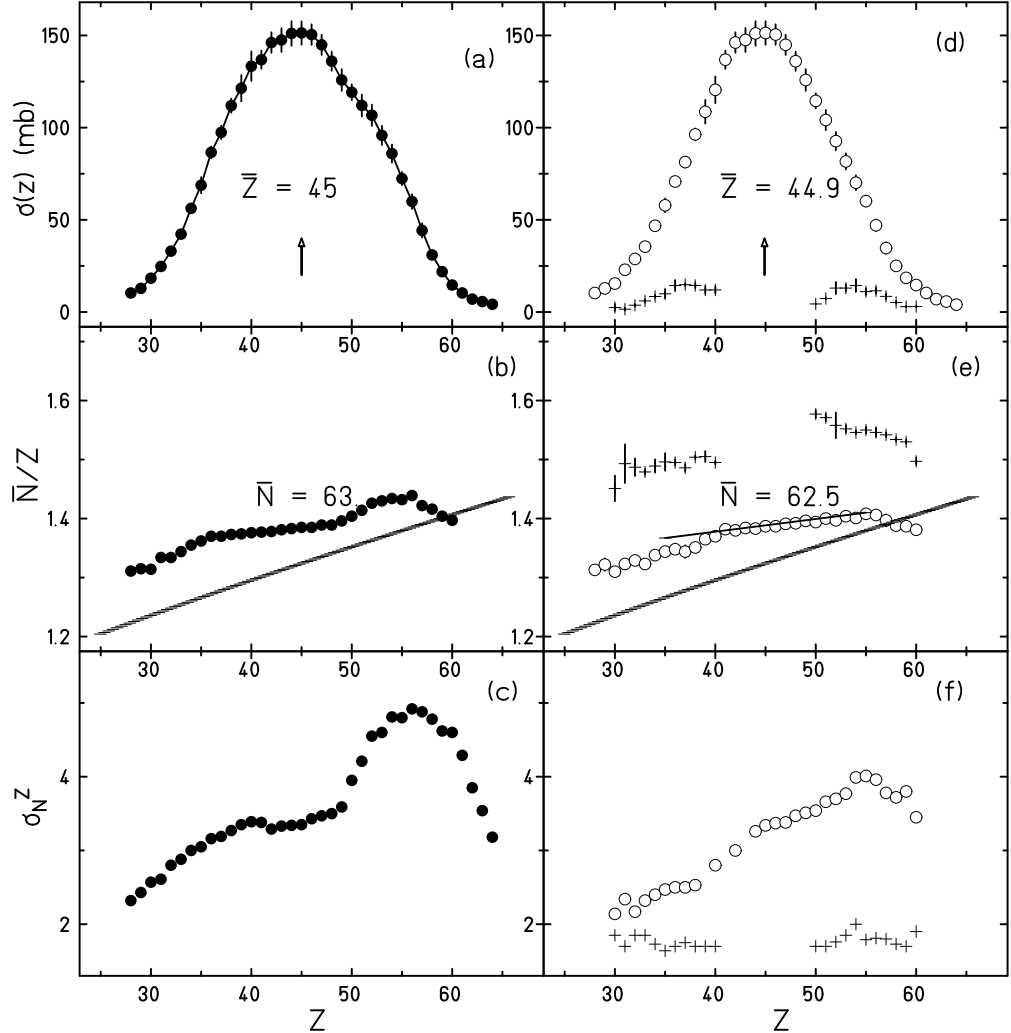} \hfill
\caption{ Integrated distributions of fission fragments from ($^{238}$U
 + p) at 1 A GeV (full points), symmetric fission (empty points)
and asymmetric fission  (crosses). a) and d): Z-distribution,
b) and e) mean neutron number divided by Z as a function of Z, and c)
and f) 
variance in neutron number  $\sigma^Z_N$ for a given Z. The short line
in frame e) corresponds to P~=~-0.04 \cite{Armb2}. The thick line in
frame b) and e) shows the position of the stable isotopes.}
\end{figure}

 The isotopic distributions of elements above Z = 50 can be decomposed 
 into fragments arising from two components: high-excitation symmetric
 fission and low-excitation asymmetric fission. A decomposition into
 two Gaussian distributions has been performed, as illustrated in 
 Fig. 15.  For the neutron-rich component the mean
 mass number A$_p$ and the dispersion $\sigma^Z_N$ are found to agree
 with the values known in asymmetric fission, see for example 
 C. Donzaud et al.
 \cite{Donz}. 
 For the light fission fragment group, the share of asymmetric fission
 does not exhibit so clearly and the known values of A$_p$ and 
 $\sigma^Z_N$ \cite{Donz} are used as further inputs.

\begin{figure}[h]
\includegraphics[scale=0.35]{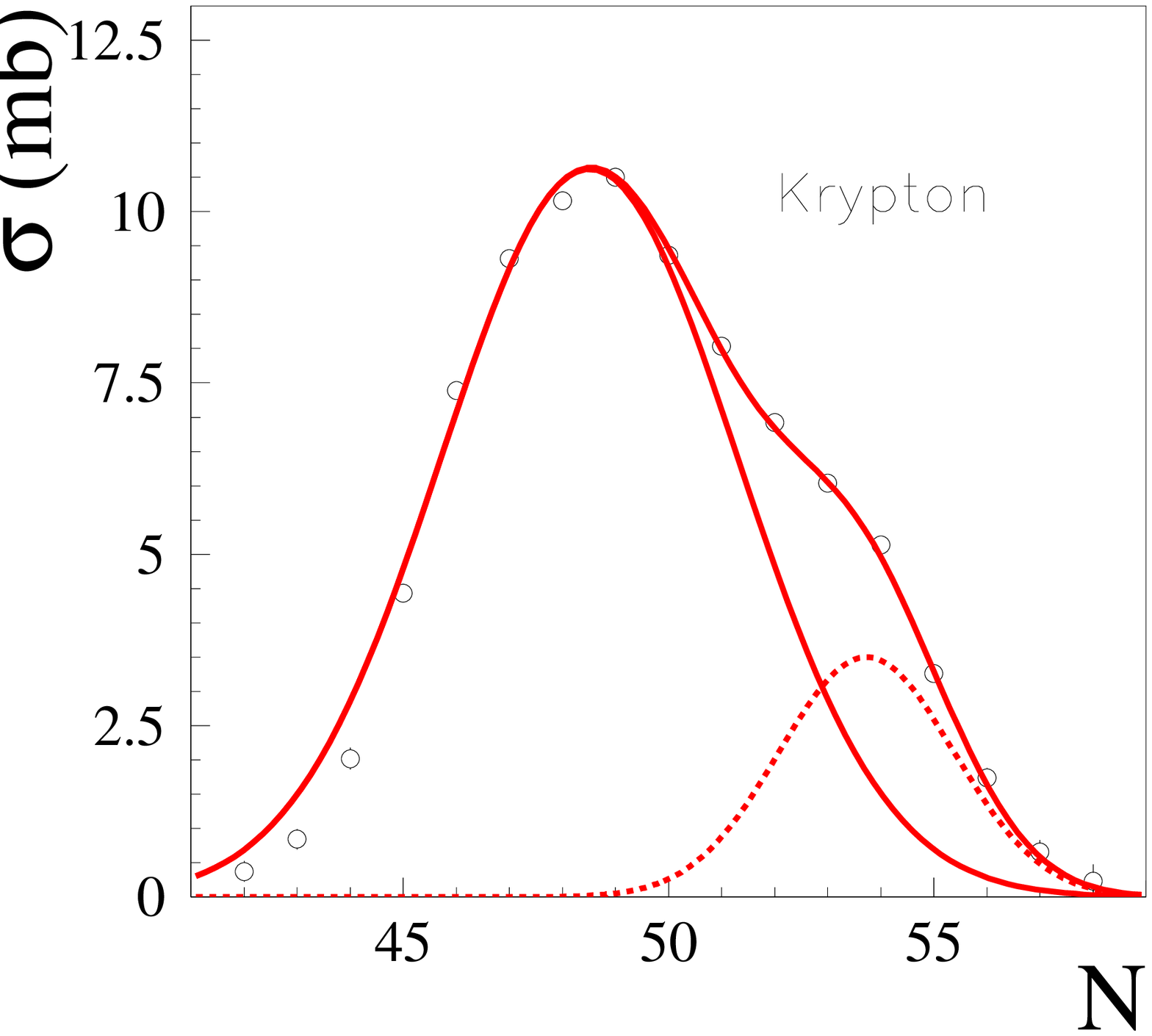} \hfill
\includegraphics[scale=0.35]{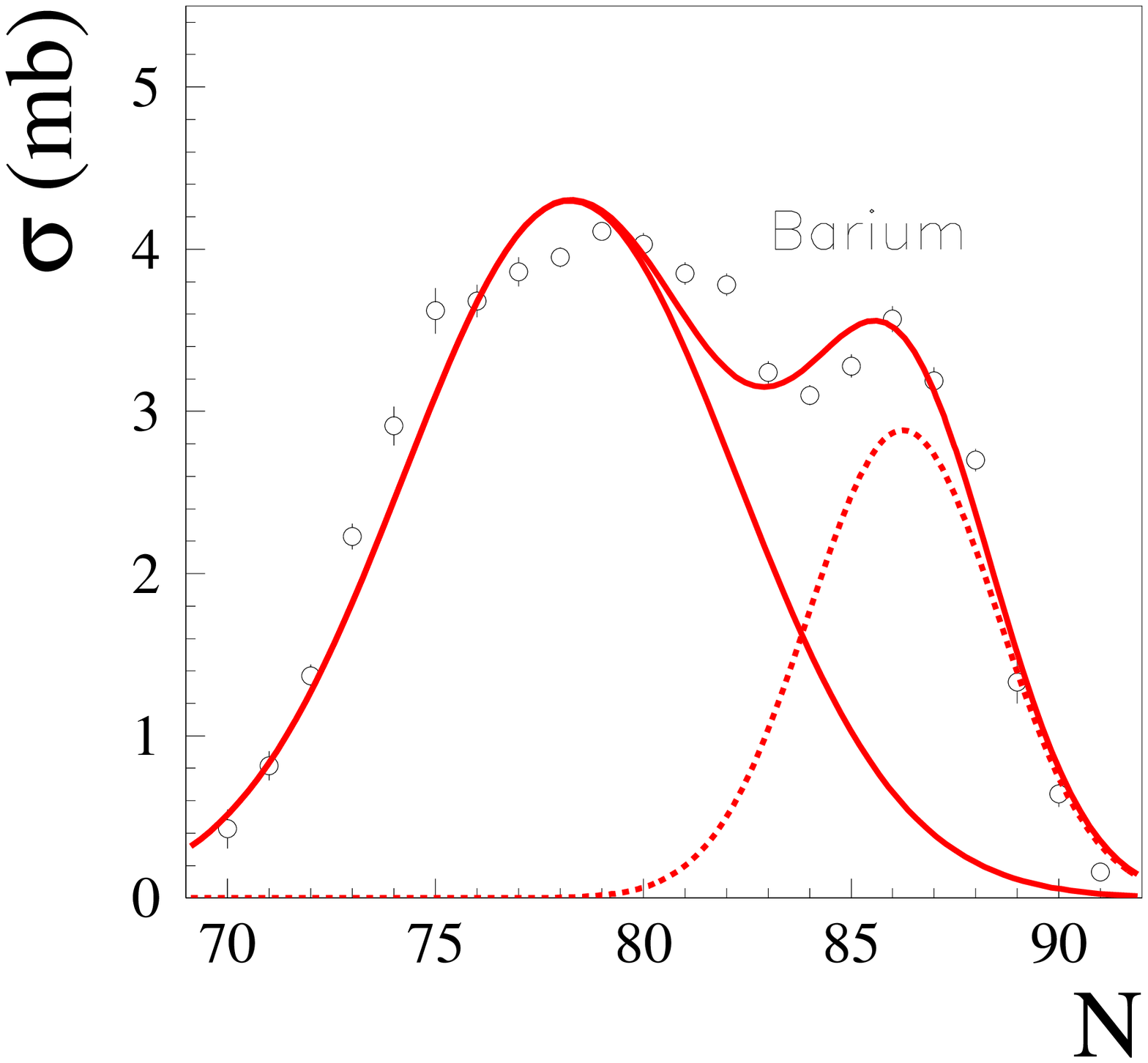} \hfill
\caption{ Examples of the shares of symmetric (full lines) and 
asymmetric (dotted lines) fission processes. The sum, resulting from the
least square fit is indicated.  }
\end{figure}  

In spite of the small yields of the
 asymmetric component, cross sections for the groups of light and heavy
 fragments are obtained. Z-integrated cross sections  are
 found the same for the two groups, as to be expected, and equal to
 (105 $\pm$ 10) mb. For symmetric fission a cross section of (1.42
 $\pm$ 0.15) b follows. The distribution of elemental cross sections
 becomes a symmetric curve  slightly shifted (empty points in 
 Fig. 14d compared to the total distribution  on Fig. 14a.
  The parameters  $\overline{N}$ / Z and variances $\sigma^Z_N$ related
 to the two modes are clearly different. Their values are compatible 
 with two mean primary fissioning nuclei : for the asymmetric mode
 $^{234}$U and  for the symmetric mode $^{221}$Th,
 with 6 post-scission neutrons added.

 The  $\overline{N}$ / Z values of the fragments are not constant , but a
 slow and regular increment of  $\overline{N}$ / Z with the Z of the
 fragment is observed for the high energetic symmetric fission process,
 see Fig 14e. An electric polarisation
 $\delta A_2^{Pol}$~= 
 -~$\delta A_1^{Pol}$~= 
 $\overline{A_2}$ - Z$_2$$\times$A$_0$/Z$_0$ $>$0 for the fission fragment
 pairs  $\overline{A_2}$, $\overline{A_1}$ is indicated.
 Beyond the regime of nuclear structure dominating asymmetric fission
 (E$^*$ $>$~40~MeV) a constant charge polarisability of P~=~-0.04 is
 expected  \cite{Armb2}. P~can be related to the slope of the
 $\overline{N}$ / Z-dependence in the range of symmetric fission
 (Z$_0$/2~=~45):
 $$ P = \frac{-\delta A_2^{pol}(Z)}{A_2 - 0.5A_0} =
 -(Z_0/2)\times(Z_0/A_0)\times \frac {d(\overline{N}/Z)} {dZ}$$
 The thin line in Fig. 14e shows a slope calculated with P~=~-0.04.


 On Fig. 14f the variance $\sigma^Z_N$ in
 the region of symmetric fission increases with the Z-value of the
 elements. On the contrary for asymmetric fission, as known for a long
 time \cite{Sist},   a narrow distribution of isotopes $\sigma_Z^A$  is
 populated with a large neutron excess $\overline{N}$/Z .
 Beyond Z = 55 the values of  $\overline{N}$ / Z and of  $\sigma^Z_N$
 decrease again and the valley of stability is crossed at Z = 58. 
  For elements still heavier, fission fragments populate the
  neutron-deficient  side of the valley.


\subsection{Mass distribution}

\begin{figure}[h]
\begin{center}
\includegraphics[scale=0.6, angle= -90]{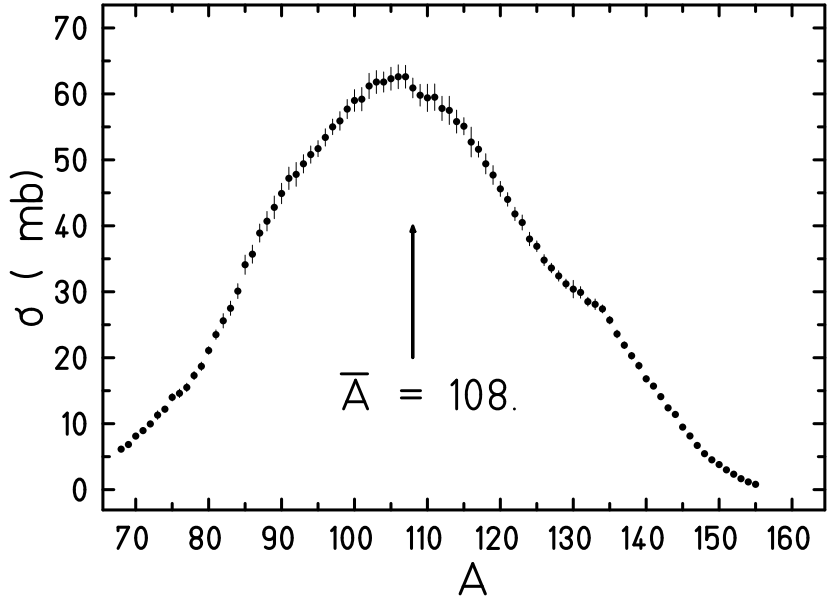} \hfill
\caption{ Mass-distributions of fission fragments for $^{238}$U (1
A GeV) + p.} 
\end{center}
\end{figure}  

The mass distribution is obtained by summing isotopic cross sections
for a value of A fixed, Fig. 16. Since isobaric yields do not depend upon 
$\beta$-decay half-lives, mass distribution are accessible also by
radiochemical methods  and spectroscopic methods 
\cite{Glor,Tita,Weng}. The distribution shows a wide 
Gaussian-like shape slightly
asymmetric: the slope is steeper on the side of light masses.

 The mean mass is $\overline{A}$ = 108.0 $\pm$ 0.3 and the variance of
 the distribution $\sigma_A$ = (17.5 $\pm$ 0.5) a.m.u. Enhancements on
 both sides result from the asymmetric fission component. Since 
 elements below Z = 28 are neglected, which contribute by about 5\% to
 the total cross section \cite{Ricc}, the value of $\overline{A}$ given
 is slightly larger than the average over all fission fragments 
 produced. We assume that*** fission proceeds from  a mean parent nucleus
 of mass A = 220 with  $\nu$ = 6 the averaged number of emitted  
 neutrons from the symmetric and asymmetric modes.
 On table 2 the cross sections, mean values of A and Z, variances and 
 local variances are presented and compared to the results of symmetric
 fission in (U + Pb) \cite{Schw}.

\input{tab3.tex}

\subsection{Kinetic energies}

\begin{figure}[h]
\begin{center}
\includegraphics[scale=0.6]{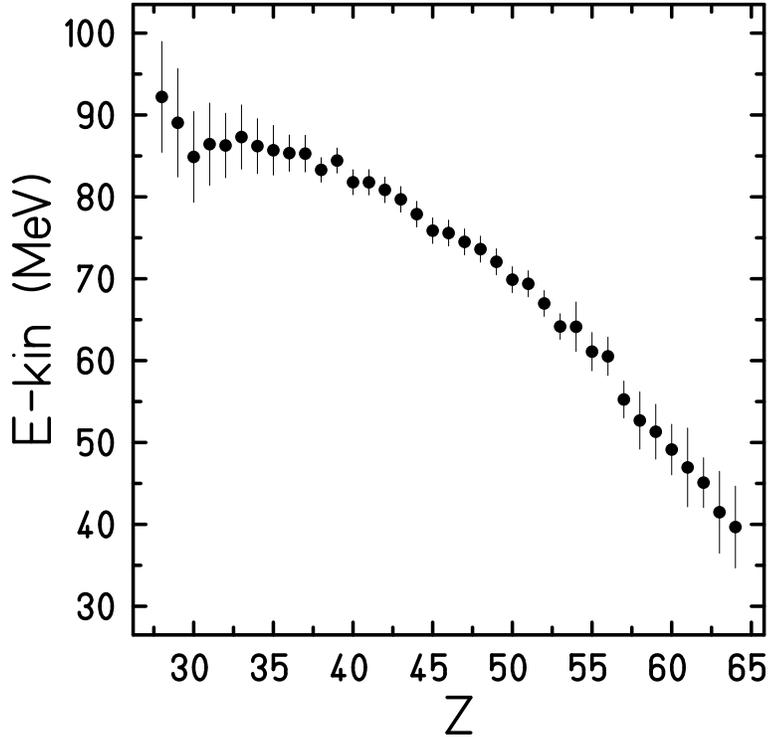} \hfill
\caption{ Mean kinetic energy of fragments from 1 A GeV $^{238}$U + p 
fission as a function of their atomic number Z.} 
\end{center}
\end{figure}  

The average kinetic energy of fission fragments as a function of their
atomic number is given by E$_{kin}$ (Z) = 
$\frac{1}{2}$.m$_0$  $\overline{A}(Z)V_f^2$ where $\overline{A}$(Z) is
the average mass number obtained from the measured
 isotopic distribution, and V$_f$ is the corresponding fragment velocity
 as shown in Fig. 9a.
 The mean kinetic energies are given in table 3 and shown in Fig. 17. 
 They decrease smoothly for higher atomic numbers and show a broad
 maximum around Z = 32. The mean value of the kinetic energy, 
 calculated by using the Z-distribution of Fig. 14a gives 
 (76 $\pm$ 3) MeV, corresponding to a mean total kinetic energy
 release of (152 $\pm$ 6) MeV.

\input{tab4.tex}


\subsection{ Comparison with simulation codes}

\begin{figure}[h]
\begin{center}
\includegraphics[scale=0.85]{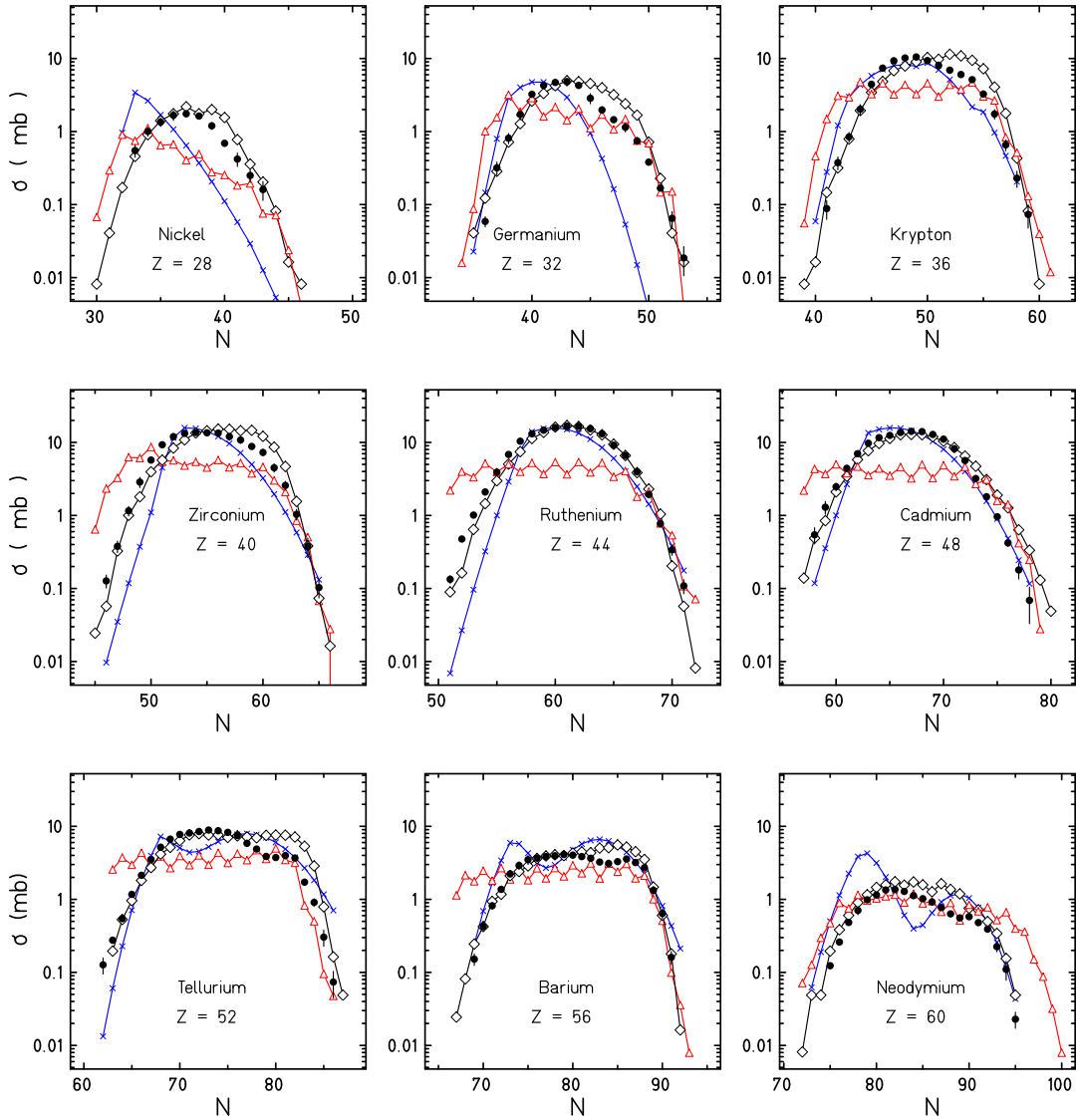} \hfill
\caption{ Comparison of measured isotopic distributions (full points) 
to calculations with the formula of Silberberg et al.  
\cite{Silb} (crosses), by LAHET \cite{LAHET} (triangles), and by
our improved codes \cite{Gaim,Jung,Benl,Boud} (diamonds) for a
selection of elements}
\end{center}
\end{figure} 

The set of isotopic distributions of fission fragments provides a real
challenge for simulation codes. A fast and simple formulation developed
by Silberberg and Tsao \cite{Silb} is shown in Fig. 18 and 19. The 
model reproduces the orders of magnitude of the isotopic distributions,
but the asymmetric fission component is overestimated in magnitude and 
symmetric components are too narrow.
Isotopes of elements at and below Z = 35 are poorly
reproduced. A jump at Z = 35 in the code produces the unrealistic gap 
in the mass distribution  at A = 90.

\begin{figure}[h]
\begin{center}
\includegraphics[scale=0.6,angle = -90]{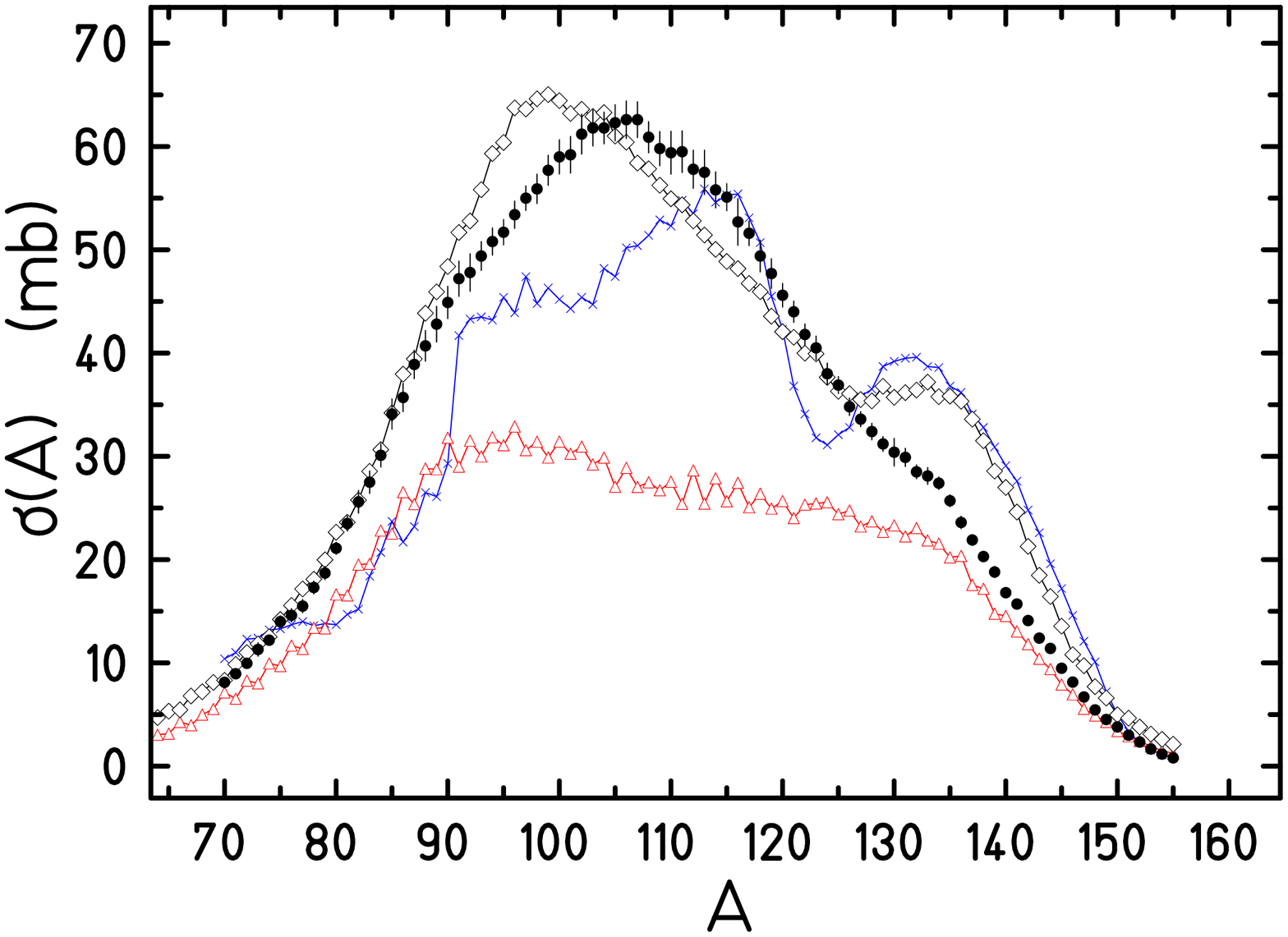} \hfill
\caption{ Mass distribution of fission fragments as a function of mass
number compared to simulations; Silberberg \cite{Silb} (crosses), 
LAHET \cite{LAHET}
(triangle) and to our improved codes \cite{Gaim,Jung,Benl,Boud} 
(diamonds). Experimental data
are shown as full symbols with error-bars.} 
\end{center}
\end{figure}  

The second code widely used is the LAHET-code  (Los Alamos \cite{LAHE}
or Oak Ridge version) \cite{LAHET}.
The agreement with this code is not satisfying. In Fig. 18 and 19
the Oak Ridge version has been used for comparison. 
For neutron-deficient isotopes, cross sections are
overestimated, and symmetric fission is underestimated. An odd-even
staggering is produced not seen in the experimental data.

Parallel to our experimental program different approaches to new
simulation codes are published or are under way.
The first
reaction step in spallation, the intranuclear cascade induced by the primary
proton collision, is simulated by the new version of the code INCL
\cite{Boud}. At this stage, a distribution of prefragment nuclei are 
predicted
with an associated excitation energy and intrinsic spin. This is then
the physical input to a system of codes simulating the second 
step of spallation, the de-excitation phase. New results on the
physics of particle evaporation were introduced in the statistical
de-excitation code ABLA \cite{Gaim,Jung} and on fission in the fission
code PROFI \cite{Benl}. 

The fission code PROFI is a semi-empirical Monte-Carlo code which
calculates the nuclide distribution of fission fragments.
It is theoretically based on the application  of the
statistical model of nuclear reactions to the concept of fission 
channels. In this model, the
population of the fission channels is assumed to be basically
determined by the number of available transition states above the
potential energy surface near the fission barrier. Several
properties, however, are finally determined at the scission time.
A full description of the model is given in \cite{Benl}.

The barrier as a function of mass asymmetry is defined by three
components. The first is the symmetric component defined by the liquid
-drop potential by means of a parabolic
function with a curvature obtained from experimental data \cite{Mulg}.
This parabola is assumed to be
modulated by two neutron shells, located at mass asymmetries 
corresponding to neutron numbers N = 82 (spherical neutron shell) 
and N = 90 (deformed neutron shell). We assume
that the mass-asymmetry degree of freedom at the fission barrier is on
the average uniquely related to the neutron number of the fragments.
The shells are represented by Gaussian
functions. These shells are associated with the fission channels 
Standard I and Standard II, respectively \cite{Broz}, while the 
liquid-drop potential is associated with the symmetric fission channel.
The population of the fission channels is proportional to the level 
density around the corresponding dips in the potential at saddle at the
available excitation energy. Shells are washed out with excitation
energy \cite{Igna}. The heights and the widths of the Gaussian curves
representing the shell effects and additional fluctuations in mass
asymmetry acquired from saddle to scission, are derived from
experimental data \cite{Benl}. The mean values of the neutron-to-proton
ratio for the channels Standard I and Standard II are deduced
from measured nuclide distributions
after electromagnetic induced fission of $^{238}$U \cite{Donz}. The
charge polarisation for the symmetric fission channel and the 
fluctuations in the neutron-to-proton ratio for all  channels are also 
considered by describing the potential in this degree of freedom again
by a parabolic function \cite{Armb2}. Assuming that the equilibration 
in this variable is fast compared to the saddle-to-scission 
time, this potential was calculated in the scission configuration.
Since the  shell effects of the nascent fragments at scission, which
are strongly deformed on the average, are not known experimentally, 
only macroscopic properties are included in the calculation 
of the charge  polarisation of the symmetric fission channel. 

Consequently, the two fission pre-fragments are obtained. Their
excitation energies are calculated from the excitation and 
deformation energy of the fissioning system at the scission point.
The fission probability and the consecutive de-excitation of the 
fission fragments via particle  evaporation are conducted by the
routines of the ABLA code \cite{Gaim,Jung}. Coupling to the cascade
code INCL \cite{Boud} gives a reasonable reproduction of our data, as
demonstrated in Fig. 18 and 19  (diamonds). Except for an excess of
asymmetric fission the simulation meets the experiment. Either INCL
provides a too small excitation energy for the most peripheral
collisions, or the excitation energy in the statistical de-excitation 
chain is underestimated or, finally the parameter set in PROFI has
to be reajusted in order to reduce the share of asymmetric fission.

\subsection{ Production of very neutron rich isotopes}

The experimental method of 1 A GeV U-projectile fission opens a very
efficient way to produce secondary beams of neutron-rich isotopes. 
In order to provide experimental cross sections for such a project,
the production of very neutron-rich fragments by 1 A GeV U+p collisions
is evaluated renormalising the values obtained from a previous
dedicated measurement for U+Be \cite{Enge}. The production of fission 
fragments differs in both systems because of a relatively larger
contribution of asymmetric fission with U + Be. Normalising the
elemental cross sections for U + Be to the corresponding element for
U + p, the production of very neutron-rich can be deduced  reliably
down to a level of 1 nb. Not only the  falls of the isotopic production
with increasing neutron excess are the same for  U + Pb and U + Be
\cite {Donz,Enge}, but also the production is  found to be the same in
the gap of overlapping masses presently covered in U + p.

\begin{figure}[h]
\begin{center}
\includegraphics[scale=0.6]{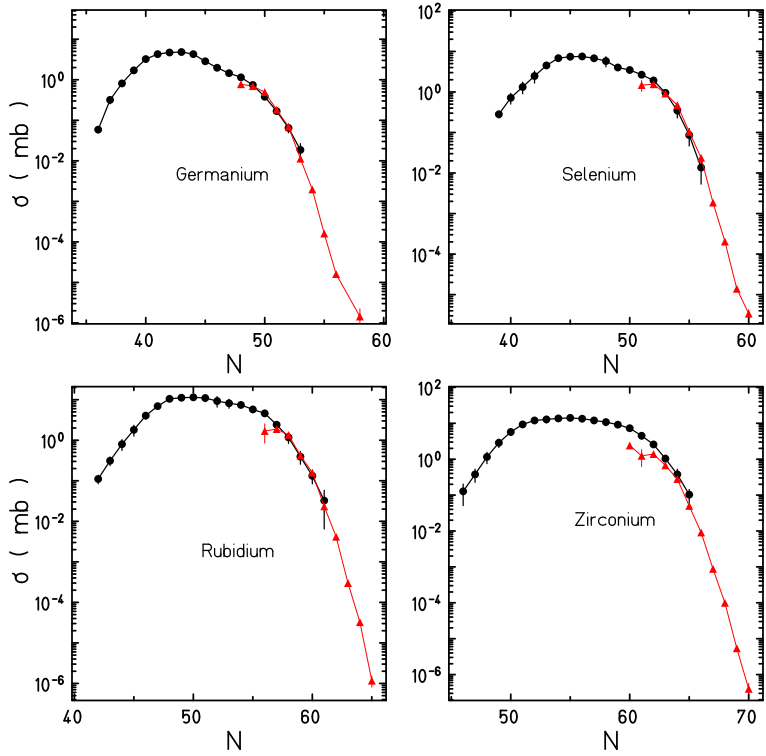} \hfill
\caption{ Comparison of cross-sections from $^{238}$U 1 A GeV + p
(points) to $^{238}$U 0.75 AGeV + Be \cite{Enge} (triangles). These
last data are renormalised in $\sigma$(Z) to the present elemental
cross sections.} 
\end{center}
\end{figure}

Fig. 20 illustrates four examples of isotopic distributions.
Such exotic beams in spite of their low intensities will be of
importance to test predictions for the vanishing of shell-effects at
N = 82
and 50, and
the possible appearance of neutron halos in the region of tin and 
nickel isotopes with large neutron excess.

\subsection{Parent fissioning nuclei}

The reconstitution of the charge Z$_0$ and mass A$_0$ of the parent
fissioning nuclei is of importance to constrain critical parameters of
the intranuclear cascade phase: the transfer of excitation
energy and of momentum to the pre-fragments, the value of the density
of intermediate states and of the viscosity of highly excited heavy
nuclei. The  present experimental findings should converge
with all other results of $^{238}$U at 1 A GeV + p
\cite{Taie,Ricc,Jura2} on an unified description of a first stage of 
nucleon-nucleus  collision described by a cascade followed by 
an evaporation stage among which fission occurs.

What is learnt about the fissioning parent nuclei in the present work is 
listed below:
\begin{itemize}
\item Isotopic distributions lead to a mean fission parent nucleus
$^{220}$Th. 
\item For a given element the  increase of the mean
velocities with the mass of the isotopes (Fig. 8) indicates that parent
elements in the range of Z$_0$ $\le$ 90  do contribute to the fission.
\item Velocities of fragments Fig. 9a show that the region of
50$\le$~Z~$\le$ 64 arises from parent elements of 
$\overline{Z_0}$ $\approx$ 90, while the region of 30~$\le$~Z~$\le$~37
indicates lower fissioning elements at $\overline{Z_0}$  $\approx$ 84.

\item The variance of the fission fragment velocities $\sigma_{Vf}$ is 
extracted from the external slopes of the observed velocity
distributions  and presented in Fig. 9b. The
value of 0.13 cm/ns, the same for all fission-fragment elements, is
correlated to  the variance
of the recoil momenta generated in the nuclear cascade for the ensemble
of the fissioning parent nuclei. 
At $\Delta$A = 18, the analysis of spallation presented by J. Taieb 
\cite{Taie}
gives  $\sigma_p$ = 350 MeV/c as the variance of the recoil momenta
of the central parent nucleus, $^{220}$Th. This variance is transferred
to the fission fragments and leads to a contribution of 0.09 cm/ns in
the fissioning system. The ensemble of parent elements and isotopes 
contributes also by the fluctuations of their TKE-values. A pair of
fission-fragments issued from a fixed parent nucleus shows a variance
of its velocities. Taking from thermal-neutron induced fission of 
$^{235}$U the measured value of $\sigma_{TKE}$ = 6.0 MeV as a lower
estimate,  a contribution of 0.02 cm/ns to the variance of fission
fragment velocities is calculated. A much larger contribution is
generated by the difference in TKE-values of the many nuclei present.
The variance in the Z$_0$-distribution of the fissioning parent nuclei
of $\sigma_{Z_0}$/Z$_0$  of 4.6\% is taken from the measurement of
B. Jurado \cite{Jura2}. Including the dispersion due to a spread over
several isotopes in each element, and converting TKE-values into 
fragment-velocities, the contribution of the fissioning ensemble is
evaluated to 0.09 cm/ns. The location straggling in the thin
H$_2$-target induces a negligible contribution of 0.01 cm/ns to this
variance.  Adding quadratically all contributions, a total variance of
0.13 cm/ns is obtained which coincides with the value extracted
from the observed  velocity distributions.
 
\item The mean recoil velocities of the parent fissioning
nuclei arise from the first phase of the reaction
and contribute to the momentum losses measured 
\cite{Taie}. The transferred momentum of
-150 MeV/c measured for the evaporation residues at
$\Delta$A = 18 should be close to the momentum of the mean fissioning
parent $^{220}$Th. This recoil momentum is equivalent to a recoil
velocity of -0.04 cm/ns, indicated by a line in Fig. 9c. The recoil
velocities presented in Fig. 9c, in the range of -(0.04-0.13) cm/ns are
on the average larger by a factor of 2 and show a trend towards larger
values for lighter elements.

\end{itemize}

\subsection{Comparison of system U + p and Pb + p}

\begin{figure}[h]
\begin{center}
\includegraphics[scale=0.5]{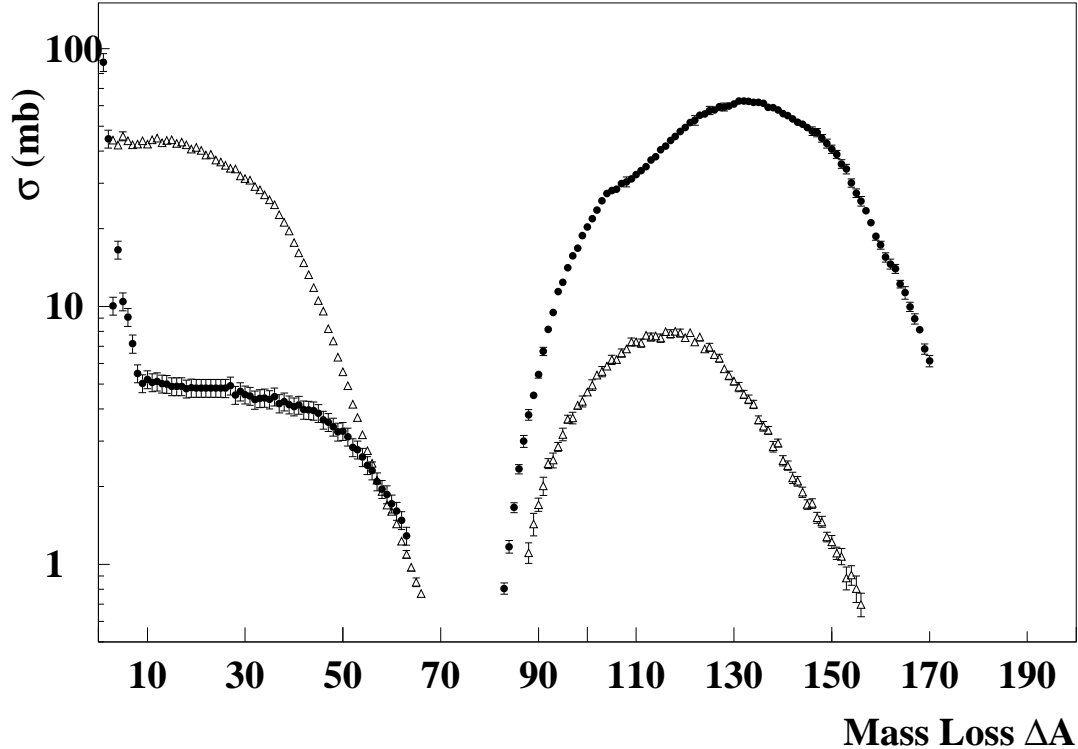} \hfill
\caption{ Comparison of cross-sections for evaporation
and fission residues 
as a function of the  mass-loss $\Delta$A  for the two collision 
systems Pb~+~p (triangles) and U~+~p (full points) at 1~A~GeV.} 
\end{center}
\end{figure}

\input{tab5.tex}

Fig. 21 and Table 4 demonstrate the basic difference between the
collision systems Pb + p and U + p. Both reactions have nearly the same
reaction cross section, 1.84 b for Pb + p and 1.99 b for U + p. 
Fragmentation is the main reaction channel for Pb+p, with
$\sigma$ = (1.68 $\pm$ 0.2) b, whereas
fission dominates the U+p reaction with a cross section $\sigma$ =
(1.53~$\pm$~0.15) b. 


Averaging over all cross sections from evaporation and fission
residues, altogether a mean neutron number of 127 $\pm$ 1 is found for
reaction products of
$^{238}$U + p at 1 A GeV, i.e. 19 $\pm$ 1 neutrons are liberated in the reaction.
This number does not include neutrons bound in the lightest elements
( Z $\le$ 7 ), and should be an upper limit for the number of neutrons  
produced in thin targets. It compares with the neutron
multiplicity of 20 measured with 1.2 GeV proton on depleted uranium 
\cite{Gali} when extrapolated to thin targets.

\section{Conclusions}

We have measured isotopic production cross-sections for about 733
fission fragments produced in 1 A GeV $^{238}$U + p collisions with a
mean accuracy of 10 \%, down to values of 20 $\mu$b for neutron-rich 
and 0.1 mb for neutron-deficient isotopes. The population of isotopes
resulting from fission is pictured in Fig. 22 on the chart of nuclei.
The isobaric slope of the production cross section towards the neutron
rich side
of the chart of isotopes is found to be the same for hydrogen as for heavier
targets investigated earlier.

\begin{figure}[h]
\begin{center}
\includegraphics[scale=0.6]{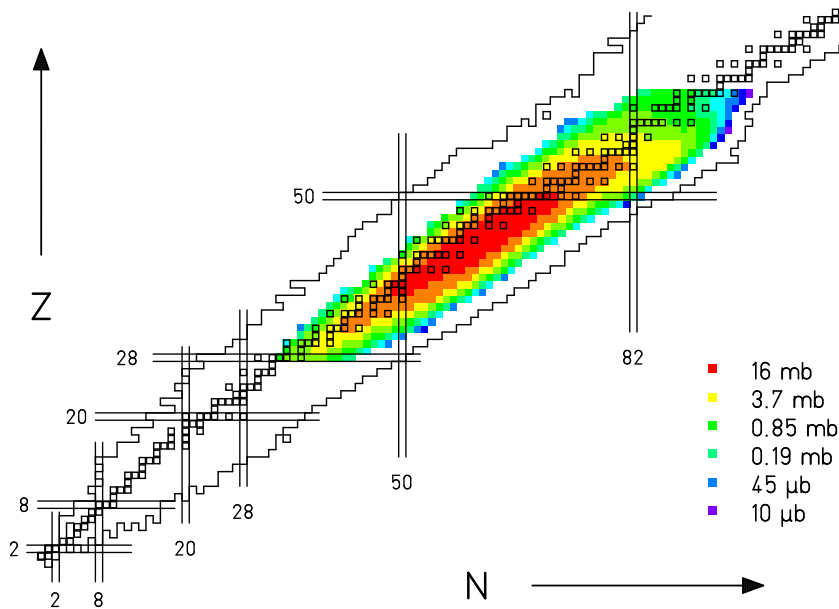} \hfill
\caption{ Two-dimensional plot of the isotopic cross sections for
fission-fragments obtained in 1~A~GeV~$^{238}$U + p shown on the chart
of isotopes with squares indicating stable isotopes. 
Colours correspond to increasing 
cross sections according to the logarithmic scale indicated.} 
\end{center}
\end{figure}

The fission cross section, including the production of very light
elements, amounts
to (1.53 $\pm$ 0.15) b which compares well with previous results 
\cite{Boch,Vais}
as with the
recent result of B. Jurado \cite{Jura2}. The total cross section
of fission plus evaporation residues amounts to (1.99 $\pm$ 0.17) b,
close to the calculated geometrical cross section of 1.96 b  
\cite{Karo,Broh} and to the INCL calculation 1.94 b \cite{Boud}.

Our results show a good quantitative agreement with previous isotopic
distributions of cross sections measured for rubidium and cesium with
on-line mass-separator techniques except for a discrepancy in cross sections
for the 6 most neutron-rich isotopes
of cesium \cite{Beli}. The coherence of the present experimental method
validates our mass-calibration relatively to all other isotopes. 
Parameters characterising 
fission and its variances are extracted. The elemental distribution
and the kinetic energies of the reaction products are of importance 
concerning chemical corrosion and the composition of nuclear waste
produced by spallation of actinides.

The analysis of the momentum and velocity spectra of fission-fragments
as function of A and Z
gives an indication
on the nature and the domain of excitation energies of the fragments
arising from the first step of the (U + p)-interaction and  undergoing
fission. Comparison with previous measurements shows that the
symmetric fission cross section (1.42$\pm$0.20) b is almost the same 
for U + p as for
U + Pb, (1.4 $\pm$ 0.2) b. In this last case the first phase of violent
abrasion leads to a wider range of more excited and lighter fragments.
The fission probabilities of these fragments are smaller than for
actinides close to the U-projectiles which are mainly produced
in (U + p)-collisions. Finally the cross section to observe symmetric
fission remains the same for both systems.

There is certainly not a perfect overlap between the distribution of
fragments
identified as evaporation-residues and the fissioning parent 
nuclei.
The parent nuclei close to U have high fissilities.
For elements of decreasing  Z numbers, fissile isotopes  
become more and more neutron-deficient. Compared
to the evaporation-residues the fissioning-parent nuclei are shifted
towards the neutron-deficient side. They are born from prefragments of
higher excitation energies and their recoil momenta may be higher 
than for those ending as evaporation residues.

Measured isotopic and mass distributions are compared to results
obtained with simulation codes commonly used. Large discrepancies are
shown, and the need for improved reaction models is underlined. 
Models developed recently in the collaboration are discussed 
\cite{Gaim,Jung,Benl,Boud} and the improvements achieved are
demonstrated.

\section{Acknowledgements}
We wish to thank F. Ameil, K. G\"unzer and M. Pravikoff for their
participation to the data taking and K. H. Behr, A. Br\"unle and 
K. Burkard for their technical support during the experiment. We are
thankful to the group of P. Chesny who built the liquid-hydrogen
target. We thank O. Yordanov and A. Kelic for their help and interest
in the late stage of our work. This work was partially supported by the
European Union under the contract ERBCHBCT940717.

\section{Appendix 1} 

\subsection{
Demonstration of equations (7) and (8)}

The transformation from the c.m.-system to the lab-system is given by:
\begin{eqnarray}
  \gamma\tan\phi = \frac{\beta_f\sin\varphi} {\beta_0 \pm
 \beta_f\cos\varphi}         
\label{A}
\end{eqnarray}
with $\pm$ signs for forward and backward angles $\varphi^F$ and
$\varphi^B$, respectively. To transform back to the c.m.-system for the
limiting case $\phi_{lim}$ = $\alpha$, we solve equ. (11) for this case:
\begin{eqnarray}
    \gamma \tan\alpha = \frac{ \sin\varphi}{ K \pm \cos \varphi}
\label{A}
\end{eqnarray}
 with K = $\beta_0$ / $\beta_f$.
It is convenient to solve equ. (12) as a quadratic equation in
$\tan(\varphi/2)$,
which has 2 solutions corresponding to $\varphi^F$ and $\varphi^B$
given for  $\phi_{lim}$ = $\alpha$ in the lab-system.
\begin{eqnarray}
 \tan (\varphi^{F,B}/2) = \frac{K \pm 1}{\sqrt{K^2 - 1}} \tan \delta/2 
\label{A}
\end{eqnarray}
with an auxiliary variable 

\begin{eqnarray}
\sin \delta =  \gamma \sqrt{K^2 - 1} \tan\alpha 
\label{A}
\end{eqnarray}
Equ. 13 shows that the angles $\varphi^F$ and$\varphi^B$ depend via
eq. (14) on all 3 variables $\alpha$, $\beta_0$ and $\beta_f$. The
transmission of each of the cones may be evaluated with the total
transmission obtained by summing T$_\Omega^F$ and  T$_\Omega^B$. This
method we applied in our evaluations until now \cite{Benl1,Enqv1,Enqv2}.
However, introducing the two solutions of equ. (13) into equ. (3) we
obtain another very compact expression for T$_\Omega$:

\begin{eqnarray}
  T_\Omega = 1 -\frac {\cos \delta}{ 1 +
(\gamma\tan\alpha)^2} 
\label{A}
\end{eqnarray}

Equ. (15) depends via the auxiliary variable $\delta$ on the fission
velocity. To illuminate the variable $\delta$ we introduce
$\varphi_\Sigma$ = ($\varphi^F$ + $\varphi^B$)/2 and  $\varphi_\Delta$
= ($\varphi^F$ - $\varphi^B$)/2, or $\varphi^F$ =
$\varphi_\Sigma$ +  $\varphi_\Delta$  and $\varphi^B$ =
$\varphi_\Sigma$ -  $\varphi_\Delta$. Equ. (13) transforms to the new
variables using the trigonometrical sum relations for
$\tan \varphi_\Sigma $ and $\tan \varphi_\Delta$. 
 
\begin{eqnarray}
 \tan\varphi_\Sigma = \frac {K}{\sqrt{K^2-1}} \tan\delta
\label{A}
 \tan\varphi_\Delta = \gamma \tan \alpha 
\label{A}
\end{eqnarray}

Angle $\varphi_\Delta$ depends only on two variables, $\gamma$ and
 $\alpha$, which represent the relativistic beam and the FRS
 spectrometer. The fission velocity $\beta_f$ enters into the angle
 $\varphi_\Sigma$.
 Between $\varphi^{F,B}$ and $\varphi_{\Sigma,\Delta}$ another 
 trigonometrical relation holds 
 $$ \cos \varphi^F + \cos \varphi^B =
 2 \cos \varphi _\Sigma \cos \varphi _\Delta $$
This relation introduced in equ. (3) gives for the transmission :

\begin{eqnarray}
 T_\Omega = 1 - \cos \varphi _\Sigma \cos \varphi _\Delta 
\label{A}
\end{eqnarray}

Comparing equ. (15) and (18) a relation between
$\varphi_{\Sigma,\Delta}$ and $\delta$ follows:

$$ \cos \delta = \frac {\cos \varphi_\Sigma}{\cos \varphi_\Delta} =
 \cos \varphi_\Sigma \sqrt {1 + (\gamma \tan\alpha)^2}$$
The three expressions equ. (3),(15) and (18) are exact, identical
representations in different angular coordinates. All expressions
derived for the transmission until now depend on $\beta_f$. The
conversion of $\beta_f$ into the measured quantity $\beta_{app}$ makes
use of equ. (6) and
(14) :

\begin{eqnarray}
 \sin\delta = \frac {\sqrt{\beta_0^2 - \beta_f^2}}{\beta_f} \gamma
\tan\alpha  =   \frac {\sqrt {\beta_0^2 -
\beta_f^2} \gamma \tan \alpha}{2 \beta_{app}} (2 - T_\Omega) 
\label{A}
\end{eqnarray}

Also equ. (6) and (19) are exact without any approximations.

For our experiment with $\gamma$ = 2.07 and $\alpha$ =
$\overline{\alpha}$ = 14.8 mrad, $\varphi_\Sigma$ is larger than
$\delta$ by only 0.1 mrad. It is a good approximation to replace
$\varphi_\Sigma$, the
mean value of $\varphi^F$ and $\varphi^B$, by the auxiliary variable
$\delta$. A next approximation for the transmission follows:

\begin{eqnarray}
 T_\Omega = 1 - \cos \varphi_\Sigma   
\label{A}
\end{eqnarray}
 which deviates from the exact value by less than 0.2 \% in the worst
 case of Z = 28.
  Neglecting terms in $\alpha^2$ and K$^{-2}$ useful approximations
  derived from eqs (16) and (17) are:
  
  $\varphi_\Sigma$ = $\delta$ and $\varphi_\Delta$ = $\gamma\alpha$ 
 
 $\varphi^F$ = $\varphi_\Sigma$ + $ \gamma\alpha$ and $\varphi^B$ =
 $\varphi_\Sigma$ - $ \gamma\alpha$

 From equ. (20) and (6) we derive equ. (7) and (8). 
 
 With  $\beta_{lim}$ $\equiv$
 $\alpha$ $\sqrt{\gamma^2 - 1}$  follows:
\begin{eqnarray}
  \tan {\varphi_\Sigma/2} = \frac {  \alpha \sqrt { \gamma ^2  -1}}
  {2 \beta_{app}} =
 \frac {\beta_{lim}}{2 \beta_{app}} 
\end{eqnarray}

Finally with equ. (21) $\beta_f$ and T$_\Omega$ can be presented as a 
function of $\beta_{app}$, see equ. (7) and (8) given in section 2.5.

$$ T_\Omega   = \frac {2} {1 + [2\beta_{app}/\beta_{lim}]^2 } $$
and 
$$ \beta_f =  \beta_{app} [ 1 + (\beta_{lim}/ 2 \beta_{app})^2] $$
\hspace{11cm}q. e. d.

\newpage
\section{Appendix 2}

\input{appendix2.tex}

\end{document}

%% file: tab1.tex
\begin{table}
\small
\begin{tabular}{|c|c|c||c|c|c||c|c|c|c|}
   \hline
     & S4-     & sec. react. & Z &S4- & sec. react.& Z & S4- & sec. react. & atomic  \\
   Z  &  shift &  in SC2     &   & shift  &  in SC2 &   & shift &  in SC2 &  charge \\
     &           &               &     & &         &   &   &         &exchange\\

\hline
 28 & 1.14  & 1.108 & 40 & 1.01  & 1.13  & 52 &1.02  & 1.149 & 1.01\\[-0.3cm]
 29 & 1.12  & 1.11  & 41 & 1.005 & 1.131 & 53 &1.03  & 1.151 & 1.01\\[-0.3cm]
 30 & 1.105 & 1.112 & 42 & 1.    & 1.133 & 54 &1.035 & 1.152 & 1.01\\ [-0.3cm]
 31 & 1.09  & 1.114 & 43 & 1.    & 1.135 & 55 &1.045 & 1.154 & 1.015\\[-0.3cm]
 32 & 1.08  & 1.116 & 44 & 1.    & 1.136 & 56 &1.055 & 1.156 & 1.02 \\[-0.3cm]
 33 & 1.065 & 1.118 & 45 & 1.    & 1.138 & 57 &1.065 & 1.157 & 1.028\\[-0.3cm]
 34 & 1.055 & 1.118 & 46 & 1.    & 1.140 & 58 &1.08  & 1.158 & 1.034\\[-0.3cm]
 35 & 1.045 & 1.122 & 47 & 1.    & 1.141 & 59 &1.09  & 1.162 & 1.04 \\[-0.3cm]
 36 & 1.035 & 1.123 & 48 & 1.    & 1.143 & 60 &1.105 & 1.164 & 1.05 \\[-0.3cm]
 37 & 1.03  & 1.125 & 49 & 1.005 & 1.144 & 61 &1.12  & 1.164 & 1.06 \\[-0.3cm]
 38 & 1.02  & 1.126 & 50 & 1.01  & 1.146 & 62 &1.14  & 1.166 & 1.07 \\[-0.3cm]
 39 & 1.015 & 1.128 & 51 & 1.015 & 1.148 & 63 &1.16  & 1.167 & 1.08 \\
\hline

\end{tabular}
\caption{   
Correction factors applied to the measured
yields to account for losses due to 1)
the shift in S4 position compared to the central position, 2) 
the secondary reactions in the SC2-scintillator at
S2, and 3)
the atomic charge exchange in target and S2-scintillator for 
Z $\ge$ 52.}
\end{table}

%% file: tab3.tex
\begin{table}
\small
\begin{tabular}[ ]{|c|c|c|c|c|c|c|c|}
\hline

reaction& $\sigma_{sym}$ & $\overline{E_k}$ &
$\overline{Z}$ & $\overline{N}$ &$\sigma_Z$&$\sigma_A$ &$\sigma_Z^A$\\[-0.2cm]
  
&(barn) & (MeV) &(a.ch.u.) &(a.m.u.)  &(a.ch.u.)&(a.m.u.)&(a.ch.u.)\\
 
\hline

U + p & 1.42$\pm$0.2&76$\pm$3 &45.0$\pm$0.1&63.0$\pm$0.15& 
7.0$\pm$ 0.2 & 17.5$\pm$0.5 &1.2$\pm$0.1 \\[-0.2cm]

U + Pb  
 &1.4$\pm$0.2&79$\pm$2 &42.9$\pm$0.6 &58.1$\pm$0.3& 6.9$\pm$0.7 
 & 17.2 $\pm$1.7 &1.3$\pm$0.3   \\

\hline
\end{tabular}
\caption{Parameters characterising fission of
$^{238}$U(1 A Gev) + p in comparison to the symmetric share of the fission 
process for $^{238}$U  + Pb at 0.75 AGeV \cite{Schw}. The small asymmetric
contribution in the first
system is neglected in evaluating mean values and variances. }
\end{table}

%% file: tab4.tex
\begin{table}
\begin{center}
\begin{tabular}{|c|c||c|c||c|c|}
\hline
   Z & $\overline{E}_{kin}$[MeV] & Z & $\overline{E}_{kin}$[MeV]  &  Z &
   $\overline{E}_{kin}$[MeV]  \\
\hline
28  & 92.2 $\pm$ 6.8 &  41  & 81.8 $\pm$1.5  & 54  & 64.1 $\pm$ 3.0\\   [-0.3cm]
29  & 89.0 $\pm$ 6.6 &  42   & 80.8 $\pm$ 1.5   & 55 & 61.1 $\pm$ 2.3\\ [-0.3cm]
30  &84.9 $\pm$ 5.5  &   43  & 79.7  $\pm$ 1.6  & 56 & 60.5 $\pm$ 2.3\\ [-0.3cm]
31  &86.4 $\pm$ 5.0  &   44  & 77.9  $\pm$ 1.5  & 57 & 55.2 $\pm$ 2.2 \\[-0.3cm]
32  &86.3 $\pm$ 3.9  &   45  & 75.9 $\pm$ 1.5  & 58 & 52.7 $\pm$ 3.5  \\[-0.3cm]
33  &87.3 $\pm$ 3.9  &   46  & 75.6 $\pm$ 1.6  & 59 & 51.3 $\pm$ 3.3  \\[-0.3cm]
34  &86.2 $\pm$ 3.3  &    47  & 74.5 $\pm$ 1.6  & 60& 49.1 $\pm$ 3.1  \\[-0.3cm]
35  &85.7 $\pm$ 3.0  &   48   & 73.6 $\pm$ 1.6  & 61& 46.9 $\pm$ 4.8 \\	[-0.3cm]
36  &85.3 $\pm$ 2.2  &   49   & 72.1 $\pm$ 1.6  & 62 & 45.1 $\pm$ 3.0 \\[-0.3cm]
37  &85.3 $\pm$ 2.2  &   50   & 69.9 $\pm$ 1.6  & 63 & 41.5 $\pm$ 5.  \\[-0.3cm]
38  &83.3 $\pm$ 1.5  &   51   & 69.4 $\pm$ 1.6  & 64 & 39.7 $\pm$ 5. \\[-0.3cm]
39  &84.4 $\pm$ 1.5  &   52   & 67.0 $\pm$ 1.6  &  &   \\	       [-0.3cm]
40  &81.8 $\pm$ 1.5  &   53   & 64.2 $\pm$ 1.6  &   &   \\      
\hline  		           
\end{tabular}
\caption{ The mean kinetic energy values of single fission fragments 
as a function of the atomic
number Z. 
   }
\end{center}
\end{table}

%% file: tab5.tex
\begin{table}
\begin{center}
\begin{tabular}[ ]{|c|c|c|c|}
\hline
reaction &$\sigma_{tot}$ &$\sigma_{fis}$ & $\sigma_{EVR}$ \\[-0.2cm]  
& (barn)  &(barn) & (barn)   \\
\hline
U + p & 1.99$\pm$ 0.17   & 1.53 $\pm$ 0.13 & 0.46$\pm$0.08  \\[-0.2cm]

Pb + p & 1.84$\pm$ 0.23 & 0.16 $\pm$ 0.07 
 &1.68$\pm$0.22  \\

\hline
\end{tabular}
\end{center}
\caption{ Comparison of reaction cross sections   
of 1 A GeV $^{238}$U + p to
 1 A GeV  $^{208}$Pb~+~p~[15].
}
\end{table}

%% file: appendix2.tex
\begin{longtable}{|c|c|c|}
   \hline
   Z & A & $\sigma$ [mb]  \\
\hline 
28  & 61 & 0.55 (10)  \\[-0.3cm] 
    & 62 & 1.00 (14)  \\[-0.3cm]
    & 63 & 1.35 (17)  \\[-0.3cm]
    & 64 & 1.67 (20)  \\[-0.3cm]
    & 65 & 1.75 (15)  \\[-0.3cm]
    & 66 & 1.67 (11)  \\[-0.3cm]
    & 67 & 1.20 (13)    \\[-0.3cm]       
    & 68 & 0.69  (5)    \\[-0.3cm]
    & 69 & 0.42  (8)    \\[-0.3cm]
    & 70 & 0.25 (4)     \\[-0.3cm]         
    & 71 & 0.16 (3)     \\
    		\hline
 29 & 63 & 0.49(12)     \\[-0.3cm]
      & 64 & 0.89(16)   \\[-0.3cm]
      & 65 & 1.24(19)   \\[-0.3cm]   
      & 66 & 1.70(17)   \\[-0.3cm]
      & 67 &  2.01(15)  \\[-0.3cm] 
      & 68 & 2.08(13)   \\[-0.3cm] 
      & 69 & 1.64(13)   \\[-0.3cm] 
      & 70 & 1.20(8)    \\[-0.3cm]
      & 71 &  0.67(12)  \\[-0.3cm]
      & 72 &  0.39(4)   \\[-0.3cm]
      & 73 &  0.23(2)   \\[-0.3cm] 
      & 74 &  0.11(5)   \\
      		\hline
   30 & 64 &  0.13(3)   \\[-0.3cm]
       & 65 & 0.46(7)   \\[-0.3cm]
       & 66 &0.94(13)   \\[-0.3cm]
       & 67 & 1.52(21)  \\[-0.3cm] 
       & 68 & 2.41(20)  \\[-0.3cm]
       & 69 & 2.86(17)  \\[-0.3cm]
       & 70 & 3.00(9)   \\[-0.3cm]
       & 71& 2.54(20)   \\[-0.3cm]
       & 72& 1.95(23)   \\[-0.3cm]
       & 73&  1.27(10)  \\[-0.3cm]
       & 74&  0.88(7)   \\[-0.3cm]
       & 75&  0.54(3)   \\[-0.3cm]
       & 76&  0.26(3)   \\[-0.3cm]
       &  77& 0.12(2)   \\[-0.3cm]  
       &  78&0.047(21)  \\
       		\hline
 31    & 67 &0.36(10)   \\[-0.3cm]   
       &  68 &0.90(18)  \\[-0.3cm]
                & 69 & 1.60(20) \\[-0.3cm]
	        &70 & 2.87(14)  \\[-0.3cm] 	 
                & 71 & 3.58(21) \\[-0.3cm]
	        & 72 & 3.68(22) \\[-0.3cm]
	        & 73 & 3.60(54  \\[-0.3cm]
	        & 74 &2.99(18)  \\[-0.3cm]
	        & 75 &2.16(37)  \\[-0.3cm]
	        & 76 &1.22(17)  \\[-0.3cm]
	        & 77 &0.82(7)   \\[-0.3cm]
	        & 78 &0.55(9)   \\[-0.3cm]
	        & 79 & 0.30(2)  \\[-0.3cm]
	        & 80 & 0.14(3)  \\[-0.3cm]
	        & 81 &0.033(14) \\
		\hline
	    32  & 69  & 0.32(6) \\[-0.3cm]
	        & 70  & 0.81(13)\\[-0.3cm]
	        & 71  & 1.71(26)\\[-0.3cm]
	        & 72  & 3.24(23)\\[-0.3cm]
	        & 73  & 4.28(21)\\[-0.3cm] 
	        & 74  & 4.70(28)\\[-0.3cm]
	        & 75  & 4.83(18)\\[-0.3cm]
	        & 76  & 4.28(30)\\[-0.3cm]
	        & 77  & 2.85(45)\\[-0.3cm]
	        & 78  & 1.97(22)\\[-0.3cm]
	        & 79  & 1.45(10)\\[-0.3cm]
	        & 80  & 1.15(16)\\[-0.3cm]
	        & 81  &  0.74(6)\\[-0.3cm]
	        & 82  &  0.38(2)\\[-0.3cm]
	        & 83  &  0.17(1)\\[-0.3cm]
	        & 84  &0.065(15)\\[-0.3cm]
	        & 85  &0.019(8) \\
		\hline
	     33 &  71 & 0.29(8) \\[-0.3cm]
	        &  72 & 0.70(14)\\[-0.3cm]
	        &  73 & 1.63(26)\\[-0.3cm]
	        &  74 & 2.84(28)\\[-0.3cm]
	        &  75 & 5.01(30)\\[-0.3cm]
	        &  76 & 5.93(36)\\[-0.3cm]
	        &  77 & 6.10(30)\\[-0.3cm]
	        & 78  & 5.33(32)\\[-0.3cm]
	        & 79  & 4.50(27)\\[-0.3cm]
	        & 80  & 3.34(30)\\[-0.3cm]
	        & 81  & 2.62(13)\\[-0.3cm]
	        & 82  & 1.88(17)\\[-0.3cm]    
	        & 83  & 1.36(13)\\[-0.3cm]     
	        & 84  & 0.74(11)\\[-0.3cm]     
	        & 85  & 0.36(4) \\[-0.3cm]       
	        & 86  & 0.16(2) \\[-0.3cm]       
	        & 87  & 0.03(20)\\
		\hline
       	34      & 73     & 0.28 (5) \\[-0.3cm] 
	        & 74     & 0.70 (14)\\[-0.3cm]
	        & 75     & 1.30 (21)\\[-0.3cm]
	        & 76     & 2.42 (39)\\[-0.3cm]
	        & 77     & 4.42 (26)\\[-0.3cm]   
	        & 78     & 6.68 (27)\\[-0.3cm]  
	        & 79     & 7.25 (43)\\[-0.3cm]    
	        & 80     & 7.37 (29)\\[-0.3cm]     
	        & 81     & 6.64 (46)\\[-0.3cm]       
	        & 82     &  5.67(80)\\[-0.3cm]     
	        & 83     &  3.93(24)\\[-0.3cm]       
	        & 84     & 3.41(14) \\[-0.3cm]       
	        & 85     & 2.61(18) \\[-0.3cm]      
	        & 86     & 1.88(11) \\[-0.3cm]        
	        &87  &  0.93 (9)    \\[-0.3cm]        
	        &88  & 0.34 (6)     \\[-0.3cm]        
	        &89  & 0.085 (25)   \\
		\hline
	 35      & 75 & 0.19(7)     \\[-0.3cm]        
	         & 76 &0.52 (10)    \\[-0.3cm]     
	        &  77   &1.08(20)   \\[-0.3cm]      
	        &  78   &2.37(38)   \\[-0.3cm]      
	        &  79    &4.22(34)  \\[-0.3cm]     
	        &  80    &6.79(20)  \\[-0.3cm]      
	        &  81    &8.23(33)  \\[-0.3cm]       
	        & 82    &8.49(51)   \\[-0.3cm]      
	        & 83    &8.02(77)   \\[-0.3cm]       
	        & 84    &7.23(72) \\  [-0.3cm]     
	        & 85    &6.13(86) \\  [-0.3cm]     
	        & 86    &4.66(56) \\  [-0.3cm]    
	        & 87    &3.80(15) \\  [-0.3cm]     
	        & 88    &2.92(23) \\  [-0.3cm]     
	        & 89      &1.95(19) \\[-0.3cm]    
	        & 90      &0.75(10) \\[-0.3cm]   
                & 91     &0.25(6)  \\ [-0.3cm] 
                & 92  &  0.06(2)   \\ 
		\hline
     36  & 77 & 0.09(3)   \\[-0.3cm]
          & 78 &0.37(7)   \\[-0.3cm]
          & 79 &0.84(13)  \\[-0.3cm]
          & 80 &2.0(3)    \\[-0.3cm]
          & 81 &4.43(40)  \\[-0.3cm]
          & 82 &7.40(44)  \\[-0.3cm]
          & 83 &9.31(56)  \\[-0.3cm]
          & 84 &10.2(7)   \\[-0.3cm]
          & 85 &10.5(9)   \\[-0.3cm]
          & 86 &9.36(94)  \\[-0.3cm]
           & 87 &8.04(96) \\[-0.3cm]
           & 88 &6.92(69) \\[-0.3cm]
           & 89 &6.03(36) \\[-0.3cm]
           & 90 &5.14(31) \\[-0.3cm]
           & 91 &3.26(16) \\[-0.3cm]
           & 92 &1.74(26) \\[-0.3cm]
          & 93 & 0.66(11) \\[-0.3cm]
          & 94 & 0.23(6)  \\[-0.3cm]
          & 95 & 0.07(2)  \\
	  \hline
      37    & 79 &0.11(3) \\[-0.3cm]
            & 80 &0.31(6) \\[-0.3cm]
            & 81 &0.80(13)\\[-0.3cm]
          & 82 & 1.80(29) \\[-0.3cm]
           & 83 &4.05(40) \\[-0.3cm]
           & 84 &6.93(41) \\[-0.3cm]
           & 85 &10.5(5)  \\[-0.3cm]
           & 86 &11.2(6)  \\[-0.3cm]
           & 87 &11.4(6)  \\[-0.3cm]
          & 88 & 11.0(10) \\[-0.3cm]
          & 89 & 9.2(14)  \\[-0.3cm]
          & 90 & 8.2(11)  \\[-0.3cm]
          & 91 & 7.4(6)   \\[-0.3cm]
          & 92 & 6.12(61) \\[-0.3cm]
          & 93 & 4.6(2)   \\[-0.3cm]
          & 94 & 2.4(2)   \\[-0.3cm]
          & 95 & 1.17(18) \\[-0.3cm]
          & 96 & 0.39(7)  \\[-0.3cm]
           & 97 & 0.13(3) \\[-0.3cm]
          & 98 & 0.031(16)\\
	  \hline
       38  & 83 &0.63(18) \\[-0.3cm]
           & 84 &1.58(27) \\[-0.3cm]
           & 85 &3.50(52) \\[-0.3cm]
           & 86  & 7.02(49)\\[-0.3cm]
	        &87    & 11.1(6) \\[-0.3cm]  
	        &88    & 11.8(6) \\[-0.3cm]  
	        & 89   & 12.0(6) \\[-0.3cm]  
	        &90    & 11.9(7) \\[-0.3cm]  
	        &91    & 12.2(12)\\[-0.3cm]  
	        &92    & 10.2(12)\\[-0.3cm] 
	        &93    &  9.05(61)\\[-0.3cm] 
	        &94    &  8.24(50)\\[-0.3cm] 
	       &95     &  6.1(5)  \\[-0.3cm] 
	       &96    &  4.36(26) \\[-0.3cm] 
	       &97    &  2.14(32) \\[-0.3cm] 
	       &98    &  0.88(18) \\[-0.3cm] 
	       &99    &  0.24(4)  \\[-0.3cm] 
	       &100   &  0.069(14)\\
	       \hline
	   39  & 85   & 0.51(14)  \\[-0.3cm] 
	       & 86   & 1.36(27)  \\[-0.3cm] 
	       & 87   & 3.21(45)  \\[-0.3cm] 
	       & 88    & 6.51(52) \\ 	    [-0.3cm]
	        & 89   & 10.3(5)  \\ 	    [-0.3cm]
	        & 90   & 12.1(6)  \\ 	    [-0.3cm]
	        & 91   & 12.4(7)  \\ 	    [-0.3cm]
	         & 92   & 12.5(9) \\ 	    [-0.3cm]
	        & 93    &12.4(9)  \\ [-0.3cm]
	        & 94   & 11.4(7)  \\ [-0.3cm]
	        & 95   & 10.5(5)  \\ [-0.3cm]
	        & 96   & 8.75(44) \\ [-0.3cm]
	        & 97    & 7.36(37)\\ [-0.3cm]
	        & 98   & 5.62(67) \\ [-0.3cm]
	        & 99   & 3.81(27) \\ [-0.3cm]
	        &100    & 1.65(26)\\ [-0.3cm]
	        &101    & 0.68(17)\\ [-0.3cm]
	        &102    & 0.19(6) \\ [-0.3cm]
	        & 103   &  0.04(2) \\
		\hline
	    40  &  86  &  0.127(25 \\[-0.3cm]
	        &  87  &  0.377(68 \\[-0.3cm]
	        &  88  &  1.15(17) \\[-0.3cm]
	        &  89  &  2.86(46) \\[-0.3cm]
	        &  90  &  5.73(45) \\[-0.3cm]
	        &  91  &  9.30(74) \\[-0.3cm]
	        &  92  &  12.0(5)  \\[-0.3cm]
	        &  93  &  13.3(7)  \\[-0.3cm]
                &  94  & 13.7(7)   \\[-0.3cm]
     & 95  & 13.5(7)    \\	     [-0.3cm]
     & 96  & 13.4(8)    \\	     [-0.3cm]
     & 97  & 12.0(13)    \\	     [-0.3cm]
     & 98  & 10.8(11)    \\[-0.3cm]
     & 99  &  8.80(44)   \\[-0.3cm]
     &100   & 7.30(73)  \\ [-0.3cm]
     &101  & 4.50(58)   \\ [-0.3cm]
     &102   & 2.57(36)  \\ [-0.3cm]
     & 103  &  1.04(18) \\ [-0.3cm]
     &104   & 0.38(11)  \\ [-0.3cm]
     & 105  & 0.10(3)  \\  
     \hline
  41  & 89  & 0.30(7)    \\[-0.3cm] 
     & 90  & 1.05(21)    \\[-0.3cm]
     & 91  & 2.38(38)    \\[-0.3cm]
    & 92  & 4.55(55)    \\[-0.3cm]
    & 93  & 7.56(38)    \\[-0.3cm]
    & 94  & 10.5(4)    \\ [-0.3cm]
   & 95   & 12.6(5)    \\ [-0.3cm]
    & 96  &  13.7(5)   \\ [-0.3cm]
    & 97  &  14.5(6)    \\[-0.3cm]
    & 98  &  14.4(9)   \\[-0.3cm]
    & 99  &  13.3(8)   \\[-0.3cm]
    &100   & 11.9(8)    \\ [-0.3cm] 
    &104   & 2.92(50)    \\[-0.3cm]
    & 105  & 1.56(44)    \\[-0.3cm]
    & 106  & 0.55(16)    \\[-0.3cm]
    &107   & 0.19(10)    \\
    \hline
 42   &92   & 0.72(19)    \\[-0.3cm]
    & 93  & 1.81(27)    \\  [-0.3cm]
    & 94  & 3.58(43)    \\  [-0.3cm]
    & 95  & 6.24(37)    \\  [-0.3cm]
    & 96  & 9.68(48)    \\  [-0.3cm]
    & 97  & 12.8(8)    \\   [-0.3cm]
    & 98  & 14.0(7)    \\    [-0.3cm]
    & 99  & 15.1(8)    \\    [-0.3cm]
    & 100  & 16.2(10)    \\  [-0.3cm]
    & 101  &  15.7(11)   \\  [-0.3cm]
      & 102     & 14.9(12)\\ [-0.3cm]
      & 103     & 13.0(8) \\ [-0.3cm]
      & 104     & 9.91(49)\\ [-0.3cm]
      & 105     & 6.18(43)\\ [-0.3cm]
      & 106     & 3.96(20)\\ [-0.3cm]
      & 107     & 1.78(19)\\ [-0.3cm]
      & 108     & 0.91(27)\\ [-0.3cm]
      & 109     & 0.28(7) \\ [-0.3cm]
      & 110     &   0.08(4)\\
      \hline
  43  & 94   &  0.68(12)  \\ [-0.3cm]
      & 95   &  1.36(31)  \\ [-0.3cm]
      & 96   &  2.69(30)  \\ [-0.3cm]
      & 97   &  5.02(35)  \\ [-0.3cm]
      & 98   &  8.11(40)  \\ [-0.3cm]
      & 99   &  11.5(7)   \\ [-0.3cm]
      & 100  &  13.3(7)   \\ [-0.3cm]
      & 101  &   15.2(8)  \\ [-0.3cm]
      & 102  &  15.9(10)  \\ [-0.3cm]
      & 103  &  16.2(11)  \\ [-0.3cm]
      & 104  &  15.8(9)   \\ [-0.3cm]
      & 105  &  14.4(11)  \\ [-0.3cm]
      & 106  &  11.6(9)   \\ [-0.3cm]
      & 107  &  8.44(51)  \\ [-0.3cm]
      & 108  &  4.88(34)  \\ [-0.3cm]
      & 109  &  2.90(20)  \\ [-0.3cm]
      & 110  &  1.31(26)  \\ [-0.3cm]
      & 111  &  0.60(18)  \\ [-0.3cm]
      & 112  &  0.17(8)   \\ 
      \hline
 44   & 95   & 0.13(2)    \\ [-0.3cm]
      & 96   & 0.47(4)    \\ [-0.3cm]
      & 97   & 1.01(7)    \\ [-0.3cm]
      & 98   & 2.10(17)   \\ [-0.3cm]
      & 99   & 3.91(23)   \\ [-0.3cm]
      & 100   &6.84(34)   \\ [-0.3cm]
      & 101   &8.77(60)   \\ [-0.3cm]
      & 102   &13.1(8)    \\ [-0.3cm]
      & 103   &14.7(7)    \\ [-0.3cm]
      & 104   &16.0(8)    \\ [-0.3cm]
      & 105   &16.7(10)   \\ [-0.3cm]
      & 106   &16.6(13)   \\ [-0.3cm]
      & 107   &15.4(11)   \\ [-0.3cm]
      & 108   &13.1(7)    \\ [-0.3cm]
      & 109 & 9.17(46)    \\ [-0.3cm]
        & 110  &6.62(40)  \\ [-0.3cm]
        & 111  &3.95(31)  \\ [-0.3cm]
        & 112  &1.94(21) \\  [-0.3cm]
        & 113  &0.77(8)  \\  [-0.3cm]
        & 114  &0.34(5)  \\  [-0.3cm]
        & 115  &0.11(2)  \\  
	\hline
      45 & 99  &  0.70(9)\\  [-0.3cm]
        & 100  &  1.52(8)\\  [-0.3cm]
        & 101  &3.00(15) \\  [-0.3cm]
        & 102  &5.28(37) \\  [-0.3cm]
        & 103  &8.62(60) \\  [-0.3cm]
        & 104  &11.5(6)  \\  [-0.3cm]
        & 105  &14.2(6)  \\  [-0.3cm]
        & 106  & 15.4(4) \\  [-0.3cm]
        & 107  & 16.6(10)\\  [-0.3cm]
        & 108  & 16.3(10)\\  [-0.3cm]
        & 109  & 15.7(9) \\  [-0.3cm]
        & 110  &13.6(14) \\  [-0.3cm]
        & 111  &  11.7(6)\\  [-0.3cm]
        &112   &  8.11(4)\\  [-0.3cm]
        &113 & 5.11(25)  \\  [-0.3cm]
        &114   & 2.60(10)\\  [-0.3cm]
        &115   & 1.40(5) \\  [-0.3cm]
        &116   & 0.54(7) \\  [-0.3cm]
        &117   & 0.20(10)\\  
	\hline
      46 & 100& 0.19(2)  \\  [-0.3cm]
         & 101& 0.54(3)  \\  [-0.3cm]
         & 102& 1.21(6)  \\  [-0.3cm] 
         & 103& 2.48(12) \\  [-0.3cm]
         & 104 & 4.36(17)\\  [-0.3cm]
        & 105  & 7.33(29)\\  [-0.3cm]
        & 106  & 10.7(7) \\  [-0.3cm]
        &107   & 13.0(6) \\  [-0.3cm]
        &108   & 14.5(6) \\  [-0.3cm]
        & 109  & 15.2(9) \\  [-0.3cm]
        & 110  & 16.2(13)\\  [-0.3cm]
        & 111  &  15.8(16)\\ [-0.3cm]
        &112   & 14.8(12) \\ [-0.3cm]
        &113   & 12.6(9) \\  [-0.3cm]
        &114   & 9.58(38)\\  [-0.3cm]
        & 115  &  6.25(25)\\  [-0.3cm]
        &116  & 3.57(14)  \\  [-0.3cm]
    & 117  & 1.91(27)  \\     [-0.3cm]
    & 118  & 0.93(11)  \\     [-0.3cm]
    & 119  & 0.37(10)  \\     [-0.3cm]
    & 120  & 0.12(2)  \\     
    \hline
 47 &103   & 0.40(4)      \\  [-0.3cm]
    &104   & 0.86(5)      \\  [-0.3cm]
     &105   & 1.81(9)     \\  [-0.3cm]
  &106   & 3.19(16)       \\  [-0.3cm]
  &107   & 5.73(23)       \\  [-0.3cm]
  &108   & 8.40(34)       \\  [-0.3cm]
  &109   & 11.4(8)        \\  [-0.3cm]
  &110   & 12.6(5)        \\  [-0.3cm]
  &111   & 14.0(7)        \\  [-0.3cm]
   &112   & 14.5(10)      \\  [-0.3cm]
   &113   & 16.0(17)      \\  [-0.3cm]
   &114   & 14.5(13)      \\  [-0.3cm]
   &115   & 13.5(8)       \\  [-0.3cm]
  &116   & 10.5(14)       \\  [-0.3cm]
    &117   & 7.73(31)     \\  [-0.3cm]
    &118   & 4.74(19)     \\  [-0.3cm]
   &119   & 2.64(18)      \\  [-0.3cm]
    &120   & 1.38(8)      \\  [-0.3cm]
    &121   & 0.65(3)      \\  [-0.3cm]
    &122   & 0.22(3)      \\  [-0.3cm]
    &123   & 0.07(1)      \\ 
    \hline
  48  &106 & 0.54(13)      \\ [-0.3cm]
      &107 & 1.30(26)      \\ [-0.3cm]
      &108 & 2.47(30)      \\ [-0.3cm]
    & 109  & 4.41(44)      \\ [-0.3cm]
    &110  &7.00(35)        \\ [-0.3cm]
    &111  &9.75(58)        \\ [-0.3cm]
    & 112  &11.6(8)        \\ [-0.3cm]
    &113  &12.5(5)         \\ [-0.3cm]
    &114  &13.7(7)         \\ [-0.3cm]
    & 115  &14.2(6)        \\ [-0.3cm]
    &116  &14.0(6)         \\  [-0.3cm]
    &117  &12.8(6)         \\  [-0.3cm]
    &118 &11.1(4)         \\   [-0.3cm]
    & 119 &8.21(33)        \\  [-0.3cm]
    &120 &5.64(23)         \\  [-0.3cm]
    & 121 &3.20(20)        \\  [-0.3cm]
         & 122 & 1.80(13)  \\  [-0.3cm]
         & 123 & 0.96(10)    \\[-0.3cm]
         & 124 & 0.42(5)     \\[-0.3cm]
         & 125 & 0.18(4)     \\[-0.3cm]
         & 126 & 0.069(34)   \\
	 \hline
      49 & 107   &0.14(3)    \\[-0.3cm] 
          & 108  &0.34(5)    \\[-0.3cm]
         & 109   &0.77(8)    \\[-0.3cm]
         & 110   &1.71(22)   \\[-0.3cm]
         & 111   &3.15(41)   \\[-0.3cm]
         & 112   &5.26(47)   \\[-0.3cm]
         & 113   &7.78(70)   \\[-0.3cm]
         & 114   &10.0(8)    \\[-0.3cm]
         & 115   &11.3(7)    \\[-0.3cm]
         & 116   &12.0(10)   \\[-0.3cm]
         & 117   &13.0(8)    \\[-0.3cm]
         & 118   &13.0(13)   \\[-0.3cm]
         & 119   &12.7(9)    \\[-0.3cm]
         & 120   &11.0(7)    \\[-0.3cm]
         & 121   &9.22(37)   \\[-0.3cm]
         & 122   &6.16(24)   \\[-0.3cm]
         & 123   &3.98(16)   \\[-0.3cm]
         & 124   &2.27(9)    \\[-0.3cm]
         & 125   &1.38(8)    \\[-0.3cm]
         & 126   &0.75(6)    \\[-0.3cm]
         & 127   &0.41(6)    \\[-0.3cm]
         & 128   &0.19(3)    \\[-0.3cm]
         & 129   &0.08(2)    \\[-0.3cm]
         & 130   &0.02(1)    \\
	 \hline
     50  & 110   &0.23(2)    \\[-0.3cm]
         & 111   &0.56(6)    \\[-0.3cm]
         & 112   &1.26(13)   \\[-0.3cm]
         & 113   &2.39(24)   \\[-0.3cm]
         & 114   &4.09(20)   \\[-0.3cm]
         & 115   &6.38(25)   \\[-0.3cm]
         & 116   &8.6(13)    \\[-0.3cm]
         & 117   &9.97(40)   \\[-0.3cm]
         & 118   &10.4(4)    \\[-0.3cm]
         & 119   &11.0(4)    \\[-0.3cm]
         & 120   &11.5(5)    \\[-0.3cm]
         & 121   &11.6(5)    \\[-0.3cm]
         & 122   &10.5(4)    \\[-0.3cm]
         &123   & 9.28(37)   \\[-0.3cm]
         &124   & 6.77(27)   \\[-0.3cm] 
        &125   & 4.82(20)    \\ [-0.3cm]
        &126   & 2.90(11)    \\	[-0.3cm]
        &127   & 2.00(8)      \\[-0.3cm]
        &128   & 1.56(6)      \\[-0.3cm]
        &129   & 1.29(5)     \\	[-0.3cm]
        &130   & 1.01(7)     \\	[-0.3cm]
        &131   & 0.66(7)     \\	[-0.3cm]
        &132   & 0.35(7)      \\[-0.3cm]
        &133   & 0.05(3)      \\
	\hline
   51 &112   & 0.16(6)     \\	[-0.3cm]
      &113   & 0.36(9)     \\	[-0.3cm]
      &114   & 0.82(12)    \\	[-0.3cm]
     &115   & 1.66(25)     \\[-0.3cm]
      &116   & 2.97(23)    \\[-0.3cm]
     &117   & 4.66(28)     \\[-0.3cm]
      &118   & 6.6(5)      \\[-0.3cm]
      &119   & 8.3(9)      \\[-0.3cm]
     &120   & 9.1(5)       \\[-0.3cm]
     &121   & 9.59(57)     \\[-0.3cm]
     &122   & 10.2(5)      \\[-0.3cm]
      &123   & 10.6(8)     \\[-0.3cm]
      &124   & 10.2(5)     \\[-0.3cm]
      &125   & 9.3(5)      \\[-0.3cm]
      &126   & 7.63(45)    \\[-0.3cm]
      &127   & 5.67(28)    \\[-0.3cm]
      &128   & 3.88(31)    \\[-0.3cm]
      &129   & 2.87(14)    \\[-0.3cm]
      &130   & 2.43(12)    \\[-0.3cm]
      &131   & 2.46(29)    \\[-0.3cm]
      &132   & 1.85(27)    \\[-0.3cm]
      &133   &1.27(23)     \\[-0.3cm]
      &134   &0.42(25)     \\[-0.3cm]
      &135   &0.12(5)      \\
      \hline
    52  & 114  & 0.12(3)   \\[-0.3cm]
      & 115  &  0.27(3)    \\[-0.3cm]
      & 116  &  0.54(4)    \\[-0.3cm]
      & 117  &  1.16(8)    \\[-0.3cm]
      & 118  &  2.11(17)   \\[-0.3cm]
      & 119  &  3.48(31)   \\[-0.3cm]
      & 120  &  5.1(5)     \\[-0.3cm]
     & 121&6.61(53) \\	     [-0.3cm]
     & 122   & 7.7(6)\\	     [-0.3cm]
     & 123   & 8.02(48)\\    [-0.3cm]
     & 124   & 8.40(42) \\   [-0.3cm]
     & 125   & 8.80(35) \\   [-0.3cm]
     & 126   & 8.61(34)  \\  [-0.3cm]
     & 127   &8.18(33)  \\   [-0.3cm]
     & 128   &7.40(30)  \\   [-0.3cm]
     & 129   &5.79(35)  \\   [-0.3cm]
     & 130   &4.84(29)  \\   [-0.3cm]
     & 131   &3.83(23)  \\[-0.3cm]  
    &132    &3.70(22)  \\ [-0.3cm]
     &133    &3.94(39)  \\[-0.3cm]
     &134    &3.64(22)  \\[-0.3cm]
    &135    &1.70(17)  \\ [-0.3cm]
     &136   &0.90(11)  \\ [-0.3cm]
     &137   &0.30(8)   \\ [-0.3cm]
     &138   &0.074(30) \\ 
     \hline
   53 &116&0.06(1) \\	  [-0.3cm]
     & 117&0.17(3)\\	  [-0.3cm]
    & 118&0.45(7)  \\	  [-0.3cm]
     &  119&0.82(8)  \\	  [-0.3cm]
     &  120&1.49(10)  \\  [-0.3cm]
     &  121&2.50(15) \\	  [-0.3cm]
      & 122&3.9(4)   \\	  [-0.3cm]
     & 123&5.4(5)   \\	  [-0.3cm]
     & 124&6.3(7)   \\	  [-0.3cm]
     & 125&6.8(4)   \\	  [-0.3cm]
     & 126&7.30(44)   \\  [-0.3cm]
     &127&7.60(30)   \\	  [-0.3cm]
     &128&7.81(39)   \\	  [-0.3cm]
     & 129&7.54(30)   \\  [-0.3cm]
    & 130&6.90(35)   \\	  [-0.3cm]
    &131 &6.06(36)   \\	  [-0.3cm]
     &132 &4.91(24)   \\  [-0.3cm]
     & 133&4.38(31)   \\   [-0.3cm]
    & 134&4.09(20)   \\    [-0.3cm]
     & 135 &4.32(21)   \\  [-0.3cm]
     &  136 &3.08(28)   \\ [-0.3cm]
     &  137 &2.29(21)   \\ [-0.3cm]
    & 138 &1.21(12)   \\   [-0.3cm]
   & 139 &0.53(13)   \\	   [-0.3cm]
     & 140 &0.16(3)   \\   [-0.3cm]
     & 141  &    0.038(8)\\ 
     \hline
 54  & 118  & 0.032(9)    \\[-0.3cm]
     & 119  & 0.118(18)   \\[-0.3cm]
     & 120  & 0.25(2)     \\[-0.3cm]
     & 121  & 0.51(5)     \\[-0.3cm]
     & 122  & 1.01(7)     \\[-0.3cm]
     & 123  & 1.82(9)     \\[-0.3cm]
     & 124  & 2.91(17)    \\[-0.3cm]
     & 125  & 4.17(25)    \\[-0.3cm]
     & 126  & 4.99(40)    \\[-0.3cm]
     & 127  & 5.68(45)    \\[-0.3cm]
     & 128  & 5.88(41)    \\[-0.3cm]
     & 129  & 6.14(37)    \\[-0.3cm]
     & 130  & 6.27(38)    \\[-0.3cm]
     & 131  & 6.39(45)    \\[-0.3cm]
     & 132  & 6.01(18)    \\[-0.3cm]
     & 133  & 5.43(33)    \\[-0.3cm]
     & 134  & 5.05(30)    \\[-0.3cm]
     & 135  & 4.66(23)    \\[-0.3cm]
     & 136  & 4.48(27)    \\[-0.3cm]
     & 137  & 4.05(16)    \\[-0.3cm]
     & 138  & 3.96(20)    \\[-0.3cm]
     & 139  & 3.15(31)    \\[-0.3cm]
     & 140  & 2.11(25)    \\[-0.3cm]
     & 141  & 0.77(9)     \\[-0.3cm]
     & 142  & 0.32(4)     \\[-0.3cm]
     & 143  & 0.06(2)     \\
     \hline
  55 &  122 &0.12(3)      \\[-0.3cm]
     &  123 &0.27(4)      \\[-0.3cm]
     &  124 &0.69(8)      \\[-0.3cm]
     &  125 &1.27(13)     \\[-0.3cm]
     &  126 &2.05(8)      \\[-0.3cm]
     &  127 &3.03(15)     \\[-0.3cm]
     &  128 &3.98(32)     \\[-0.3cm]
     &  129 &4.60(32)     \\[-0.3cm]
     &  130 &4.86(24)     \\[-0.3cm]
     &  131 &4.94(15)     \\[-0.3cm]
     &  132 &5.05(25)     \\[-0.3cm]
     &  133 &5.24(26)     \\[-0.3cm]
     &  134 &5.19(15)     \\[-0.3cm]
     &  135 &4.74(24)     \\[-0.3cm]
     &  136 &4.43(22)     \\[-0.3cm]
     & 137  & 3.89(27)  \\  [-0.3cm]
      & 138  & 3.33(20)  \\[-0.3cm]
      & 139  & 3.55(21)  \\[-0.3cm]
      & 140  & 3.34(20)  \\[-0.3cm]
      & 141  & 3.40(13)  \\[-0.3cm]
      & 142  & 2.05(8)   \\[-0.3cm]
      & 143  & 1.04(3)   \\[-0.3cm]
      & 144  & 0.40(4)   \\[-0.3cm]
      & 145  & 0.12(2)   \\
      \hline
   56 & 125   & 0.16(3)  \\[-0.3cm]
      & 126   & 0.42(5)  \\[-0.3cm]
      & 127   & 0.82(8)  \\[-0.3cm]
      & 128   & 1.37(9)  \\[-0.3cm]
      & 129   & 2.23(17) \\[-0.3cm]
      & 130   & 2.91(34) \\[-0.3cm]
      & 131   & 3.50(50) \\[-0.3cm]
      & 132   & 3.68(37) \\[-0.3cm]
      & 133   & 3.86(34) \\[-0.3cm]
      & 134   & 3.95(24) \\[-0.3cm]
      & 135   & 4.11(21) \\[-0.3cm]
      & 136   & 4.03(20) \\[-0.3cm]
      & 137   & 3.85(20) \\[-0.3cm]
      & 138   & 3.65(29) \\[-0.3cm]
      & 139   & 3.24(22) \\[-0.3cm]
      & 140   & 3.10(18) \\[-0.3cm]
      & 141   & 3.28(23) \\[-0.3cm]
      & 142   & 3.57(28) \\[-0.3cm]
      &  143  &  3.19(16)\\[-0.3cm]
      &  144  & 2.70(19) \\[-0.3cm]
      &  145  & 1.33(17) \\[-0.3cm]
      &  146  & 0.65(10) \\[-0.3cm]
      &  147  & 0.16(3)  \\
      \hline
   57 &127    &  0.10(3) \\[-0.3cm]
      &128    &  0.22(6) \\[-0.3cm]
      &129    &  0.55(8) \\[-0.3cm]
      &130    &  0.99(10)\\[-0.3cm]
      &131    &  1.56(14)\\[-0.3cm]
      &132    &  2.16(17)\\[-0.3cm]
      &133    &  2.73(22)\\[-0.3cm]
      &134    &  3.12(19)\\[-0.3cm]
      &135    &  3.24(23)\\[-0.3cm]
      &136    &  3.22(22)\\[-0.3cm]
     & 137  & 3.21(16) \\    [-0.3cm]      
    & 138  & 3.17(16)  \\    [-0.3cm]    
    & 139  & 2.97(24)             \\[-0.3cm]
    & 140  & 2.49(15)             \\[-0.3cm]
    & 141  & 2.24(16)             \\[-0.3cm]
    & 142  & 2.23(16)             \\[-0.3cm]
    & 143  & 2.20(18)             \\[-0.3cm]
    & 144  & 2.27(20)             \\[-0.3cm]
    & 145  & 2.31(18)             \\[-0.3cm]
    & 146  & 1.88(15)             \\[-0.3cm]
    & 147  & 1.25(19)             \\[-0.3cm]
    & 148  & 0.53(11)             \\[-0.3cm]
    & 149  & 0.20(5)             \\ [-0.3cm]
    & 150  & 0.03(1)             \\        
    \hline
 58 & 130  & 0.17(6)             \\ [-0.3cm]
    & 131  & 0.33(6)             \\ [-0.3cm]
    & 132  & 0.61(5)             \\ [-0.3cm]
    & 133  & 0.96(9)             \\ [-0.3cm]
    & 134  & 1.52(9)             \\ [-0.3cm]
    & 135  & 1.96(20)             \\[-0.3cm]
    & 136  & 2.20(20)             \\[-0.3cm]
    & 137  & 2.34(16)             \\[-0.3cm]
    & 138  & 2.34(14)             \\[-0.3cm]
    & 139  & 2.28(11)             \\[-0.3cm]
    & 140  & 2.27(11)             \\[-0.3cm]
    & 141  & 2.08(12)             \\[-0.3cm]
    & 142  & 1.82(9)             \\ [-0.3cm]
    & 143  & 1.60(10)             \\[-0.3cm]
    & 144  & 1.52(9)             \\ [-0.3cm]
    & 145  & 1.45(10)             \\[-0.3cm]
    & 146  & 1.45(9)             \\ [-0.3cm]
    & 147  & 1.33(9)             \\ [-0.3cm]
    & 148  & 1.19(10)             \\[-0.3cm]
    & 149  & 0.75(4)             \\ [-0.3cm]
    & 150  & 0.44(4)             \\ [-0.3cm]
    & 151  & 0.11(4)             \\ [-0.3cm]
    & 152  & 0.02(1)       \\  
    \hline
 59 & 131  &   0.053(13)         \\ [-0.3cm]    
    & 132  &   0.104(21)      \\      [-0.3cm]
    & 133  &  0.214(38)         \\   [-0.3cm]
    & 134  &  0.362(36)            \\[-0.3cm]
    & 135  &  0.69(10)            \\ [-0.3cm]
   &136 &1.09(11)             \\     [-0.3cm]
   &137 &1.41(10) \\		     [-0.3cm]
   &138 &1.66(12) \\		     [-0.3cm]
   &139 &1.77(11) \\[-0.3cm]
   &140 &1.73(14) \\[-0.3cm]
   &141 &1.69(8)  \\[-0.3cm]
   &142 &1.59(8)  \\[-0.3cm]
   &143 &1.50(9)  \\[-0.3cm]
   &144 &1.43(7)  \\[-0.3cm]
   &145 &1.26(9)  \\[-0.3cm]
   &146 &1.13(12) \\[-0.3cm]
   &147 &0.94(10) \\[-0.3cm]
   &148 &0.84(7)  \\[-0.3cm]
   &149 &0.88(7)  \\[-0.3cm]
   &150 &0.73(7)  \\[-0.3cm]
   &151 &0.54(3)  \\[-0.3cm]
   &152 &0.27(2)  \\[-0.3cm]
   &153 &0.066(19)\\[-0.3cm]
   &154 &0.012(6) \\
   \hline
 60  &135 & 0.12(1)\\[-0.3cm]
     &136 & 0.26(2)\\[-0.3cm]
     &137 & 0.48(3)\\[-0.3cm]
     &138 & 0.71(9)\\[-0.3cm]
     &139 & 0.99(6)\\[-0.3cm]
     &140 & 1.15(6)\\[-0.3cm]
     &141 & 1.34(7)\\[-0.3cm]
     &142 & 1.37(5)\\[-0.3cm]
     &143 & 1.28(9)\\[-0.3cm]
     &144 & 1.12(7)\\[-0.3cm]
     &145 & 1.02(4)\\[-0.3cm]
     &146 & 0.92(5)\\[-0.3cm]
     &147 & 0.77(4)\\[-0.3cm]
     &148 & 0.63(4)\\[-0.3cm]
     &149 & 0.56(2)\\[-0.3cm]
     &150 & 0.58(3)\\[-0.3cm]
     &151 & 0.48(3)\\[-0.3cm]
     &152 & 0.39(2)\\[-0.3cm]
      &153 &0.22(3) \\     [-0.3cm]
        &154 &0.11(3)  \\  [-0.3cm]
        &155 &0.023(6) \\ 
	\hline
    61  &137 & 0.086(26)\\ [-0.3cm]
        &138 & 0.20(3) \\[-0.3cm]
        &139 & 0.32(6) \\[-0.3cm]
        &140 & 0.50(4) \\[-0.3cm]
        &141 & 0.73(7) \\[-0.3cm]
       &142  & 0.86(9) \\[-0.3cm]
       &143  & 1.01(12)\\[-0.3cm]
       &144  & 1.02(15)\\[-0.3cm]
       &145  & 0.88(5) \\[-0.3cm]
       &146  & 0.84(4) \\[-0.3cm]
       &147  & 0.75(5) \\[-0.3cm]
       &148  & 0.70(3) \\[-0.3cm]
       &149  & 0.58(8) \\[-0.3cm]
       &150  & 0.51(12)\\[-0.3cm]
       &151  & 0.45(7) \\[-0.3cm]
       &152  & 0.33(8) \\[-0.3cm]
      &153  & 0.26(3)  \\[-0.3cm]
      &154  & 0.186(22)\\[-0.3cm]
      &155  & 0.117(12)\\[-0.3cm]
      &156 &  0.062(10)\\
      \hline
   62 &142  &0.33(6)  \\  [-0.3cm]
      &143  &0.47(8)   \\ [-0.3cm]
      &144  &0.65(10)  \\ [-0.3cm]
      &145  &0.70(7)   \\ [-0.3cm]
      &146  &0.75(7)   \\ [-0.3cm]
      &147  &0.700(28) \\[-0.3cm]
      &148  &0.645(45) \\[-0.3cm]
      &149  &0.572(40) \\[-0.3cm]
      &150  &0.51(8)   \\[-0.3cm]
      &151  &0.436(22) \\[-0.3cm]
      &152  &0.35(3)   \\  [-0.3cm]
      &153  &0.267(11)  \\ [-0.3cm]
      &154  &0.203(16)  \\ [-0.3cm]
      &155   & 0.15(2)   \\  [-0.3cm]      
      &156   & 0.10(2)   \\  [-0.3cm]      
      &157   & 0.054(15)      \\ [-0.3cm]
      &158   & 0.029(7)       \\ 
      \hline
 63 &144  &   0.26(4)         \\ [-0.3cm]
   &145  & 0.40(5)           \\	 [-0.3cm]
   &146  &   0.514(26)         \\[-0.3cm]
   &147 &    0.535(21)         \\[-0.3cm]
   &148 &    0.560(34)         \\[-0.3cm]
    &149 &   0.567(34)         \\[-0.3cm]
    &150 &   0.558(33)         \\[-0.3cm]
   &151 &    0.527(42)         \\[-0.3cm]
   &152 &    0.491(24)         \\[-0.3cm]
   &153 &    0.387(23)      \\	 [-0.3cm]
   &154 &    0.315(16)         \\[-0.3cm]
   &155 &    0.221(15)         \\[-0.3cm]
    &156 &   0.154(11)         \\[-0.3cm]
   &157 &    0.108(11)         \\[-0.3cm]
   &158 &    0.064(23)         \\[-0.3cm]
   &159 &    0.044(15)         \\[-0.3cm]
    &160 &  0.019(11)          \\
\hline
 64 &147 & 0.26(4)             \\[-0.3cm]
           &148 & 0.34(4)      \\[-0.3cm]
           &149 & 0.429(34)    \\[-0.3cm]
          &150 &  0.454(45)    \\[-0.3cm]
         &151 &   0.47(12)     \\[-0.3cm]
        &152 &   0.476(24)     \\[-0.3cm]
        &153  &   0.45(5)      \\[-0.3cm]
        &154 &   0.34(5)       \\[-0.3cm]
        &155 &   0.29(3)       \\[-0.3cm]
        &156 &   0.24(3)       \\[-0.3cm]
        &157 &    0.17(2)      \\[-0.3cm]
        &158 &  0.12(2)        \\[-0.3cm]
        &159 &    0.077(14)    \\[-0.3cm]
        &160 &   0.052(7)   \\	 [-0.3cm]
        &161 &   0.028(5)   \\[-0.3cm]
        &162 &   0.012(7) \\ 
\hline		      
\caption { Fission fragment isotopic cross sections measured 
in the present work. Statistical uncertainties are given on the last
significant numbers, excluding the 10\%  systematical uncertainties. }
\end{longtable}

%% file: art_fission.bbl
\newpage

\begin{thebibliography}{99}


\bibitem{Frie} G. Friedlander, L. Friedman, B. Gordon,and L. Yaffe,
 Phys. Rev. 129 (1963) 1809
\bibitem{Klap} R. Klapisch, Annual Review of Nuclear Sciences, 
Vol 19 (1969)  33  \\
and R. Klapisch, J. Chaumont, C. Philippe, I. Amarel, R. Fergeau, R. Salome
and R. Bernas, Nucl. Inst. Methods 53 (1967) 216 
\bibitem{Beli} B.N. Belyaev, V.D. Domkin, Yu.G. Korobulin, L. N. Androneko
and G. E. Solyakin,
Nucl. Phys. A348 (1980) 479
\bibitem{Prok} A. V. Prokovief, 
Nucl. Inst. Methods A 463 (2001) 557
\bibitem{Glor} M. Gloris, R. Michel, F. Sudbrock, U. Herpers,
P. Malmborg, B. Holmqvist,
Nucl. Instrum. and Methods  A 463 (2001) 593 
\bibitem{Tita} Yu. E. Titarenko, O. V. Shvedov, M. M. Igumnov, S. G.
Mashnik, E. I. Karpikhin, V. D. Kazaritsky, V. F. Batyaev,
A. B. Koldobsky, V. M. Zhivun, A. N. Sosnin, R. E. Prael, M. B. Chadwick, T.
A. Gabriel and M. Blann.
Nucl. Instrum. and Methods  A 414 (1998) 73
\bibitem{Bern} M. Bernas, C. Engelmann, P. Armbruster, S. Czajkowski,
F. Ameil, C. B\"ockstiegel, Ph. Dessagne, C. Donzaud, H. Geissel,
A. Heinz, Z. Janas, C. Kozhuharov, C. Mieh\'e, G. M\"unzenberg,
M. Pf\"utzner,
W. Schwab, C. St\'ephan, K. S\"ummerer, L. Tassan-Got and B. Voss,
Phys. Lett B 415 (1997) 111
\bibitem{Enge} Ch. Engelmann, F. Ameil, P. Armbruster, M. Bernas 
S. Czajkowski, Ph. Dessagne, C. Donzaud, H. Geissel, A. Heinz,Z. Janas,
C. Kozhuharov, Ch. Mieh\'e, G. M\"unzenberg, M. Pf\"utzner, C. R\"ohl,
W. Schwab, C. St\'ephan, K. S\"ummerer, L. Tassan-Got and B. Voss,
 Zeit. Phys. A352 (1995) 351
\bibitem{Rubb}  F. Carminati, R. Klapisch, J. P. Revol, Ch. Roche,
 J. A. Rubio, C. Rubbia. CERN Report  CERN/AT/93-47(ET), 1993.
\bibitem{pate}P. Armbruster, H. Geissel, G. M\"unzenberg, (GSI)
Darmstadt, M. Bernas, (IPN) Orsay\\
Patentschrift Deutsches Patentamt P 44 10 587.8.33, GSI Darmstadt 
27.03.1994\\
Brevet de l' Institut national de la propri\'et\'e industrielle,
Paris, France 95 03 499, GSI Darmstadt 24.03.1995  
\bibitem{Must} B. Mustapha, PhD Thesis Paris XI University (1999)
IPNO-T-99-05
\bibitem{Rejm} F. Rejmund, B. Mustapha, P. Armbruster, J. Benlliure,
M.~Bernas, A. Boudard, J. P. Dufour, T. Enqvist, R. Legrain, S. Leray,
K.-H. Schmidt,  C. St\'ephan, J. Taieb, L. Tassan-Got and C. Volant,
Nucl. Phys. A683 (2001) 481
\bibitem{Benl1} J. Benlliure, P. Armbruster, M. Bernas, A. Boudard,
J.-P.~Dufour, T. Enqvist, R. Legrain, S. Leray, B. Mustapha,
F. Rejmund, K.-H. Schmidt, C. St\'ephan, L. Tassan-Got and C. Volant,
Nucl. Phys. A683 (2001) 513
\bibitem{Wlaz} W. Wlazlo, T. Enqvist, P. Armbruster J. Benlliure,
M. Bernas, A. Boudard, S. Czajkowski, R. Legrain, S. Leray,
B. Mustapha, M. Pravikoff, F. Rejmund, K.-H. Schmidt, C. St\'ephan,
J. Taieb, L. Tassan-Got and C. Volant,
Phys. Rev. Lett. 84 (25)  (2000) 5736
\bibitem{Enqv1} T. Enqvist, W. Wlazlo, P. Armbruster, J. Benlliure,
M. Bernas, A. Boudard, S. Czajkowski, R. Legrain, S. Leray,
B. Mustapha, M. Pravikoff, F. Rejmund, K.-H. Schmidt, C. St\'ephan,
J. Taieb, L. Tassan-Got and C. Volant,  
Nucl. Phys. A686 (2001) 481
\bibitem{Enqv2} T. Enqvist, P. Armbruster, J. Benlliure, M. Bernas,
A. Boudard, S. Czajkowski, R. Legrain, S. Leray, B. Mustapha,
M. Pravikoff, F. Rejmund, K.-H. Schmidt, C. St\'ephan, J. Taieb,
L. Tassan-Got, F. Viv\`es, C. Volant and W. Wlazlo,
Nucl. Phys. A 703 (2002) 435
\bibitem{Taie} J. Taieb, PhD Thesis Paris XI University (2000)
IPNO-T-00-10 and \\
J. Taieb, P. Armbruster, J. Benlliure, M. Bernas, A. Boudard,
E. Casajeros, S. Czajkowski, T. Enqvist, R. Legrain, S. Leray,
B. Mustapha, M. Pravikoff, F. Rejmund, K.-H. Schmidt, C. St\'ephan,
L. Tassan-Got, C. Volant and W. Wlazlo,
submitted to Nucl. Phys. A
\bibitem{Ruiz} E. Casarejos Ruiz, PhD Thesis, 
University de Santiago de Compostela (2001), GSI Rep. Diss.2002-04
\bibitem{Armb} P. Armbruster, M. Bernas, S. Czajkowski, H. Geissel,
T. Aumann, Ph. Dessagne, C. Donzaud, E. Hanelt, A. Heinz, M. Hesse,
C. Kozhuharov, Ch. Miehe, G. M\"unzenberg, M. Pf\"utzner,
K.-H. Schmidt, W. Schwab, C. St\'ephan, K. S\"ummerer, L. Tassan-Got
and B. Voss,
Z. Phys. A355 (1996) 191
\bibitem{Donz} C. Donzaud, S. Czajkowski, P. Armbruster, M. Bernas, C.
B\"ockstiegel, Ph. Dessagne, H. Geissel, E. Hanelt, A. Heinz, C.
Kozhuharov, Ch. Mieh\'e, G. M\"unzenberg, M. Pf\"utzner, W. Schwab, C.
St\'ephan, K. S\"ummerer, L. Tassan-Got, B. Voss,
Eur. Phys. J. A1, 407-426 (1998)
\bibitem{Schw} W. Schwab, M. Bernas, P. Armbruster, S. Czajkowski,
Ph.~Dessagne, C. Donzaud, H. Geissel, A. Heinz, C. Kozhuharov,
C. Mieh\'e, G. M\"unzenberg, M. Pf\"utzner, C. St\'ephan, K. S\"ummerer
, L. Tassan-Got and B. Voss,
Eur. Phys. J. A2 (1998) 179
\bibitem{Enqv3} T. Enqvist, J. Benlliure, F. Farget, K.-H. Schmidt, P.
Armbruster, M. Bernas, L. Tassan-Got, A. Boudard, R. Legrain,
C. Volant,
C. B\"ockstiegel, M. de Jong and J.-P. Dufour,
Nucl. Phys. A 658 (1999) 47
\bibitem{Enge2} Ch. Engelmann, Thesis Universit\"at T\"ubingen 
(1998), GSI Rep. Diss. 1998-15
\bibitem{LAHE} R. E. Prael, H. Liechtenstein, Los Alamos-UR-89-3014,
1989
\bibitem{Gaim}  J.-J. Gaimard and K.-H. Schmidt,
Nucl. Phys. A 531 (1991) 709.
\bibitem{Jung} A. R: Junghans, M. de Jong, H.-G. Clerc, A. V. Ignatyuk,
G.A. Kudyaev, K.-H. Schmidt,
Nucl. Phys. A 629 (1998) 635
\bibitem{Benl}  J. Benlliure, A. Grewe, M. de Jong, K.-H. Schmidt and
S. Zhdanov, 
Nucl. Phys. A 628 (1998) 458
\bibitem{Boud} A. Boudard, J. Cugnon, S. Leray and C. Volant,
Phys. Rev. C 66 (2002) 044615
\bibitem{Jura} B. Jurado, K.-H. Schmidt and K.-H. Behr,
Nucl. Inst. Methods A483 (2002) 603 
\bibitem{Ches} P. Chesny,A. Forges, J.M. Gheller, G. Guiller,
P. Pariset, L. Tassan-Got, P. Armbruster, H.-H. Behr, J. Benlliure,
K. Burkard,  A. Br\"unle, T. Enqvist, F. Farget, K.-H. Schmidt,
GSI Annual Report 1996, GSI 1997-1 p. 190.
\bibitem{Sche} C. Scheidenberger, Th. St\"ohlker, W. E. Meyerhof, H.
Geissel, P. H. Mokler and B. Blank, 
Nucl. Inst. and Methods 142 (1998) 441
\bibitem{Geis} H. Geissel, P. Armbruster, K.-H. Behr, A. Br\"unle,
K. Burkard, M. Chen, H.Folger, B. Franczak, H. Keller, O. Klepper,
B. Langenbeck, F.~Nickel, E. Pfeng, M. Pf\"utzner, E. Roeckl,
K. Rykaczewski, I. Schall, D. Schart, C. Scheidenberger, K.-H. Schmidt,
A. Schr\"oter, T. Schwab, K. S\"ummerer, M. Weber, G. M\"unzenberg,
T. Brohm, H.-G. Clerc, M. Fauerbach, J.-J. Gaimard, A. Grewe,
E. Hanelt, B. Kn\"odler, M. Steiner, B. Voss, J. Weckenmann,
C. Ziegler, A. Magel, H. Wollnik, J.-P. Dufour, Y. Fujita,
D. J. Vieira, B. Sherill,
Nucl. Inst. and Methods B 70 (1992) 286
\bibitem{Pfut} M. Pf\"utzner, H. Geissel, G. M\"unzenberg, F. Nickel,
C. Scheidenberger, K.-H. Schmidt, K. S\"ummerer, T. Brohm, B. Voss
and H. Bichsel,
Nucl. Inst. and Methods B86 (1994) 213
\bibitem{Voss}  B. Voss, T. Brohm, H.-G. Clerc, A. Grewe, E. Hanelt, A.
Heinz, M. de Jong, A. Junghans, W. Morawek, C. R\"ohl,
S. Steinh\"auser, C. Zeigler, K.-H. Schmidt, K.-H. Behr, H. Geissel,
F. Nickel, C. Scheidenberger, K. S\"ummerer and A. Magel,
Nucl. Inst. and Methods A 364 (1995) 150
\bibitem{Boch} B.A. Bochagov, V.S. Bychenkov, V.D. Dmitriev,
S.P. Mal'tsev, A.I. Obukhov, N.A. Perfilov and O. E. Shigaev,
Sov. J. Nucl. Phys. 28 (2) (1978) 291
\bibitem{Pere} J. Benlliure, J. Pereira-Conca and K.-H. Schmidt,
Nucl. Instrum. and Methods A478 (2002) 493
\bibitem{Schw3} Th. Schwab, GSI report 91-10 (1991)
\bibitem{Wilk}  B.D. Wilkins, E.P. Steinberg, R.R. Chasman,
Phys. Rev. C14 (1976) 1832
\bibitem {Bock} C. B\"ockstiegel, S. Steinh\"auser, J. Benlliure, H.-G.
Clerc, A. Grewe, A. Heinz, M. de Jong, A.R. Junghans, J. M\"uller,
K.-H. Schmidt,
Phys. Lett. B 398 (1997) 259
\bibitem{Pere2} J. Pereira-Conca, PhD Thesis, in progress,
Univ. Santiago de Compostela, private communication.
\bibitem{Paol} P. Napolitani, P. Armbruster, M. Bernas and
L. Tassan-Got,
to be submitted to Nucl. Phys. A
\bibitem{Karo} P.J. Karol,
Phys. Rev. C 11 (1975) 1203  
\bibitem{Chau} J. Chaumont, PhD Thesis Univ. of Paris XI (1970)\\
csnwww.in2p3.fr/AMDC/experimental/th-chaumont.pdf
\bibitem{Ricc}
 M. V. Ricciardi, K.-H. Schmidt, J. Benlliure, T. Enqvist, 
 F. Ameil,  P. Armbruster, M. Bernas, A. Boudard, S. Czajkowski, 
 R. Legrain, S. Leray, B. Mustapha, M. Pravikoff, C. Stephan, 
 L. Tassan-Got, C. Volant, proceedings of the XXXIX Int. Winter Meeting on
 Nucl. Phys., Bormio, Italy (2001). \\
 M. V. Ricciardi, PhD Thesis in progress, GSI-Darmstadt.
\bibitem{Vais} L.A. Vaishnene, L.N. Andronenko, G.G. Kovshevny, A.A.
Kotov, G.E. Solyakin and W. Neubert,
Z. Phys. A 302 (1981) 143
\bibitem{Jura2}  B. Jurado, PhD Thesis, Compostela  (2002), 
Diss. GSI Rep. 2002-10
\bibitem{Broh}  T. Brohm, PhD thesis, TU-Darmstadt, (1994) 
\bibitem{Armb2} P. Armbruster,
Nucl. Phys. A140, (1970) 385  
\bibitem{Sist} K. Sistemich, P. Armbruster, J. Eidens and E. Roeckl,
 Nucl. Phys. A139 (1969) 289  
\bibitem{Weng} H. U. Wenger, F. Botta, R. Chawla, M. Daum, D. Gavillet,
Z.~Kopajtic, D. Ledergerber, H. P. Linder and S. R\"ollin,
Annals of Nucl. Energy 26 (1999) 141
\bibitem{Silb} R. Silberberg, C. H. Tsao, A. F. Barghouty, 
Astrophys. J. 501 (1998) 911 and (Tsao@nrl.navy.mil)
Phys. Report 191, 351 (1990)
\bibitem{LAHET} L. Dresner, ORNL-TM-7882, Oak Ridge National Laboratory, 
1981.
\bibitem{Mulg} S.I. Mulgin, K.-H. Schmidt, A. Grewe, S. V. Zhdanov,
Nucl. Phys. A 640 (1998) 375.
\bibitem{Broz} U. Brosa, S. Grossmann, A. M\"uller,
Phys. Rep. 197 (1990) 167 
\bibitem{Igna} A.V. Ignatyuk, G. N. Smirenkin and A. S. Tishin,
Sov. J. Nucl. Phys. 21 (1975) 255.
\bibitem{Gali} D. Hilscher, U. Jahnke, F. Goldenbaum, L. Pienkowski,
J. Galin and B. Lott,
Nucl. Inst. and Methods A414 (1998) 100
\end{thebibliography}
